\documentclass[longauth]{aa}

\usepackage{graphicx}
\usepackage{natbib}
\usepackage{txfonts}
\usepackage{array}
\usepackage{multirow}
\usepackage{longtable}
\usepackage{moresize}
\usepackage{gensymb}
\usepackage[dvipsnames]{xcolor}
\usepackage{lscape}
\usepackage[version=4]{mhchem}
\usepackage{caption}
\usepackage[skip=0.5ex]{subcaption}
\usepackage{lipsum}
\usepackage{silence}
\usepackage{tabularx}
\usepackage{orcidlink}

\newcommand{\kms}{km s$^{-1}$}
\newcommand{\msun}{M$_{\odot}$}
\newcommand{\lsun}{L$_{\odot}$}
\newcommand{\lbol}{L$_{\mathrm{bol}}$}
\newcommand{\um}{$\mu$m}

\newcommand{\hi}{H{\sc i}}
\newcommand{\hii}{H{\sc ii}}
\newcommand{\chiiicn}{CH$_3$CN (12--11)}
\newcommand{\chiiitcn}{CH$_3^{13}$CN (12--11)}
\newcommand{\hiico}{H$_2$CO}
\newcommand{\chiiioh}{CH$_3$OH}
\newcommand{\dcn}{DCN (3 -- 2)}
\newcommand{\ocs}{OCS (19 -- 18)}
\newcommand{\hciiin}{HC$_3$N(24 -- 23)}
\newcommand{\so}{SO (6$_5$ -- 5$_4$)}
\newcommand{\sio}{SiO (5 -- 4)}
\newcommand{\tco}{$^{13}$CO (2--1)}
\newcommand{\ceo}{C$^{18}$O (2--1)}
\newcommand{\vlsr}{$v_{\rm LSR}$}
\newcommand{\gcmtwo}{g cm$^{-2}$}
\newcommand{\asec}{$^{\prime \prime}$}
\newcommand{\deltacirc}{$\Delta^{\rm circ} _{5\sigma}$}
\newcommand{\qhull}{$Q^{\rm hull} _{5\sigma}$}
\newcommand{\fres}{\textit{\sc 7m+tm2+tm1}}
\newcommand{\ires}{\textit{\sc 7m+tm2}}
\newcommand{\lres}{\textit{\sc 7m}}
\newcommand{\n}{"near"}
\newcommand{\f}{"far"}
\newcommand{\cpear}{$r_P$}

\newcommand{\nhigal}{915}
\newcommand{\nrms}{98}
\newcommand{\nall}{1013}
\newcommand{\nvlim}{77}
\newcommand{\ndnf}{24}
\newcommand{\ndfn}{39}
\newcommand{\ndfiveh}{95}
\newcommand{\nkdadefnear}{7}

\newcommand{\samplen}{538}
\newcommand{\samplef}{479}

\begin{document}

%\WarningsOff[hyperref]

\title{ALMAGAL I. The ALMA evolutionary study of high-mass protocluster formation in the Galaxy} 
\subtitle{Presentation of the survey and early results} 

%\include{authorslist}
% Start authors list

\author{
\orcidlink{0000-0002-9826-7525}S. Molinari\inst{\ref{iaps}} \and 
\orcidlink{0000-0003-2141-5689}P. Schilke\inst{\ref{u-koln}} \and 
\orcidlink{0000-0002-6073-9320}C. Battersby\inst{\ref{u-conn}} \and
\orcidlink{0000-0002-3412-4306}P.T.P. Ho\inst{\ref{asiaa}} \and
%
% Steering Group
\orcidlink{0000-0002-3078-9482}\'A.~S\'anchez-Monge\inst{\ref{icecsic}, \ref{ieec}} \and
\orcidlink{0000-0003-1665-6402}A. Traficante\inst{\ref{iaps}}\and
B.~Jones\inst{\ref{u-koln}} \and
\orcidlink{0000-0003-3315-5626}M.~T.~Beltr\'an\inst{\ref{arcetri}} \and
\orcidlink{0000-0002-1700-090X}H. Beuther\inst{\ref{mpia}}
\orcidlink{0000-0001-8509-1818}G.~A.~Fuller\inst{\ref{u-man}, \ref{u-koln}} \and
\orcidlink{0000-0003-2384-6589}Q.~Zhang\inst{\ref{cfa}} \and
\orcidlink{0000-0002-0560-3172}R.S. Klessen\inst{\ref{u-hei},\ref{ralf-2},\ref{cfa},\ref{cfa-rad}}\and
\orcidlink{0000-0001-6941-7638}S.~Walch\inst{\ref{u-koln}, \ref{datacologne}} \and
\orcidlink{0000-0002-0675-276X}Y.-W. Tang\inst{\ref{asiaa}}\and
%
% SG and TechWG more active people
%
\orcidlink{0000-0002-3597-7263}M.~Benedettini\inst{\ref{iaps}} \and
\orcidlink{0000-0002-9120-5890}D.~Elia\inst{\ref{iaps}} \and
\orcidlink{0000-0001-8239-8304}A.~Coletta\inst{\ref{iaps},\ref{sapienza}} \and
\orcidlink{0000-0002-2974-4703}C.~Mininni\inst{\ref{iaps}} \and
\orcidlink{0000-0003-1560-3958}E.~Schisano\inst{\ref{iaps}} \and
\orcidlink{0000-0002-2562-8609}A. Avison\inst{\ref{skao}}\and
\orcidlink{0000-0003-1964-970X}C.~Y.~Law\inst{\ref{arcetri}} \and
\orcidlink{0009-0005-9192-5491}A. Nucara\inst{\ref{iaps},\ref{roma2}}\and
\orcidlink{0000-0002-0294-4465}J.D. Soler\inst{\ref{iaps}}
\orcidlink{0000-0002-4935-2416}, G. Stroud\inst{\ref{u-man}}\and
J.~Wallace\inst{\ref{u-conn}}\and
\orcidlink{0000-0002-3643-5554}M.R.A. Wells\inst{\ref{mpia}}\and 
\orcidlink{0000-0003-4037-5248}A.~Ahmadi\inst{\ref{leiden}} \and
\orcidlink{0000-0002-6558-7653}C.~L.~Brogan\inst{\ref{nraoCH}} \and
\orcidlink{0000-0001-6492-0090}T.~R.~Hunter\inst{\ref{nraoCH}} \and
S.-Y.~Liu\inst{\ref{asiaa}} \and
\orcidlink{0000-0001-7852-1971}S.~Pezzuto\inst{\ref{iaps}} \and
Y.-N.~Su\inst{\ref{asiaa}} \and
B. Zimmermann\inst{\ref{u-koln}}\and
T. Zhang\inst{\ref{u-koln}}\and
F.~Wyrowski\inst{\ref{mpifr}} \and
F. De Angelis\inst{\ref{iaps}}\and
S. Liu\inst{\ref{iaps}}\and
%
% alphabetical order general consortium members
%
\orcidlink{0000-0001-9751-4603}S. D. Clarke\inst{\ref{asiaa}}\and
\orcidlink{0000-0003-0348-3418}F. Fontani\inst{\ref{arcetri},\ref{mpe},\ref{lerma}} \and
\orcidlink{0000-0001-9443-0463}P.D. Klaassen\inst{\ref{roe}}\and
\orcidlink{0000-0003-2777-5861}P. Koch\inst{\ref{asiaa}}\and
\orcidlink{0000-0003-4509-1180}K.~G.~Johnston\inst{\ref{lincoln}} \and
\orcidlink{0000-0001-8060-1890}U. Lebreuilly\inst{\ref{saclay}}\and
T. Liu\inst{\ref{shanghai}}\and
\orcidlink{0000-0001-5748-5166}S. L. Lumsden\inst{\ref{leeds}}\and
\orcidlink{0000-0002-9277-8025}T.~Moeller \inst{\ref{u-koln}} \and
\orcidlink{0000-0002-8517-8881}L. Moscadelli\inst{\ref{arcetri}}\and
\orcidlink{0000-0003-2309-8963}R. Kuiper\inst{\ref{u-duisb}}\and
\orcidlink{0000-0002-0500-4700}D. Lis\inst{\ref{jpl}}\and
\orcidlink{0000-0002-6893-602X}N. Peretto\inst{\ref{u-card}}\and
\orcidlink{0000-0002-5003-4714}S. Pfalzner\inst{\ref{juelich}}\and
\orcidlink{0000-0002-3351-2200}A. J. Rigby\inst{\ref{leeds}}\and
\orcidlink{0000-0002-7125-7685}P.~Sanhueza\inst{\ref{meguro}, \ref{naoj}} \and
\orcidlink{0000-0003-4146-9043}K.~L.~J.~Rygl\inst{\ref{ira}}
\orcidlink{0000-0002-8942-1594}F. van der Tak\inst{\ref{sron}, \ref{u-gron}}\and
H. Zinnecker\inst{\ref{u-auton-chile}}\and
F. Amaral\inst{\ref{u-koln}}\and
\orcidlink{0000-0001-8135-6612}J. Bally\inst{\ref{u-colo}}\and
\orcidlink{0000-0002-9574-8454}L. Bronfman\inst{\ref{u-chile}}\and
\orcidlink{0000-0002-2430-5103}R. Cesaroni\inst{\ref{arcetri}}\and
%V. Chen\inst{\ref{}}\and
K. Goh\inst{\ref{u-koln}}\and
\orcidlink{0000-0003-2684-399X}M. G. Hoare\inst{\ref{leeds}}\and
\orcidlink{0000-0003-0946-4365}P. Hatchfield\inst{\ref{jpl}}\and
\orcidlink{0000-0002-0472-7202}P. Hennebelle\inst{\ref{saclay}}\and
T. Henning\inst{\ref{mpia}}\and
K. T. Kim\inst{\ref{kassi}}\and
W.-J. Kim\inst{\ref{u-koln}} \and 
%Y.-J. Kuan\inst{\ref{}}\and
\orcidlink{0000-0002-7675-3565}L. Maud\inst{\ref{eso}}\and
M. Merello\inst{\ref{u-chile}, \ref{pontificia}}\and
\orcidlink{0000-0001-5431-2294}F. Nakamura\inst{\ref{naoj}}\and
\orcidlink{0000-0002-6482-8945}R. Plume\inst{\ref{u-calg}}\and
S.-L. Qin\inst{\ref{u-yunnan}}\and
\orcidlink{0000-0002-8502-6431}B. Svoboda\inst{\ref{nraoSO}}\and
\orcidlink{0000-0003-1859-3070}L. Testi\inst{\ref{u-bo}}\and
V.S. Veena\inst{\ref{u-koln},\ref{mpifr}}\and
D. Walker\inst{\ref{u-man}}
}

%%%%% Institutes

\institute{
Istituto Nazionale di Astrofisica (INAF)-Istituto di Astrofisica e Planetologia Spaziale, Via Fosso del Cavaliere 100, I-00133 Roma, Italy \label{iaps}
\and 
Physikalisches Institut der Universit\"at zu K\"oln, Z\"ulpicher Str. 77, D-50937 K\"oln, Germany \label{u-koln} 
\and
University of Connecticut, USA\label{u-conn} 
\and
Institute of Astronomy and Astrophysics, Academia Sinica, 11F of ASMAB, AS/NTU No.\ 1, Sec.\ 4, Roosevelt Road, Taipei 10617, Taiwan \label{asiaa}
\and
Institut de Ci\`encies de l'Espai (ICE, CSIC), Campus UAB, Carrer de Can Magrans s/n, E-08193, Bellaterra (Barcelona), Spain\label{icecsic}
\and
Institut d'Estudis Espacials de Catalunya (IEEC), E-08860, Castelldefels (Barcelona), Spain\label{ieec}
\and
Max Planck Institute for Astronomy, K\"onigstuhl 17, 69117 Heidelberg, Germany\label{mpia}
\and
SKA Observatory, Jodrell Bank, Lower Withington, Macclesfield, SK11 9FT, UK\label{skao}
\and
Jodrell Bank Centre for Astrophysics, Department of Physics \& Astronomy, The University of Manchester, Oxford Road, Manchester M13 9PL, UK\label{u-man}
\and
Universit\"{a}t Heidelberg, Zentrum f\"{u}r Astronomie, Institut f\"{u}r Theoretische Astrophysik, Albert-Ueberle-Straße 2, D-69120 Heidelberg, Germany \label{u-hei} 
\and
Universit\"{a}t Heidelberg, Interdisziplin\"{a}res Zentrum f\"{u}r Wissenschaftliches Rechnen, Im Neuenheimer Feld 205, D-69120 Heidelberg, Germany \label{ralf-2} 
\and
Center for Data and Simulation Science, University of Cologne, Germany\label{datacologne}
\and
Harvard-Smithsonian Center for Astrophysics, 60 Garden Street, Cambridge, MA 02138, U.S.A. \label{cfa} 
\and
Radcliffe Institute for Advanced Studies at Harvard University, 10 Garden Street, Cambridge, MA 02138, U.S.A. \label{cfa-rad} 
\and
National Radio Astronomy Observatory, 520 Edgemont Road, Charlottesville VA 22903, USA\label{nraoCH}
\and
Dipartimento di Fisica, Università di Roma Tor Vergata, Via della Ricerca Scientifica 1, I-00133 Roma, Italy\label{roma2}
\and
Leiden Observatory, Leiden University, PO Box 9513, 2300 RA Leiden, The Netherlands\label{leiden}
\and
Max-Planck-Institut f\"ur Radioastronomie, Auf dem H\"ugel 69, D-53121 Bonn, Germany\label{mpifr}
\and
Istituto Nazionale di Astrofisica (INAF), Osservatorio Astrofisico di Arcetri, Largo E. Fermi 5, Florence, Italy \label{arcetri} 
\and
Max-Planck-Institute for Extraterrestrial Physics (MPE), Garching bei M\"unchen, Germany \label{mpe} 
\and
Laboratoire d’\'Etudes du Rayonnement et de la Mati\`ere en Astrophysique et Atmosph\`eres (LERMA), Observatoire de Paris, Meudon, France \label{lerma} 
\and
UK Astronomy Technology Centre, Royal Observatory Edinburgh, Blackford Hill, Edinburgh EH9 3HJ, UK\label{roe}
\and
School of Engineering and Physical Sciences, Isaac Newton Building, University of Lincoln, Brayford Pool, Lincoln, LN6 7TS, UK\label{lincoln}
\and
School of Physics and Astronomy, University of Leeds, Leeds LS2 9JT, UK\label{leeds}
\and
Universit\'e Paris-Saclay, Universit\'e Paris-Cit\'e, CEA, CNRS, AIM, 91191 Gif-sur-Yvette, France\label{saclay}
\and
Shanghai Astronomical Observatory, Chinese Academy of Sciences, 80 Nandan Road, Shanghai 200030, China\label{shanghai}
\and
INAF-Istituto di Radioastronomia \& Italian ALMA Regional Centre, Via P. Gobetti 101, I-40129 Bologna, Italy
\label{ira}
\and
Faculty of Physics, University of Duisburg-Essen, Lotharstraße 1, D-47057 Duisburg, Germany\label{u-duisb}\and
J\"ulich Supercomputing Centre, Forschungszentrum J\"ulich, Wilhelm-Johnen-Straße, J\"ulich, 52428, NRW, Germany\label{juelich}\and
Jet Propulsion Laboratory, California Institute of Technology, 4800 Oak Grove Drive, Pasadena, CA 91109, USA\label{jpl}\and
Department of Earth and Planetary Sciences, Tokyo Institute of Technology, Meguro, Tokyo, 152-8551, Japan\label{meguro}
\and
Cardiff Hub for Astrophysics Research \& Technology, School of Physics \& Astronomy, Cardiff University, Queens Buildings, The Parade, Cardiff CF24 3AA, UK \label{u-card}\and
National Astronomical Observatory of Japan, National Institutes of Natural Sciences, 2-21-1 Osawa, Mitaka, Tokyo 181-8588, Japan\label{naoj}
\and
National Radio Astronomy Observatory, PO Box O, Socorro, NM 87801, USA\label{nraoSO}
\and
SRON Netherlands Institute for Space Research\label{sron}
\and
Kapteyn Astronomical Institute, University of Groningen, Landleven 12, 9747 AD Groningen, The Netherlands\label{u-gron}
\and
\label{u-auton-chile}Universidad Autonoma de Chile, Pedro de Valdivia 425, Providencia, Santiago de Chile, Chile\and
Department of Astrophysical and Planetary Sciences, University of Colorado, Boulder, CO 80389, USA\label{u-colo}\and
Dipartimento di Fisica e Astronomia, Alma Mater Studiorum - Universit\`a di Bologna\label{u-bo}\and
Korea Astronomy and Space Science Institute, 776 Daedeokdae-ro, Yuseong-gu, Daejeon 34055, Republic of Korea\label{kassi}
European Southern Observatory, Karl-Schwarzschild Str.\ 2, 85748 Garching bei M\"unchen, Germany\label{eso}\and
Department of Physics and Astronomy, University of Calgary, 2500 University Drive NW, Calgary, Alberta T2N 1N4, Canada \label{u-calg}\and
\label{u-chile}Departamento de Astronomía, Universidad de Chile, Casilla 36-D, Santiago, Chile
\and
\label{pontificia}Centro de Astro-Ingeniería (AIUC), Pontificia Universidad Católica de Chile, Av. Vicuña Mackena 4860, Macul, Santiago, Chile\and
\label{u-yunnan}Department of Astronomy, Yunnan University, Kunming 650091, People’s Republic of China \and
Dipartimento di Fisica, Sapienza Universit\`a di Roma, Piazzale Aldo Moro 2, I-00185, Rome, Italy \label{sapienza}
}

% End authors list

\titlerunning{ALMAGAL I. Presentation of the survey and early results.}
\authorrunning{S. Molinari et al.}

\abstract 
{A large fraction of stars form in clusters containing high-mass stars, which subsequently influences the local and galaxy-wide environment.} 
{Fundamental questions about the physics responsible for fragmenting molecular parsec-scale clumps into cores of a few thousand astronomical units (au)  are still open, that only a statistically significant investigation with ALMA is able to address; for instance: the identification of the dominant agents  that determine the core demographics, mass, and spatial distribution as a function of the physical properties of the hosting clumps, their evolutionary stage and the different Galactic environments in which they reside. The extent to which fragmentation is driven by clumps dynamics or mass transport in filaments also remains elusive.}
{With the ALMAGAL project, we observed the 1.38~mm continuum and lines toward more than 1000 dense clumps in our Galaxy, with $M\,\geq$\,500\,\msun, $\Sigma\geq 0.1$\,\gcmtwo\ and $d$\,$\leq$ 7.5\,kiloparsec (kpc). Two different combinations of ALMA Compact Array (ACA) and 12-m array setups were used to deliver a minimum resolution of $\sim$1000 au over the entire sample distance range. The sample covers all evolutionary stages from infrared dark clouds (IRDCs) to \hii\ regions from the tip of the Galactic bar  to the outskirts of the Galaxy. With a continuum sensitivity of 0.1 mJy, ALMAGAL enables a complete study of the clump-to-core fragmentation process down to $M\sim 0.3\,${\msun} across the Galaxy. The spectral setup includes several molecular lines to trace the multiscale physics and dynamics of gas, notably CH$_3$CN, H$_2$CO, SiO, CH$_3$OH, DCN, HC$_3$N, and SO, among others.}
{We present an initial overview of the observations and the early science product and results produced in the ALMAGAL Consortium%. A detailed description of the data reduction process and the generation of the data products is given in the companion paper of \cite{SanchezMonge+2025}. Here we present 
, with a first characterization of the morphological properties of the continuum emission detected above 5$\sigma$ in our fields. We used "perimeter-versus-area" and convex\,hull-versus-area metrics to classify the different morphologies. We find that more extended and morphologically complex (significantly departing from circular or generally convex) shapes are found toward clumps that are relatively more evolved and have higher surface densities.}
{ALMAGAL is poised to  serve as a game-changer for a number of specific issues in star formation: clump-to-core fragmentation processes, demographics of cores, core and clump gas chemistry and dynamics, infall and outflow dynamics, and disk detections. Many of these issues will  be covered in the first generation of papers that closely follow on the present publication.
}
%  with similar linear resolution. , mapping the temperature and the local/global infall velocity patterns of the hosting clumps. 
%ALMAGAL publicly accessible data cubes and catalogs will be an invaluable legacy of ALMA, that will allow numerous community follow-up studies.


\maketitle

\section{Introduction}
\label{intro}

Most stars form in clusters \citep{Lada+Lada03, PSZ2010, Adams2010, Adamo2020} and about 50\% of them, including the Sun itself (and the fraction probably was even higher in the early Universe), form in very rich clusters of at least 1000 stars, %that contain 
containing at least one 10~\msun\ star. High-mass stars (M$\geq$8\msun) influence their immediate environment through gravitational (ejection of stars from the cluster, disk truncation, etc.), mechanical (winds, outflows),  radiative interaction (e.g., radiative heating, photo-ionization, and radiation pressure), and, eventually, through their supernovae explosions. Hence, the formation process of low-mass stars and their associated planets in such clusters will 
%be vastly different 
vastly differ from star formation in isolation or %in 
small clusters. However, the latter scope is where most low-mass star formation studies have taken place so far, mostly because such objects are close by and easier to study. However, they are not representative of the most significant fraction of star formation.

High-mass stars dominate the energy input into the ISM through feedback processes, not only locally, but also on galactic scales \citep[see, for example,][]{veilleux2005,Bolatto+2013}. They also enrich the ISM with heavy elements, which, in turn, influences  subsequent events of star formation  \citep{klessen2016}. Yet, their formation processes differ 
%from low-mass stars in significant ways
significantly from low-mass stars: while the Kelvin-Helmholtz timescale of low-mass stars is significantly longer than the time required to assemble them, for any reasonable accretion rate, it is shorter for high-mass stars. Hence, high-mass stars will continue accreting up to and even after reaching the main sequence \cite[e.g.,][]{Maeder2000, KW2006, Zinnecker2007}. The combination of being deeply embedded in dusty cores through most of their evolution, their scarcity and correspondingly high average distance, together with evolution in clusters with many sources close together, make these observations challenging.

Cluster formation involves cores (linear scales $\leq$0.1 pc) forming inside molecular clumps (linear scales of 0.5-1.0 pc), as indeed observed at high spatial resolution at mm wavelengths \citep{Zhang+2015, Beuther+2018, Svoboda+19, Sanhueza+19, Traficante+2023, Avison+2023}. A full understanding of the fragmentation process and its role in allowing high-mass stars to collect material up to their final mass is still elusive. Furthermore, thermal \citet{Jeans1902} mass fragmentation leads to low-mass objects \cite[][]{Rees1976, Larson1985} %which are 
that are less massive than the observed masses of high-mass stars (M$\geq$8\msun). Mechanisms to stop fragmentation and delay collapse, allowing the collection of enough gas in existing fragments to form massive stars, are a possible solution. Turbulent support (e.g., \citealt{mck03}) or magnetic and radiation feedback (e.g., \citealt{Krumholz+2009c,Comm+2011}) have recently been invoked. In this model family (hereafter monolithic collapse), the final stellar mass of the emerging stars is pre-assembled in the cores, and we would expect to find high-mass monolithic prestellar cores. Yet, the search for high-mass protostellar clumps without any sign of star formation has only resulted in very few candidates (e.g., \citealt{Tackenberg+2012, Beuther+2015, Nony+18, Motte+2018}); so, this is unlikely to be a dominant path to high-mass star formation.  

In the more dynamical scenario of competitive accretion, cores compete \citep{Bonnell+2007} for gas from the cloud mass reservoir that is not initially local to the core itself \citep{Klessen2000, BonnellBate2006,  Peters2011, Girichidis2012}, whereas high-mass monolithic prestellar cores should not feasibly exist \citep{Smith+2009}. Infall motions would be dominated by accretion from the cloud onto the core. %And 
Indeed, observations reveal large-scale infall motions in massive star-forming regions, showing that high-mass clumps are not isolated from the cloud mass reservoir and they are shown, instead, to globally accrete while star formation is internally ongoing (e.g., \citealt{Wu+2003, Rygl+2013, Klaassen+2012, He+2015, Traficante+2017, Contreras+2018, Traficante+2020}) also via continuous mass flow along filaments to stars forming at filamentary hubs \citep{Peretto+2013, Chen+2019, Wells+2024}. Theoretical studies have shown that infall motions are crucial both for initiating the formation of high-mass stars and in subsequent evolutionary stages for maintaining accretion flows to increase the stellar mass (e.g., \citealt{Jijina+1996, yor02, Gong+2009, Peters2010, Vazquez2019}). 

To make fundamental progress in our understanding of the formation of high-mass stars and their surrounding clusters, we want to investigate two key issues in a statistically significant way: 1) what  physical processes govern the fragmentation of cluster-forming clumps and how they evolve with time; and 2) how cores gain mass and how  this process is influenced by internal feedback from the cores into the clump gas. 
%\item {\sf \textcolor{red}{This needs to be resolved: Juan talks about cluster-forming clumps, and you talk about cluster forming cores. We should probably somewhere in the introduction bring our definitions of clouds/clumps/cores, and then stick with it throughout the paper. At the moment, there seems to be a mix, which is confusing}}
%To address this, we need to measure the velocity dispersion between the cores, the mass-flow along and onto filaments in the clump, and the mass distribution between large-scale clump, filaments and cores.

So far, $\sim$300 dense and massive clumps have been observed at interferometric spatial resolution in total over fewer than a dozen different observing programs. In particular, early-stage, 70~\um-dark, or IRDC-like targets have received considerable attention. \cite{Svoboda+19} targeted 12 such massive clumps at a resolution of $\sim$3000~au, revealing fragmentation with separations comparable to the thermal Jeans length, as previously proposed by \cite{Palau+2015}. Likewise, ASHES \cite{Sanhueza+19} observed a sample of similar size at slightly lower resolutions ($\sim$4500~au), confirming fragmentation in these early stages of evolution, Jeans-compatible fragmentation lengths, and also suggesting sub-clustering in the distribution of the fragments. \cite{MAnderson+21} observed with ALMA six hub-filament systems in IRDCs at resolutions $\sim$6000~au; combining them with data from 29 clumps previously observed by \cite{Csengeri+2017} offered evidence of clump-fed accretion. A similarly sized sample of IRDCs was recently studied by \cite{Rigby+2024}, displaying gas kinematics that is also consistent with the latter scenario.

At the other end of the evolutionary path, \cite{Beuther+2018} observed 20 relatively more evolved massive clumps with luminosities in excess of $10^4$~\lsun using NOEMA with high resolutions ($\sim$1000~au), revealing various levels of fragmentation, again compatible with thermal Jeans length and typical fragment separations showing no dependence on core masses. More recently, ATOMS \citep{Liu+2020} targeted 146 evolved UC/HC-HII regions, while ASSEMBLE \citep{Xu+2024} observed 11 evolved massive clumps at $\sim$2000 au resolution. Similarly, \cite{Ishihara+2024} observed 30 hot cores at $\sim$1000 au resolution confirming that thermal Jeans fragmentation may be at work. Other surveys have a wider coverage in the evolutionary stage of targeted clumps, with TEMPO \citep{Avison+2023} reporting (with resolutions $\sim$3000--4000~au) Jeans-compatible fragmentation distance only in a fraction of the clumps; or, as in SQUALO \citep{Traficante+2023}, suggesting that the mass and minimum distance between fragments, respectively, increase or decrease with evolution. 

The significance and robustness of these results %is 
are hampered by the relatively low number of targets in each study, as well as the different selection criteria that prevent trustworthy conclusions from studies of composite samples that, in principle, could be assembled. In addition, the %difference 
different frequency setups and different linear resolutions used make a quantitative comparison difficult. Large ALMA projects like ALMA-IMF \citep{Motte+2022} partially overcome these limitations by imaging 15 well-known large star formation complexes (e.g., the W43 Galactic starburst), harbouring regions of very different mass and evolution with extensive spectroscopic coverage, to address the issue of the emergence of the IMF. %These regions however, e.g. the W43 Galactic starburst, are radiatively and possibly dynamically dominated by the feedback from massive ZAMS stars that considerably complicate the interpretation of the results hampering the potential to assess the role of different physical agents in determining the fragmentation outcome in star formation.
%However, these are generally very active regions, such as the W43 Galactic starburst, are radiatively and possibly dynamically dominated by the feedback from young high-mass stars, considerably complicating the interpretation of the results and hampering the assessment of the different physical agent's role in determining the clump fragmentation.
%ALMA-IMF is specifically designed to address the emergence of the IMF by sampling completely entire star-forming complexes.

%Millimeter interferometry has already started to tackle several of these issues by observing relatively small samples of intermediate and high-mass star forming regions with ALMA \citep{Peretto+2013, Brogan+2016, Csengeri+2017, Cesaroni+2017, Zhang+2015, Motte+2018} covering several tens of objects from the early stage InfraRed Dark Clouds (IRDCs) to \hii\ regions. Multiplicity of fragments in resolved clumps are found to vary considerably in different regions \citep{Beuther+2017}, and the notion of a Core Mass Function (CMF) resembling the stellar IMF \citep{Konyves+2010} has been challenged in a very high mass star formation region W43-MM1 \citep{Motte+2018}. Chemical diversity is found in 5 cores in which the G28.34+0.06 P1 clump is resolved \citep{Zhang+2015}, that can be traced to different evolutionary stages, similarly in about 20 cores in NGC6334I \citep{Brogan+2016}. Ultimately, the data obtained so far are not conclusive, due to the lack of a statistically relevant sample of high spatial resolution observations of high-mass clumps.

Answering the fundamental questions outlined above requires  a number of crucial observables to be measured in a statistically significant way and compared to predictions from numerical simulations over a wide variety of evolutionary stages and Galactic environments: i) the spatial distribution of dense cores as a function of mass within dense clumps; ii) their fragmentation properties, such as average distances and their relation to Jeans masses or filament fragmentation scales; iii) their evolution both in morphology and in total number; iv) the temperature and density distribution of gas in the clumps; v) the velocity field of the clumps gas and the presence of global infall motions; and vi) the dynamical state of the compact fragments.

\begin{figure}[h]
\begin{center}
\includegraphics[width=0.43\textwidth]{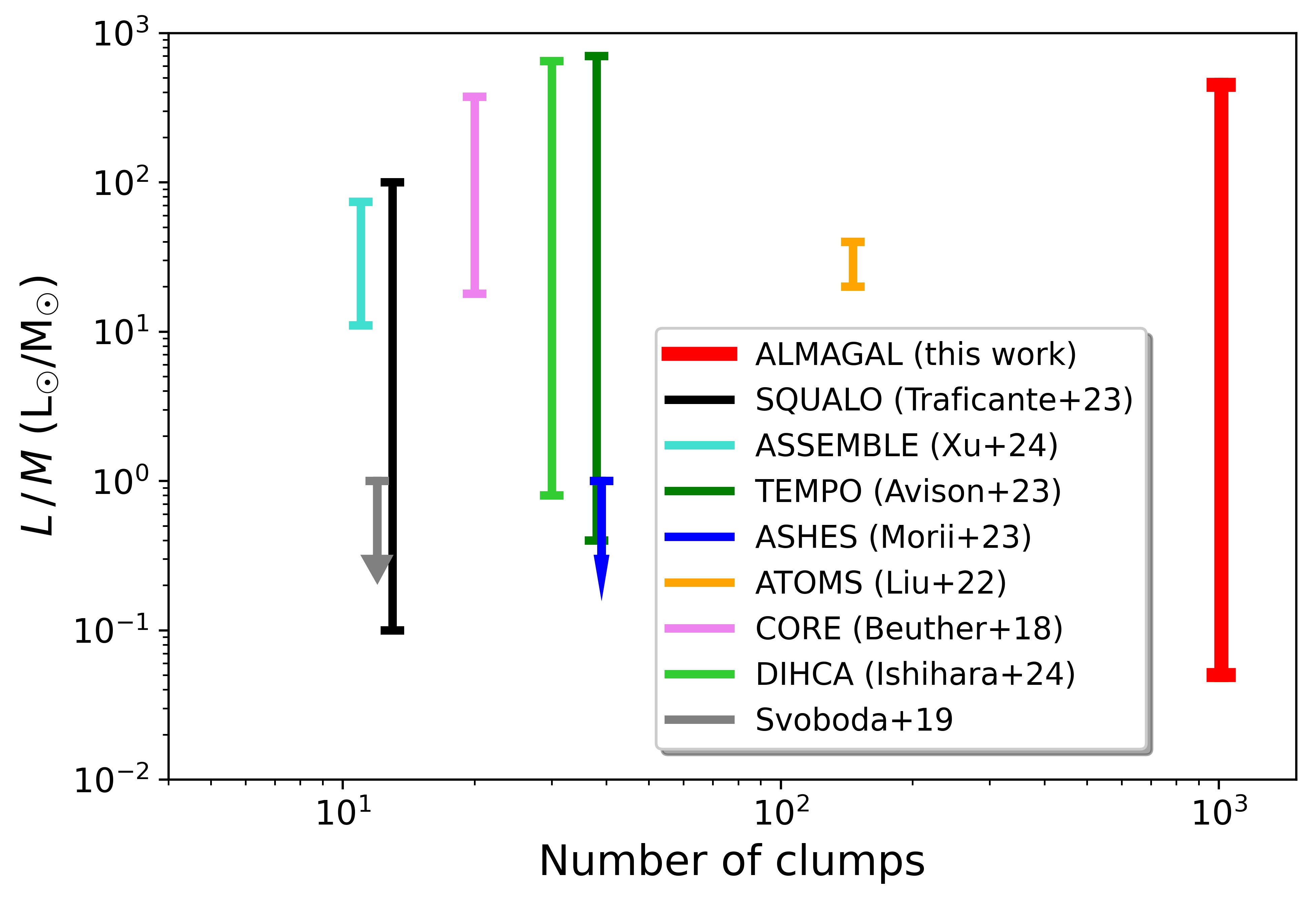} 
\includegraphics[width=0.43\textwidth]{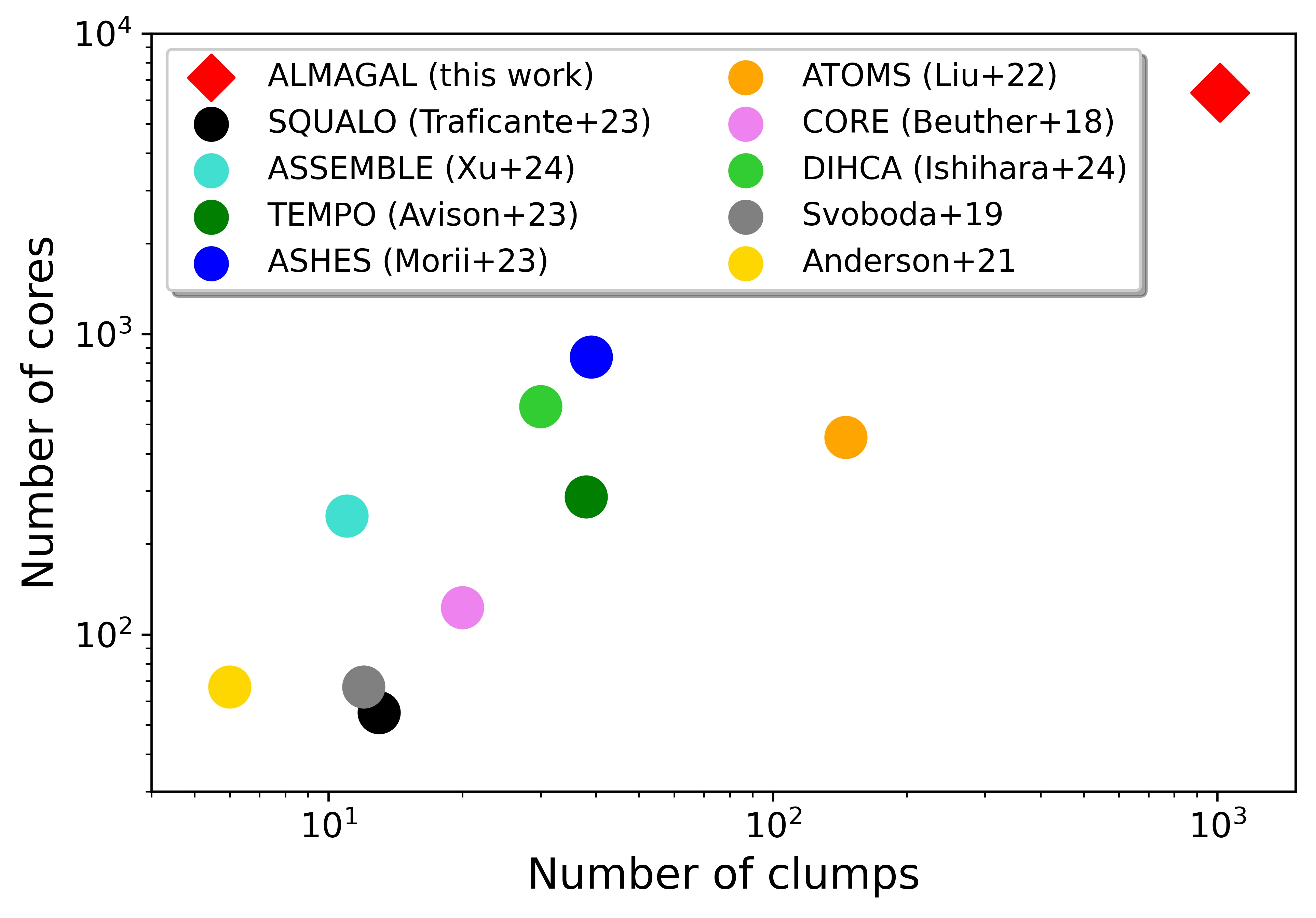} 
\includegraphics[width=0.43\textwidth]{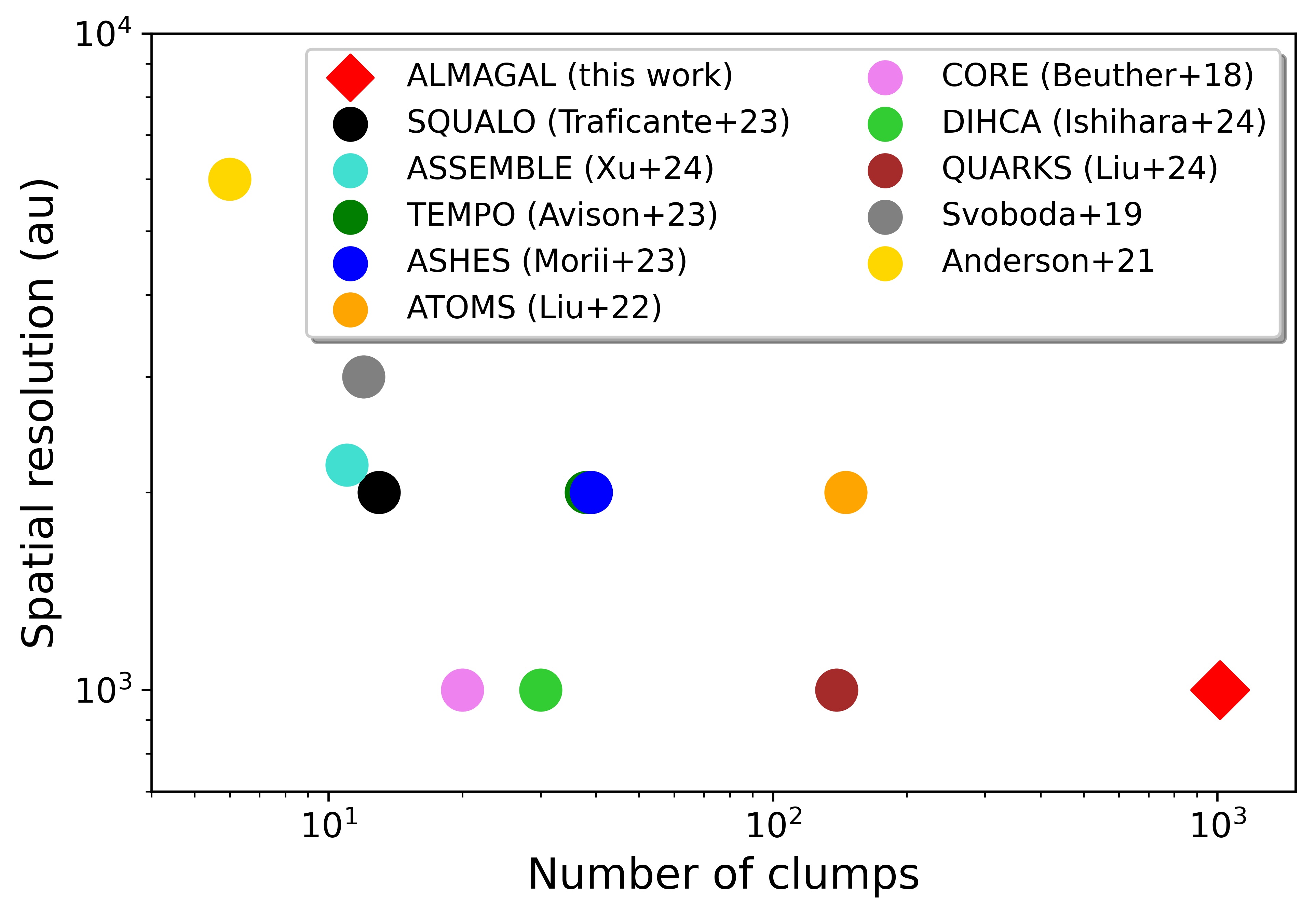} 
\caption{ALMAGAL in the context of other interferometric surveys of star-forming regions in terms of number of target clumps vs. evolutionary stage (top), number of detected fragments (middle, from \cite{Coletta+2025}), and spatial resolution achieved (bottom). The panels show surveys for which the necessary information was immediately available in the papers and for which observations consist of single pointings or relatively small mosaics. For the latter reason, ALMA-IMF is not included in the figure as the program does relatively large maps of 15 star-forming complexes.}
\label{sample_comparison}
\end{center}
\end{figure}

The ALMAGAL Large Program was specifically designed to deliver this key science by mapping more than 1000 intermediate and high-mass dense clumps with ALMA in band 6 (Sect. \ref{sample}) at a spatial resolution of 1000 au. Figure~\ref{sample_comparison} shows that ALMAGAL is a game-changer with respect to coupling the ability to map the variance of physical, evolutionary, and environmental conditions of the targets (Fig. \ref{sample_fig}) with an unprecedented resolution and statistical significance. 

%{\sf \textcolor{red}{We are missing a paragraph: This paper is structured as follows .... or: We start with a discussion of ... followed by ...}}

The paper is structured as follows. In Sect. \ref{obs_setup}, we summarize the characteristics of the ALMAGAL observations, deferring a much more thorough discussion to the companion paper by \cite{SanchezMonge+2025}. In Sect. \ref{sample}, we discuss in detail the properties of the observed clumps and the process of revision of source distances following the ALMAGAL observations and the homogeneization of the way in which physical parameters (e.g., masses and luminosities) are derived. In Sect. \ref{continuum}, we present a characterization of the continuum emission detected with ALMA in the context of clump-integrated properties (Sect. \ref{alma_higal}) and in terms of the morphology of the emission (Sect. \ref{cont_morph}). In Sect. \ref{outlook}, we outline the potential of the ALMAGAL science in a number of areas, anticipating the dedicated papers that are anticipated in the near future. Finally, in Sect. \ref{concl}, we summarize our conclusions.

\section{Observational setup}
\label{obs_setup}

The ALMA Band 6 217 to 220 GHz frequency range is ideally suited to provide access to the J=2-1 lines of $^{13}$CO and C$^{18}$O (for column density and velocity structure), \sio\  (to trace outflows), three \hiico\ lines (temperature and density, infall and outflow tracers), \chiiioh\ (temperature estimates), the K-ladder of the \chiiicn\ and \chiiitcn\ lines (temperature, velocity structure), and several other species (\dcn, \ocs, \hciiin\, etc.). The correlator was configured to produce two 1.875-GHz wide windows, providing  $\sim1.4$\,\kms\ resolution, plus two higher resolution windows with a $\sim0.34$\,\kms\ resolution. To be able to address our science goals, it is imperative to recover signal from the minimum target 1000~au scale up to the clump-size scale. To achieve this goal in an optimal way, given the large distance range of our sources, all the targets have been observed in single-pointing, with the ACA plus two configurations of the 12-m array with two different antenna configuration combinations, depending on the source distances; the C-5/C-2/ACA setup was chosen for sources with d<4.7 kpc (hereafter the \n\  subsample, amounting to \samplen\ objects), while the C-6/C-3/ACA setup was adopted for sources with d>4.7 kpc (the \f\ subsample, with \samplef\  objects). In band 6, these combinations provide maximum angular resolutions of 0.13\arcsec (C-6 for the \f\ subsample) and 0.24\arcsec (C-5 for the \n\ subsample), along with a largest recoverable scale of about 30\arcsec. %The "near" and "far" subsamples are nearly equal in size (546 "near" and 471 "far" sources). 
More details about the observations setup are given in the companion paper \citep{SanchezMonge+2025}. 

We will refer to as "\lres" the data taken with the ACA, "TM2" the data taken in short baseline (hence lower resolution) configurations of the 12-m array (C-2 for the \n\ sample, and C-3 for the \f\ sample), and "TM1" the data taken in long baseline (hence higher resolution) configurations (C-5 for the \n\ sample, and C-6 for the \f\ sample). A fundamental step in data processing was the production of continuum images and spectral cubes by jointly deconvolving the combined data taken with ACA \lres\ array and the different 12-m array configurations. In a first set of products, we combined ACA and the TM2 into what is called here and in all ALMAGAL papers the \ires\ products. In a subsequent step,  the TM1 data were also combined to obtain the full resolution \fres\ products. 

We required a 0.1~mJy continuum sensitivity for all targets irrespective of clump properties, which should (in principle) allow for detections at the 3$\sigma$ level of compact cores with a mass of $\geq$0.3\msun\ (assuming dust temperature of 20K and opacity from \cite{OH1994} with index $\beta=1.75$) at a distance of 7.5kpc; this is sufficient to reliably sample the peak of the core mass function in a star-forming region such as Aquila \citep{Konyves+2010}, but at up to 7.5 kpc from the Sun. This estimate is based on pure sensitivity requirements, while the effective detection of compact objects in interferometric maps critically depends on local background conditions and residuals from the CLEANing procedures. More details are given in \cite{SanchezMonge+2025, Coletta+2025}. The ALMAGAL Large Program (2019.1.00195.L) was approved in ALMA Cycle 7, for a total granted observing time of 117.7 hours for all the 12-m array configurations, and 88.4 hours for the ACA observations \citep{SanchezMonge+2025}.

%%%%%

\section{The ALMAGAL sample and observations}
\label{sample}

\subsection{Initial selection and survey planning}
\label{sample_orig}

The ALMAGAL sample initially consisted of 1017 targets with declination $\delta \leq 0\degree$, distributed from the near tip of the Galactic bar to the third quadrant, spanning a large range of clump masses, evolutionary stages and Galactocentric distances. 

\begin{figure*}[h]
        \begin{center}
                \includegraphics[width=0.45\textwidth]{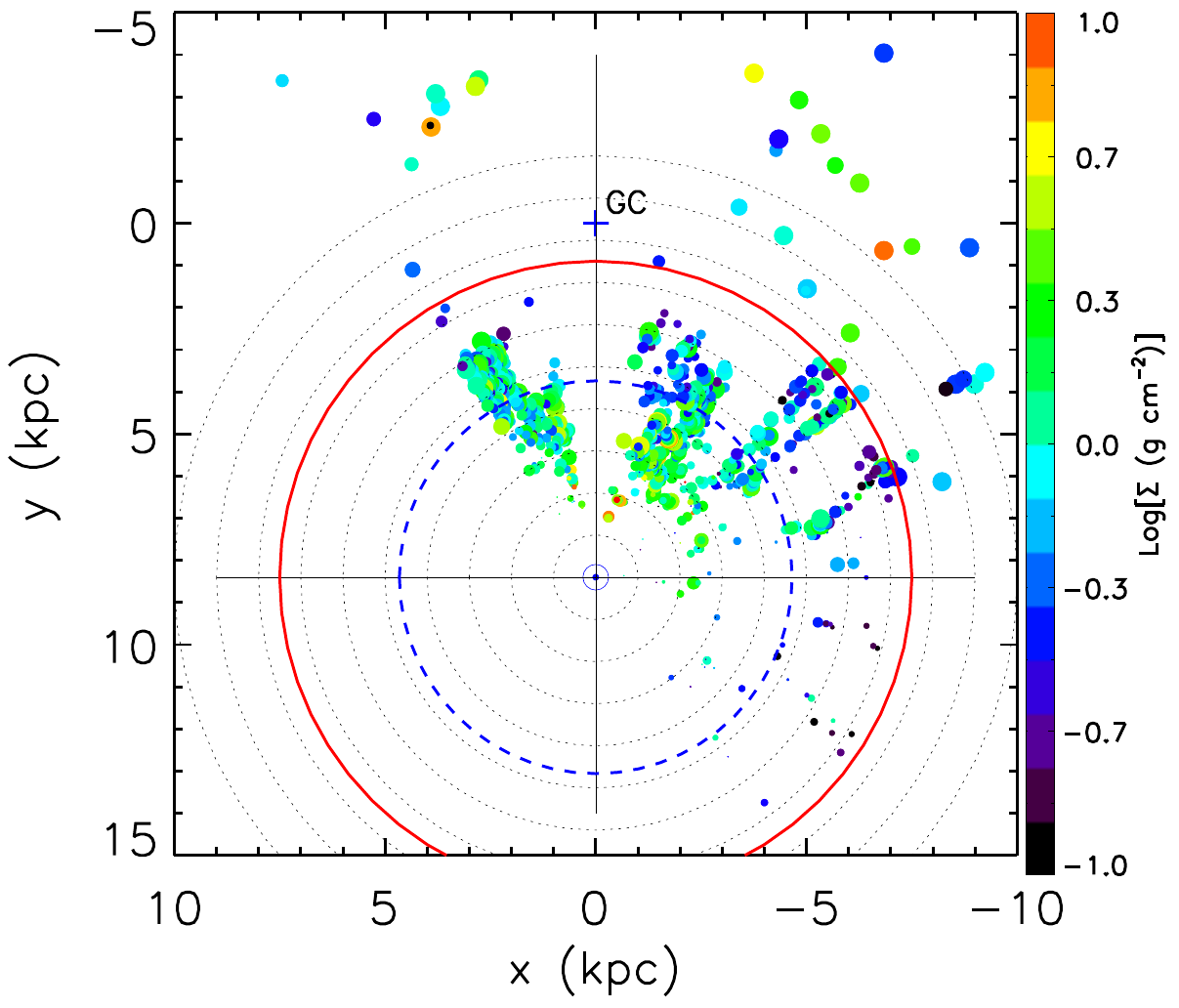} 
                \includegraphics[width=0.54\textwidth]{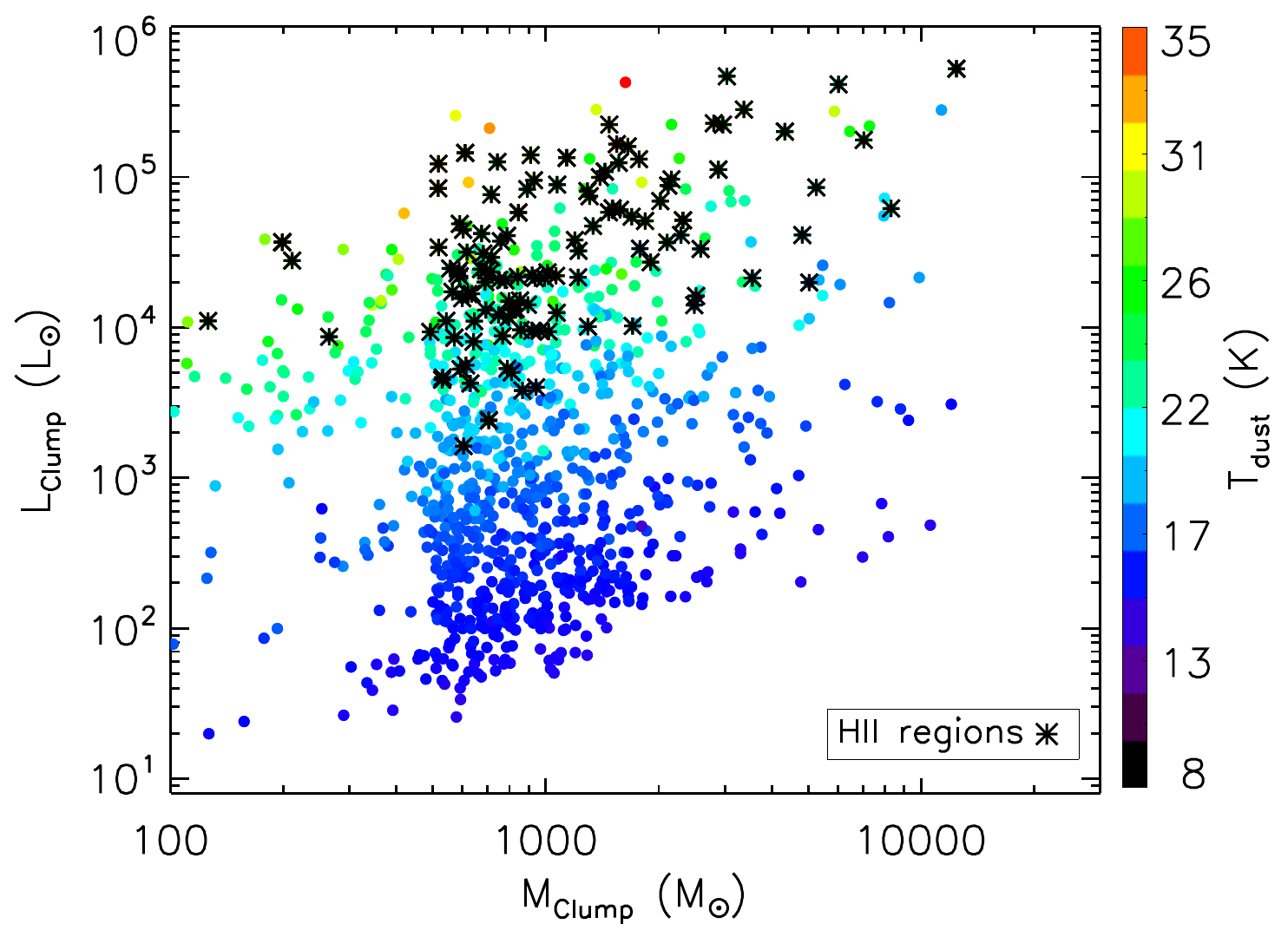} 
                \caption{(a) Galactic distribution of ALMAGAL target clumps are shown on the left, with a symbol size of $\propto$Log(M$_c$) (clump mass) and color coded by $\Sigma_c$ (surface density), as determined from the Hi-GAL data (non-beam-deconvolved), reflecting the updated distances and physical parameters described in Sect. \ref{revision} and reported in Table 1, available at the CDS. Dotted circles represent 1 kpc distance interval centered on the Sun and the plus symbol marks the location of the Galactic Center. The blue dashed circle marks the distance threshold used to split the target sample among the two ALMA antenna configurations designed to provide a minimum 1000 au linear resolution. The red circle marks the 7.5 kpc distance originally used as an upper limit for source selection. b)  L$_{bol}$/M$_{clump}$ plot for the 1017 selected clumps (color-coded by T$_{dust}$) is shown on the right. 
  %; the black lines are the evolutionary tracks for embedded cluster population synthesis models in \cite{Molinari+2019}. 
  Asterisks indicate Hi-GAL sample sources associated with \hii\ regions from the surveys of CORNISH \citep{Purcell2013}, and CORNISH-S \citep{Irabor+2023}. 
  %The magenta and cyan lines are the thresholds ${L/M}=1$ for CH$_3$C$_2$H detection, and ${L/M}=10$ above which ${T}_{gas} \propto {L/M}$ \citep{Molinari+2016c}.
  }     \label{sample_fig}

        \end{center}
\end{figure*}

The larger fraction of the sample (\nhigal\ objects) was selected from the complete catalog of dense clumps from \cite{Elia+2017} based on the Hi-GAL 70-500\um\ photometry \citep{Molinari+2016a} complemented with ancillary data at $\lambda\leq$24~\um\  and $\lambda$>~500\um\, and with a distance determined after cross-comparison with an extensive suite of CO surveys of the Galactic plane, extinction maps, and \hi\ line profile analysis, following \cite{Mege+2021}. The following selection criteria were adopted: i) distance < 7.5~kpc from the Sun to be able to resolve the target 1000~au spatial scale with the selected configurations setup (Sect. \ref{obs_setup}); ii) clump masses > 500~\msun\ in the inner and 250~\msun\ in the outer Galaxy; and iii) surface densities $\Sigma\geq$ 0.1~g~cm$^{-2}$ threshold that is critical for high-mass star formation \citep{Kauffmann+Pillai2010, Krumholz+2014, Tan+2014, Traficante+2020}. These selection criteria, adopted at the time of the proposal, are no longer 100\%\ fulfilled, following a revision of target physical parameters using the ALMAGAL data itself (see Sect. below). Overall, more than 90\%\ of targets fulfill the initial selection requirements. Target sources are spread over 3-14~kpc in Galactocentric distance from the near tip of the Galactic bar to the outskirts of the Milky Way (Fig. \ref{sample_fig}a). They span the full evolutionary path from the IRDC to the \hii\ region stage (Fig. \ref{sample_fig}b). They reside in very different environments, such as arm and inter-arm regions, and proximity to triggering agents, such as OB associations and expanding \hii\ region bubbles.

Sources in the inner 20\degree{} around the Galactic Center were excluded because this range  subtends the central regions of the Galaxy. These regions are heavily affected by non-circular motions that make distance estimates unreliable \cite[e.g.,][]{Hunter2024}. In addition, our main goal is to characterize the variance of the environmental conditions in the Galactic disc. In this sense, the central regions show peculiar conditions not representative of the disc and they are typically studied with dedicated programs such as ACES \citep{Nonhebel+2024}.

We visually inspected the PACS 70-160\,\um\ and the SPIRE 250-350-500\,\um\ \textit{Herschel} maps of all candidate targets to confirm the presence of compact emission at the position of the catalog source. The position of the 250~\um\ peak was adopted for the ALMA observations. From the evolutionary viewpoint, both the $L/M$ and the shape of the SED for $\lambda\leq$70~\um\ have been used as a broad evolutionary classification of the clumps \citep{Molinari+2008, Duarte-Cabral13, Molinari+2016c, Molinari+2019, Traficante+2018, Merello19, Wells+2022}. The values of clump $L/M$ go from $\sim$0.05, which is typical of early-stage IRDC-like clumps, to $\sim$450, which is common to IR-bright clumps hosting actively forming protostellar objects often associated with HII regions (e.g., \citealt{Cesaroni+2015, Elia+2021}). 
%As the wavelength coverage of the Hi-GAL survey pivots at $\lambda \sim 100--200$~\um, it is not surprising that the sources selected from Hi-GAL are biased toward the relatively early fraction of the evolutionary path.

To extend the coverage of the sample toward relatively more evolved \hii\ regions, we also included \nrms\ objects drawn from the Red Midcourse Space Experiment Source (RMS)  catalog of star-forming regions \citep{Lumsden+2013}, using the same distance cut and with luminosity in excess of 3000~\lsun. The RMS initial selection being based on the the mid-IR 8-21~\um\ wavelength range allows it to be more representative of relatively late stages; the sources selected  do indeed display 1$\leq $L/M$ \leq$ 200. The distribution of the final ALMAGAL sources sample in the Galaxy and in the L versus M evolutionary plot is presented in Fig. \ref{sample_fig}, where the asterisk-symbols mark the location of the target clumps with \hii\ counterparts \citep{Purcell2013, Irabor+2023} in the CORNISH and CORNISH-S 5 GHz radio continuum surveys and occupy mostly the region above ${L/M}=10$. This confirms that the ALMAGAL sample spans the entire path from IRDC-hosted clump to the \hii\ region phase.

During the analysis stage, after the observations were taken, it was realized that the template WR stellar system $\eta$~Car, which has the target ID '653755' in the name of the data files in the ALMA Archive was included by error in the sample. We also overlooked that three fields (with target IDs '615590', 'G348.7342-01.0359B', and 'G323.7410-00.2552C' for reference in the ALMA Archive) had  coordinates closer than 10\asec\ to other targets in the sample and they are therefore  considered to be duplications. Since they have larger noise levels compared to the fields they duplicate, we removed them from the ALMAGAL sample data considered in the rest of the paper and, thus, from further scientific analysis (amounting to 1013 targets). Figure \ref{sample_fig} reports the Galactocentric distribution and the L-M plot for the sample targets, with distances and physical parameters updated based on the new ALMAGAL observations, as explained in the next subsection below.

%The size of the targets sample, and the range of physical parameters and evolutionary stages it spans, set ALMAGAL completely apart from any other ALMA approved project, making it the program with the largest  potential to draw statistically significant conclusions when it comes to intermediate-high mass clumps hosting protoclusters in the Galaxy.

%This is a comfortable size to reduce the variance of derived parameters enough to be able to draw statistically relevant conclusions. The lower-mass bins are sufficiently populated to also control the effects of different geometries, viewing angles and different histories in the sense of interactions with their environment, which add additional dimensions to the sampling space. 
%The sample includes 180 well-studied mid-IR bright MYSOs also present in the RMS survey \citep{Lumsden+2013}, representing the stage just prior to the UC\hii\ region phase. 

\subsection{Revision of ALMAGAL targets physical properties}
\label{revision}

A critical aspect in ALMAGAL science is relating the physical properties of ALMA detected cores, as well as the properties of the dense gas at all sampled spatial scales, to the physical properties of the parent dense clumps (\citealt{Coletta+2025, Mininni+2025}, Elia et al., in prep, Jones et al., in prep.). To this end, we decided to revise and update the target clump distances (also based on ALMAGAL itself) and physical properties, for three reasons:
\begin{itemize}
    \item The methods by which the distance to the targets is estimated may have changed due to new evidence. For the fraction of the sample selected from the Hi-GAL survey, a new source catalog with new distance estimates and revised physical properties was released \citep{Elia+2021}. For the RMS-based selected targets, their properties are constantly updated on the project's website.\footnote{http://rms.leeds.ac.uk}
    \item The \vlsr\ of the bulk gas traced by high-critical density lines in the ALMAGAL spectral cubes provide a much more reliable input for kinematic distance estimates, especially in cases where CO (even with its rarer isotopologs) shows multiple gas components along the line of sight (Sect. \ref{dist_update}).
    \item A meaningful and reliable science analysis of the ALMAGAL data requires that clumps physical properties (in particular mass and luminosity) are estimated in a homogeneous way for the Hi-GAL and the RMS-based fractions of the sample (Sect. \ref{sample_homo}).
\end{itemize}

We report these updates in the following subsections.
%that were carried out about source distances and physical properties, leading to a final consolidated set of source properties.

\subsubsection{Revision of targets distance}
\label{dist_update}

Assigning a heliocentric distance to a source detected in 2D thermal continuum images is a mandatory critical step for any subsequent physical parameter estimate. The typical backbone workflow of any distance determination recipe can be summarized as follows \citep{Urquhart+2018, Mege+2021} in order of precision and priority: i) check source position against literature catalogs of maser or stellar parallax that offer the most reliable of all estimates; ii) correlate source position and velocity against literature catalogs of optical, radio, or absorption line \hii\ regions, or in YSOs where the distance has already been determined via, for instance, optical spectroscopy; iii) correlate source position with submm/radio molecular line surveys, use a Galactic rotation curve to derive the distance and (in cases where this places the source in the inner Galaxy) to try to solve the kinematic distance ambiguity (KDA) using additional evidence such as an association with IRDCs, \hi\ absorption features, and so on. 
A similar methodology was used to derive distances for the Hi-GAL and RMS catalogs, with only small differences. 

\begin{figure*}[h]
\begin{center}
\includegraphics[width=0.43\textwidth]{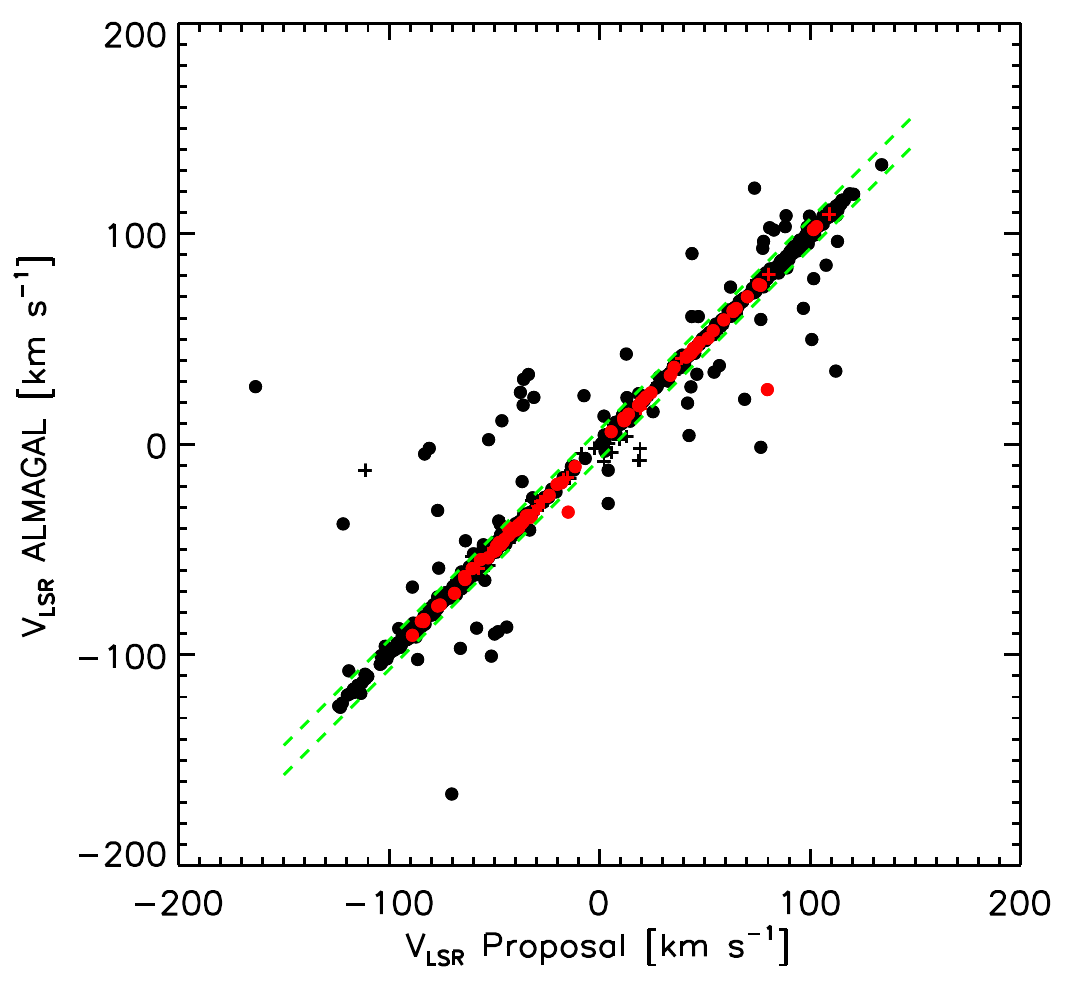} 
\includegraphics[width=0.56\textwidth]{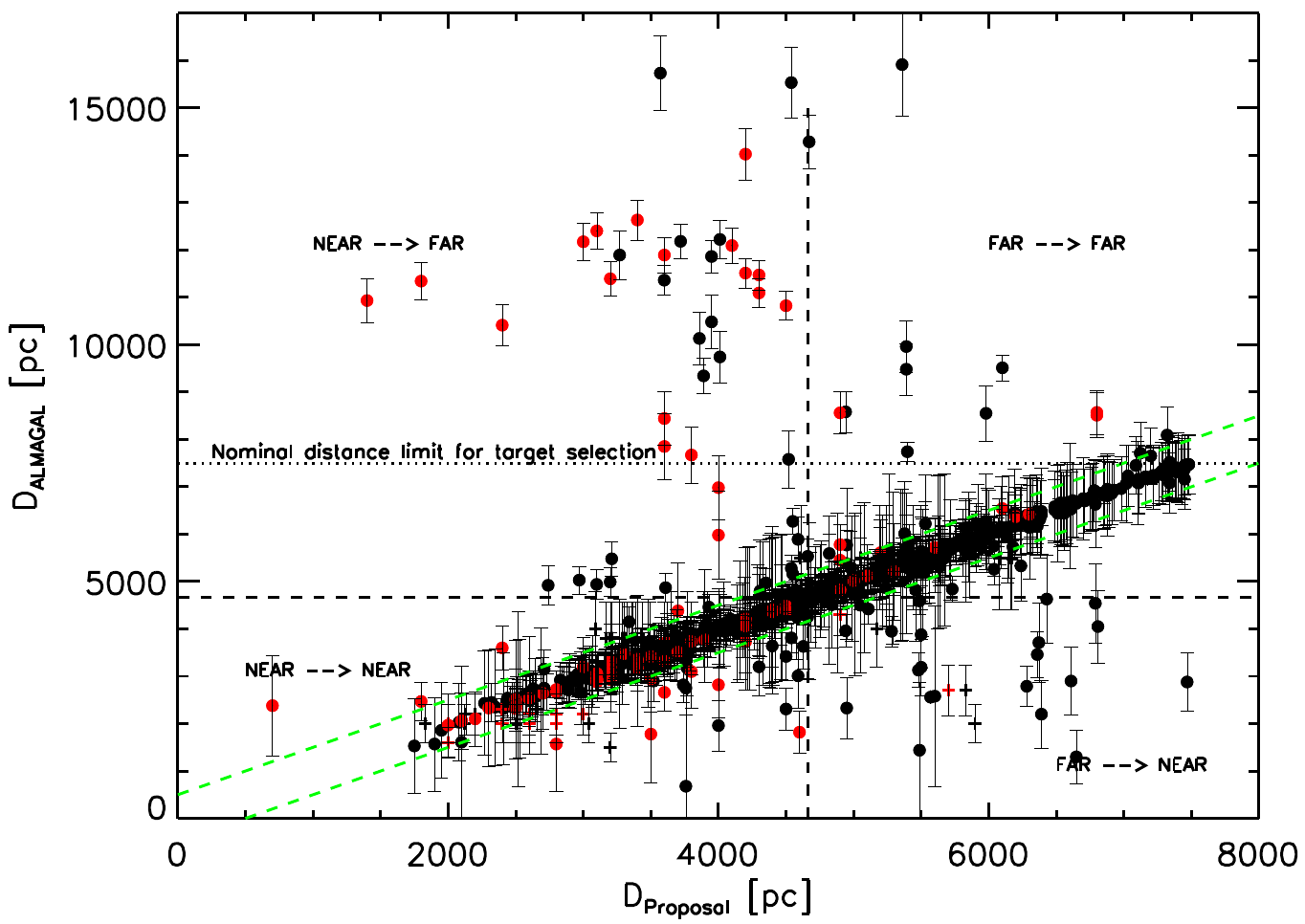} 
\caption{Physical parameters derived from the ALMAGAL observations.
%a) (left) 
General properties of the ALMAGAL targets. {\it Left:} Plot of the ALMAGAL-determined \vlsr\ vs the values in the original catalogs at the time of the proposal. Black and red symbols identify sources extracted from the Hi-GAL and RMS survey catalogs. %, respectively. 
The formal uncertainty in the determination (1.4\,\kms\, see text) is smaller than the size of the points, while green dashed lines mark the extent of a $\pm$7\,\kms\ \vlsr\ variation typically assigned to local non-circular motions (see text). 
 %b) (right) 
{\it Right:} New distances determined from the ALMAGAL-derived \vlsr\ vs the original values (colors as in left panel). Green dashed lines mark a $\pm$500\,pc distance variation. The dashed lines mark the distances that were adopted to generate the \textit{"near"} and \textit{"far"} ALMAGAL subsamples, with text showing quadrants where sources would change subsample membership (see text for more details). }
\label{distance_fig}
\end{center}
\end{figure*}

To obtain a homogenized set of properties for the sample of target clumps and to take advantage of the ALMAGAL spectroscopy that offers the ideal database to improve existing kinematic distance estimate, in particular, for those objects (the majority), where only CO and its isotopologues were previously used in the literature, we revised the distances by measuring the \vlsr\, from several high critical density transitions and applied the final  module of the \citet{Mege+2021} algorithm (previously used for deriving distances for the Hi-GAL catalog) that converts \vlsr\, into kinematic distance. The details of this work will be reported in Benedettini et al. (in prep.). In summary, for each ALMAGAL source, we extracted subcubes from the ALMAGAL \lres\ datacubes over large velocity ranges ($\pm$180\kms) centered on \chiiioh, \hiico, \dcn, \so, \hciiin, \sio, \ceo,\ and \tco\ lines. For each line, the cubes were co-added over the emitting region, defined by selecting the spatial pixels where the moment zero map is above three times its noise level, to obtain integrated spectra that were fed to an automatic 3$\sigma$-clipping line detection and Gaussian fitting algorithm. The \vlsr\ of the targets were then derived averaging the fitted line centers of all the detected lines with a good fit among the transitions with critical density higher than 10$^5$ cm$^{-3}$; namely,  \chiiioh, \hiico, \dcn, \so, \hciiin\ and \sio. \ceo,\ and \tco\ were used only for 207 targets where none of the other higher critical density lines were detected. We assigned an uncertainty to the derived \vlsr\ of 1.4\,\kms, which is the spectral resolution of the data. The formal uncertainty obtained from averaging the \vlsr\ of the individual lines used is much lower than this. However, since the resolution element is sampled with only 2 channels and the linewidths are generally below $\sim$2\kms, we decided to adopt the resolution as a conservative uncertainty. It is worth noting that in 257 ALMAGAL targets we found that more than one emission feature is present in the spectrum centered toward the \ceo\ frequency, indicating that multiple components of gas are present in the same ALMA field of view (FOV). In these cases we selected the densest component choosing the velocity where line emissions were detected also in the transitions with higher critical densities (> 10$^5$ cm$^{-3}$). We also have 12 targets with two components both revealed also in the high density tracers for which the most intense component was adopted.
Figure \ref{distance_fig}a shows the extent of the variations found in the \vlsr\, derived from ALMAGAL data with respect the ones used at the time of the proposal and that were used to define the two subsamples observed with different 12m-array configurations (see Sect. \ref{obs_setup}). In \nvlim\ sources, the new \vlsr\ values differ from the original ones by more than 7~\kms\ in absolute value. This is a threshold that we consider here, along with common literature (e.g., \citet{Mege+2021} and references therein), a reliable upper limit for local gas motions above which \vlsr\ variations are likely to reflect different kinematic distances. We  therefore chose to adopt this limit as an uncertainty in the determination of the \vlsr\ of a target. Propagating this uncertainty on distance greatly depends on the Galactic longitude of the source, as we discuss  below. The differences (sometimes very large) that we found in \vlsr\ with respect to the initial values at target selection are due to an incorrect choice of the CO component that was associated with the sources in the Hi-GAL/RMS samples, among the multiple components found along the sources' line of sight. This resulted in  a substantial revision of distances and physical properties for these \nvlim\ sources (see below).

The newly determined ALMAGAL-based \vlsr\ values were converted to distances using the final module of the \cite{Mege+2021} algorithm, with one important exception. In cases where there was no way to help resolve the KDA for Hi-GAL sources in the inner Solar circle, \cite{Mege+2021} adopted (by default) the far distance, while in similar conditions, RMS sources (Urquhart et al. 2018) were put at the near distance. To provide as a homogeneous as possible table of source properties, we chose for this study to adopt the same approach of the RMS catalog and place the Hi-GAL sources for which KDA could not be resolved (\nkdadefnear\ out of the \nhigal\ Hi-GAL sources in the ALMAGAL sample) at the \textit{near} distance. In estimating the error associated with the distance we considered possible deviations from purely circular motions and proper motions as anticipated above, by applying an additional offset to the velocity of 7\,\kms\ as in \citet{Mege+2021}; the corresponding offset recovered in the distance to each source will of course be dependent on the Galactic longitude of the object, and is adopted as the distance uncertainty as reported in Table 1, available at the CDS. Error propagation transfers twice the distance relative uncertainty to source luminosity and masses. No uncertainty originating from distance affects distance-independent quantities such as $L/M$ or clump surface densities. 

In Fig.~\ref{distance_fig}b, we compare the newly obtained distances with the original ones at the time of the proposal. We find that for \ndfiveh\ objects out of the total \nall\ sources sample, the new distance differs by more than 0.5~kpc from the original one. For \ndnf\ objects the change in distance is such that the source would switch from the "near" to the "far" ALMAGAL subsample. Having been observed with the "near" C-5/C-2 ALMA antenna configuration, the target 1000~au linear scale was not reached for these objects. Conversely, for \ndfn\  objects, the change in distance is such that the source would switch from the "far" to the "near" ALMAGAL subsample; having been observed with the "far" C-6/C-3 ALMA antenna configuration, the linear scale achieved for these objects was smaller than the target 1000~au. Finally, for 44 objects, the new distance would put them beyond the 7.5~kpc distance limit adopted for the initial source selection; for these sources as well, the target resolved 1000~au linear scale was not reached.

\subsubsection{A homogeneous methodology for SED fitting.}
\label{sample_homo}

Of the ALMAGAL targets extracted from the RMS sample, 59 were also present in the Hi-GAL clumps catalogs of \cite{Elia+2017, Elia+2021}, where the requirement for inclusion was that a source should have a convex SED with at least 4 adjacent \textit{Herschel} bands. For these sources (the "G" sources in col. 3 of Table 1 with "no sed" flag in the last column) the set of physical parameters reported in Table 1 were adopted from the Hi-GAL catalogs. For the rest of the RMS sources in the sample the above criteria were failed and hence they required a custom analysis of their SED to estimate parameters with the same approach as in Hi-GAL (described in detail in the above cited Elia's papers). The different situations encountered can be grouped as follows, identified by a flag that is also reported in Table 1:

\begin{itemize}
        \item \textbf{-sed\_irr}: the RMS source had an entry with at least four adjacent bands in the Hi-GAL band-merged photometric catalog, but the \textit{Herschel} SED was "irregular" in the sense that it showed changes in concavity directions. For the present work, the processing to derive the mass, \lbol, T$_{dust}$, and T$_{bol}$, was also applied to these sources.
 %applied to these sources as well.
        \item \textbf{-sed\_rebuilt}: the RMS source had multiple "$<$4-bands" entries in the Hi-GAL band-merged photometric catalog. A typical occurrence would be a RMS source with two counterparts at 250\um, where one of them would be linked to a $70-160$\um\ SED branch, and the other to another 350-500\um\ SED branch. Both branches would feature less than four adjacent bands and  would not pass the criteria to be included in the Hi-GAL physical parameters catalog. For these cases, in the present work the partial branches were reconnected (e.g., in the example above, forcing a single 250\um\ source to be extracted).
        \item \textbf{-sed\_sat}: the RMS source is associated with saturated pixels in \textit{Herschel} 250 or 350\um\ bands. As such, the fluxes for the source could not be extracted in four adjacent bands. For the present work we forced the Hi-GAL processing allowing for "holes" in the SED.
        \item \textbf{-sed\_noband}: the RMS source has a 3-band 70-160-250\um\ counterpart in the Hi-GAL band-merged catalog, vanishing with at 350 and 500\um. For the present work we forced the Hi-GAL processing with 3-bands \textit{Herschel} SEDs.
        \item \textbf{-sed\_noprops}: the RMS source has a counterpart only at 70\um\ and, hence, no Hi-GAL-like processing can be applied to derive physical parameters.

\end{itemize}

Finally, in addition to the evolutionary information provided by the $L/M$ ratio, we also report the classification according to the method from \cite{Urquhart+2022} who define classes as quiescent, protostellar, YSO, and HII regions based on visual inspection at different wavelengths. Each region is  visually inspected at 70, 24 and 8~$\mu$m. Roughly 65\% of the ALMAGAL sources were cross-matched with the ATLASGAL catalog and the classification adopted from there. The remaining 35\% of sources were classified by hand using the same methodology.

Table 1, available at the CDS, reports the consolidated properties of the ALMAGAL targets as revised above. Uncertainties are explicitly given in the table for the distance to each source. Uncertainties on masses and luminosities are easily derived from those and are not explicitly reported. Columns explanation is available in Appendix \ref{appendix_clump_prop}.

\section{Properties of continuum emission}
\label{continuum}

\begin{figure*}
\centering
    \includegraphics[width=0.95\linewidth]{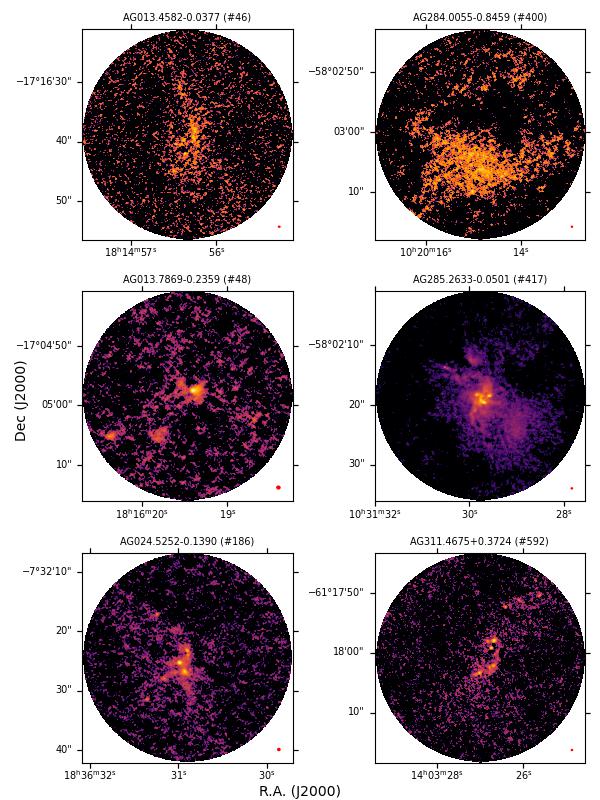}
    \caption{ALMA 1.38mm continuum \fres\ images of a selection of ALMAGAL fields showing the large variety of fragmentation levels and extended emission found. The target AG name and running number are from cols. 2 and 1 of Table 1. The small red ellipse in the bottom-right corner represents the synthesized beam.}
\label{continuum_maps}
\end{figure*}
\addtocounter{figure}{-1}
\begin{figure*}
\centering
    \includegraphics[width=0.95\linewidth]{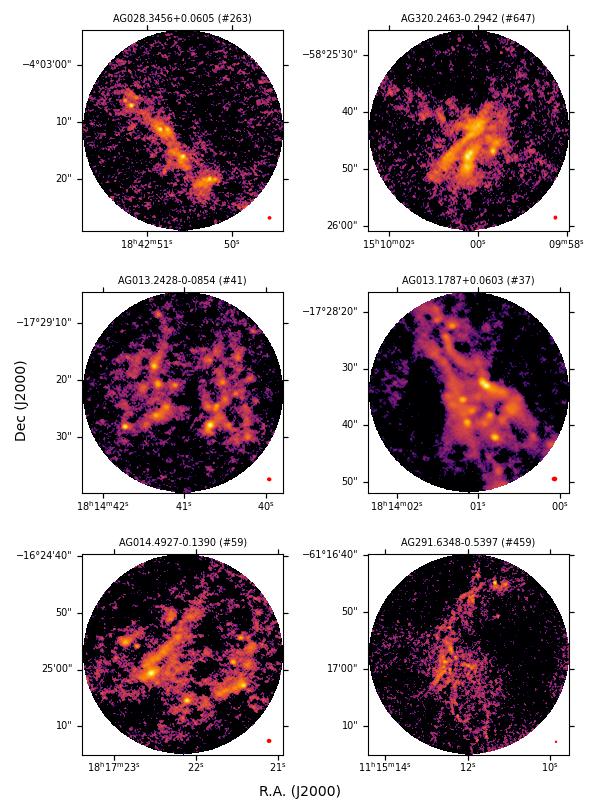}
    \caption{Continued.}
\end{figure*}

The ALMAGAL target sample spans large ranges in several physical and environmental parameters (Sect. \ref{sample}) and this is reflected in the wide variety of properties and morphologies of the 1.38~mm continuum emission. While a complete gallery of 1.38~mm continuum images is available on the project's website\footnote{www.almagal.org}, in Fig. \ref{continuum_maps}, we illustrate 12 continuum emission maps that illustrate the whole spectrum of morphologies in the ALMAGAL sources. Fields AG013.4582-0.0377 (\#46) and AG284.0055-0.8459 (\#400) shows relatively extended emission that is mostly resolved at the 1000 au scale of the full-resolution images. We can then see fields with variable level of fragmentation, from single main cores as in AG013.7869-0.2359 (\#48), to small clusters more (AG285.2633-0.0501, \#417),  or less compact as in AG024.5252-0.1390 (\#186) and AG311.4675+0.3724 (\#592), to larger clusters embedded in sometimes considerable extended emission with either distinctively filamentary, as in AG028.3456+0.0605 (\#263), AG320.2463-0.2942 (\#647), or more irregular (the remaining panels) shapes. 

\begin{figure}[h]
\begin{center}
\includegraphics[width=0.47\textwidth]{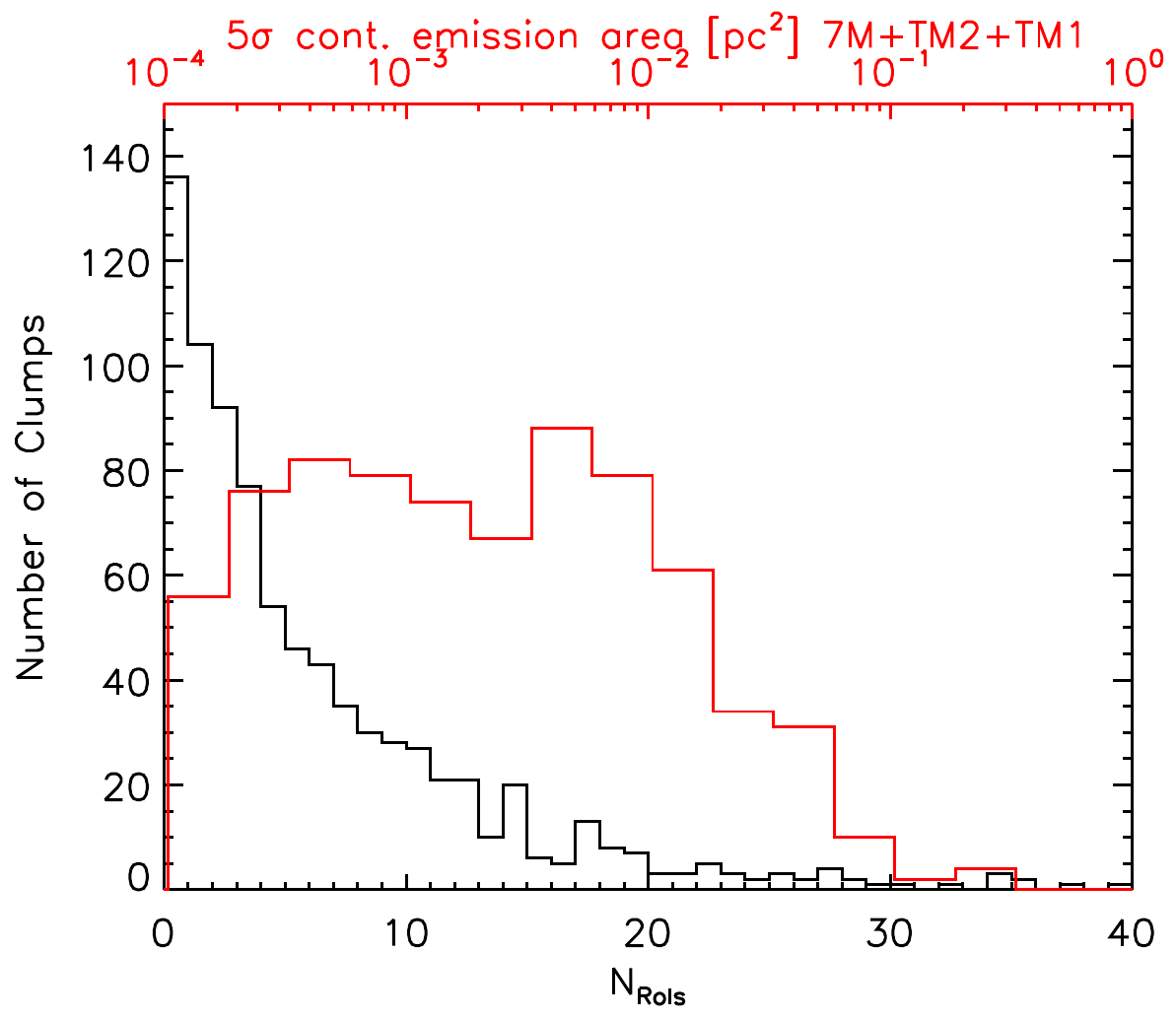} 
\caption{Distribution of the number of RoIs (in black, bottom X axis) and of their area in pc$^2$ (in red, top X axis) revealed for the 1.38mm continuum at 5$\sigma$ per clump.}
\label{stat_rois}
  \end{center}
\end{figure}

In this section, we provide a characterization of the continuum emission properties in our ALMA images in relationship to the integrated properties of the target clumps. The analysis is based on the 1.38~mm continuum emission detected over 5$\sigma$ level over the root mean square (rms). This noise level is computed in each field using the residual map outside of the largest CLEAN mask used in the ALMAGAL pipeline processing \citep{SanchezMonge+2025} and it is stored in the keyword AGSTDREM of the maps FITS Header. The  5$\sigma$ threshold was adopted to minimize the inclusion of noise residuals in the morphological analysis. For each of the \nall\ independent fields we produce masks containing the pixels of the continuum images with flux above the 5$\sigma$ level, which we  call RoI (Region of Interest). Each cluster of connected pixels constitutes an individual RoI, so that each field may have many RoIs. For each RoI, we compute the area and the perimeter, and all RoIs smaller than the beam are discarded from further analysis. For each valid RoI, we computed the fluxes (peak, total, and median fluxes) by projecting the RoI mask onto the primary beam-corrected images (PBCOR). Figure \ref{stat_rois} reports the distribution of the number of RoIs, and their total area, per imaged clump where emission was detected. It is dominated by fields with relatively low number of RoIs, showing that the threshold level is adequate to recover areas with  reliable emission. The distribution of areas is instead relatively flat, implying that large areas of emission do not necessarily result from the total contribution of many RoIs in the field.

In addition, for each field, we also computed the perimeter and area of the convex hull for each of the 5$\sigma$ emission contours, which is the smallest convex figure containing the contour (see Fig. \ref{area-roi-chull} for an example). Any of these geometrical properties of the dust emission in the observed fields can then be related to other integrated field properties (e.g., clump mass, surface density, $L/M$ as estimated from \textit{Herschel} measurements), or can be compared among different fields in the sample.

\subsection{ALMA 1.38~mm emission vs \textit{Herschel} clump properties}
\label{alma_higal}

It is interesting to compare the ALMA 1.38 mm emission properties with global clump parameters mostly based on \textit{Herschel} observations and listed in Table 1, particularly as  parameters such as the far-IR (FIR) flux and surface density have been used as the selection criteria (Sect. \ref{sample}). 

One initial aspect we want to quantify is how the thermal dust emission at different spatial scales is conserved. Interferometers introduce a spatial filtering of the emission that has a spatial scale larger than the one sampled by the shorter antenna baselines. The selection of the ALMAGAL targets sample has been based on properties directly (flux) or indirectly (mass, luminosity, surface density) estimated from the \textit{Herschel} thermal far infrared emission. The comparison of these properties with similar properties derived from the millimeter thermal dust emission from ALMA can provide interesting insight concerning the spatial distribution of the cold dust within the target clumps. For this comparison, we chose to use the \ires\ continuum images with primary-beam-corrected fluxes (PBCOR). 

\begin{figure}[t]
\begin{center}
\includegraphics[width=0.47\textwidth]{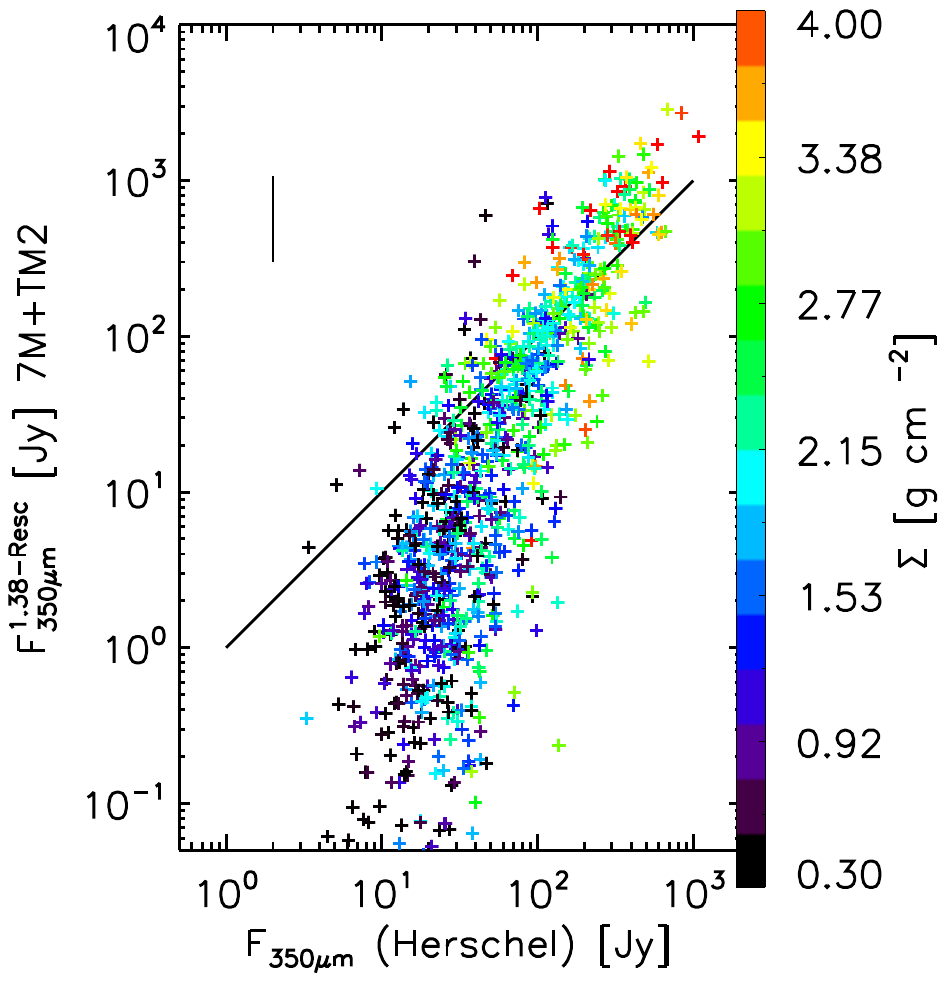} 
\caption{Total flux within the 4$\sigma$ contours of the \ires\ PBCOR images rescaled to 350~\um, as a function of the 350\um\ flux from \textit{Herschel}. For the rescaling of the ALMA 1.38~mm flux to 350~\um\ we use optically thin assumptions with $\beta=1.75$ and the dust temperature from the \textit{Herschel} SED fitting. The black line is the identity line, and the color scale is proportional to the clump surface $\Sigma$, again from \textit{Herschel}. The vertical black line shows the spread of values that the 350~\um-rescaled 1.38~mm flux can assume for different dust temperatures and opacities.}
\label{ftot_tm2_f350}
  \end{center}
\end{figure}

Figure \ref{ftot_tm2_f350} shows the relationship between the \textit{Herschel} 350~\um\ flux of the ALMAGAL target fields and the 350~\um-rescaled total flux obtained from the total 1.38~mm continuum in the 4$\sigma$ contour of the PBCOR images in the \ires\ configuration, provided that the area of the contour is larger than the synthesized beam. In particular, we extrapolated the mm flux to 350\um\ using a modified black-body assuming different dust opacities and temperatures (see below). A 4$\sigma$ was specifically adopted here (instead of the 5$\sigma$ use in the rest of the analysis) to make sure we could recover as much as possible the emission area for a most reliable assessment of a missing flux problem. Signal is detected toward 941 out of 1013 fields ($\sim$93\%), and in 336 out of 941 fields ($\sim$36\%),  the discrepancy between the fluxes is less than 50\%. For a fraction of the fields the rescaled 350~\um\ flux is above the \textit{Herschel} measured flux, but that is compatible with the uncertainties due to the assumptions in terms of dust temperature and opacities that we are forced to make to rescale the 1.38~mm flux to 350~\um. Indeed, the vertical bar in the figure represents the extent of the overall spread of 350~\um-rescaled fluxes obtained varying the temperature between 20~K and 50~K, dust spectral index $\beta$ between 1.5 and 2 (a range typical for diffuse ISM and dense clumps), and assuming opacities from \cite{preib93}, \cite{OH1994} and \cite{Draine2003}.

The figure shows that the distribution is strongly skewed toward low \ires-rescaled fluxes. For about 64\%\ of the fields the rescaled \ires\ recovered flux is much less than 50\%\ of the \textit{Herschel} flux, and for these fields this is a clear indication of missing emission. The extent of the emission in the 350\,\um\ \textit{Herschel} images is generally larger than the ALMA FOV, so that the discrepancy could result simply from the different sizes of the areas sampled. However, the beam of \textit{Herschel} at 350\,\um\ is $\sim$80\%\ of the ALMA FOV and therefore the far-IR emission could plausibly come from smaller areas within the ALMA FOV. In particular, for 90\% of the targets the measured clump FWHM from \textit{Herschel} is smaller than the ALMA FOV; the percentage rises to 95\% if the beam-deconvolved \textit{Herschel} size is adopted. We conclude that it was only in 10\% of the targets that the discrepancy in recovered flux could be attributed to differences in the areas of the sky sampled by \textit{Herschel} and ALMA. 

Instead, the color scale of the points shows that these fields have the lowest surface density, as estimated from \textit{Herschel};  hence, the dust distribution is expected to be less compact and more prone to spatial filtering by the interferometer. This indicates that for a significant fraction of targets we may be missing flux from larger angular scales, and the use of physical parameters such as the total clump mass that could be derived from integrated ALMA millimeter fluxes is not reliable for the majority of the fields.

\begin{figure}[t]
\begin{center}
\includegraphics[width=0.47\textwidth]{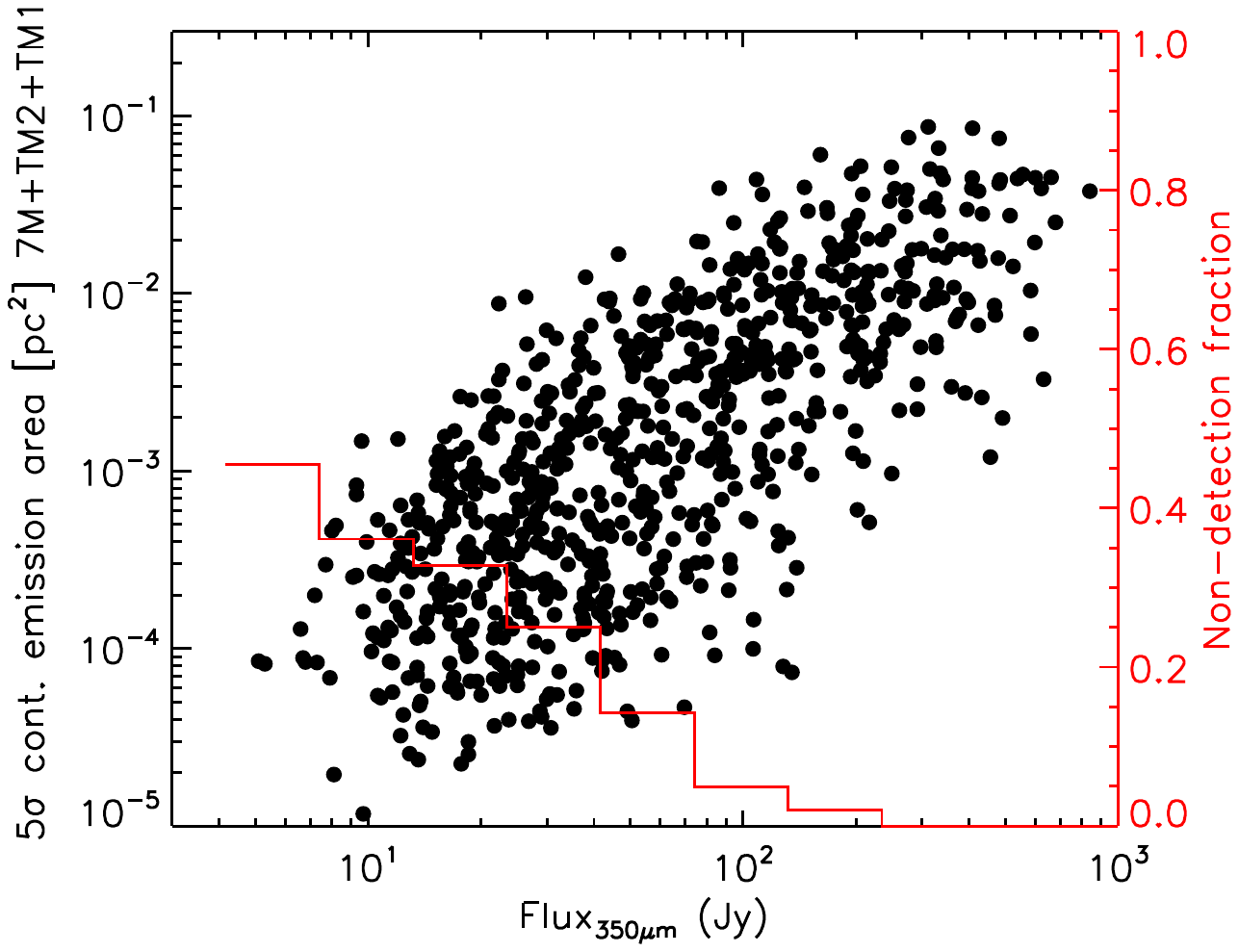} 
\caption{ALMA 1.38-mm total emission area above 5$\sigma$ in the \fres\ images as a function of the Hi-GAL 350~\um\ flux (\cpear=0.74). The red histogram reports the fractional distribution of the 350~\um\ flux for ALMAGAL sources with no 5$\sigma$ detection (to be read on the red right $y$-axis).}
\label{roi-area_f350}
\end{center}
\end{figure}

We wanted to concentrate on the morphology of continuum emission for this work. Figure \ref{roi-area_f350} shows the relationship between the 1.38~mm total emission area above 5$\sigma$ level in the ALMA \fres\ image as a function of the \textit{Herschel} 350~\um\ flux. The plot outlines a very good correlation among the two parameters  that is not influenced by the different distances of the targets as neither of the two parameters depends on it (figs. \ref{roi-area_dist},\ref{f350_dist}). The Pearson linear correlation coefficient is \cpear=0.74.
%If one choses to use distance-corrected 350\um\ fluxes (i.e. the monochromatic luminosity) the relationship is maintained (see Fig. \ref{roi-area_lum350}).

The fraction of fields with no detected emission above 5$\sigma$ in ALMA is concentrated toward the low end of the 350~\um\ flux distribution, but there is no indication of an exact far-IR flux threshold for the 1.38mm detection. The figure confirms that the 350~\um\ flux cut adopted for the selection of the ALMAGAL target clumps is well matched to the chosen ALMAGAL target sensitivity. A fraction of targets below 10\%\ is not detected with ALMA also for relatively large 350\,\um\ fluxes. The average properties of non-detected fields are discussed in Elia et al. (in prep.). 
%It is worth noting that the emitting area here does not seem significantly affected by missing flux problem outlined instead by the total flux in Fig. \ref{ftot_tm2_f350}, which leads to the conclusion that most of the missing flux is lost from within the emitting areas.

\begin{figure}[t]
\begin{center}
\includegraphics[width=0.47\textwidth]{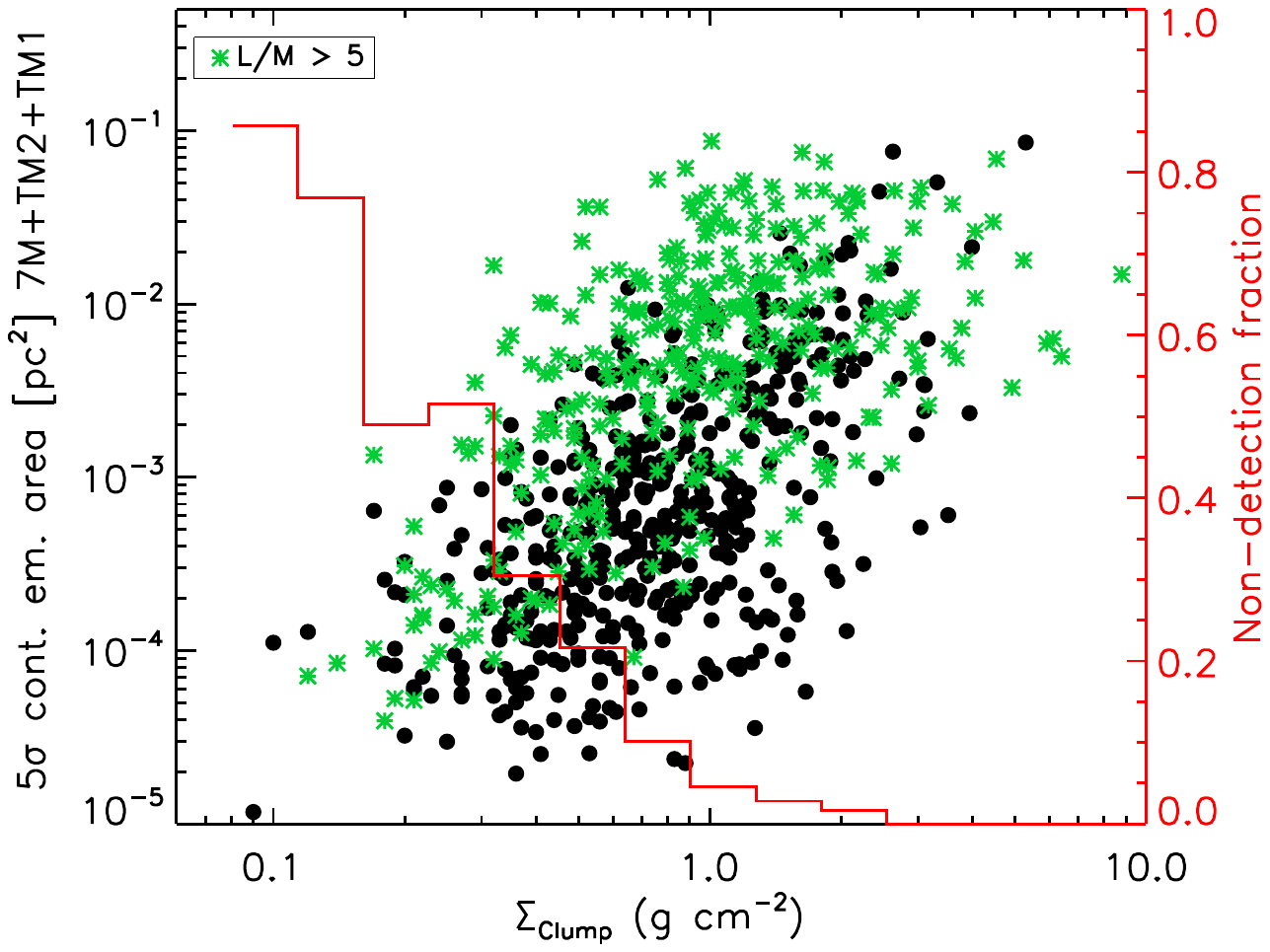} 
\caption{ALMA 1.38~mm total emission area above 5$\sigma$ in \fres\ images as a function of the clump surface density from Table 1. The green symbols are for target clumps with $L/M\geq$5. The red histogram reports the fractional distribution of the surface density  for ALMAGAL sources with no 5$\sigma$ detection (to be read on the red right $y$-axis).}
\label{roi-area_surf_d}
\end{center}
\end{figure}

Figure \ref{roi-area_surf_d} shows the relationship between the ALMA 1.38~mm 5$\sigma$ area and the clump surface density estimated from Hi-GAL data. As in the previous figure, the red histogram shows the $\Sigma_{\mathrm{Clump}}$ distribution for fields without 5$\sigma$ detection. Here again a correlation is clearly apparent, with \cpear=0.57 suggesting indeed a good correlation, although with a large scatter in the $y$-axis distribution. 

The problem of possible missing flux may play a role in determining the scatter of the points; however this should affect only the relatively lower surface density clumps (see Fig. \ref{ftot_tm2_f350}), while the scatter we see in the Y direction in Fig. \ref{roi-area_surf_d} is independent from the range of surface density. The points reported in green color identify the clumps with $L/M \geq 5$, showing that they predominantly populate the upper range of the points distribution. Therefore, it is  the clump evolutionary stage that introduces most of scatter. 

The relation between the Hi-GAL $\Sigma_{\mathrm {clump}}$ and the ALMAGAL 1.38-mm 5$\sigma$ area is interesting for two key reasons. First, it again reinforces the adopted choice of parameters to drive the ALMAGAL source selection. The 1.38~mm emitting area, under the optically thin assumption, is a proxy for the dust column density spatially resolved in the ALMAGAL maps; the surface density estimated from \textit{Herschel} observations is also a measurement of column density that is, however, an integrated quantity averaged over the extent of the clump. It is then reassuring to verify that in spite of the missing flux problem and taking into account evolutionary effects, the two parameters are indeed correlated. 
In addition, the fractional distribution of $\Sigma_{\mathrm{Clump}}$ for non-detected fields in ALMA is heavily skewed toward the lower end of the range and this suggests that the choice of the target sensitivity adopted for the ALMAGAL observations is very well matched to the average properties of selected targets.  

Second, since the millimeter continuum in interferometric observations traces denser and compact rather than diffuse and extended ISM (as suggested in Fig. \ref{roi-area_surf_d}),  the degree of mass concentration as traced by the clump-averaged $\Sigma_{\mathrm{Clump}}$ has a role in determining the amount of dense material that is found at smaller spatial scales probed by ALMA. Furthermore the correlation present in Fig. \ref{roi-area_surf_d} completely disappears if we use the clump mass instead of its surface density (see Fig. \ref{roi_area_mass}), confirming that is not simply the available mass that drives the process by which the clump ISM is shaped at smaller and smaller scales, but the shape of the gravitational potential. This was anticipated by, for instancem, \cite{MKT2002} and will be  explored in more detail in Elia et al. (in prep.). To illustrate how we excluded the possible presence of a bias introduced by the different distances of the targets, in Fig. \ref{roi-area-surf_d_bins} we report the same two quantities of Fig. \ref{roi-area_surf_d}, but in separate distance bins, showing that the relationship holds irrespectively of the distance to the sources.

\begin{figure}[h]
\begin{center}
\includegraphics[width=0.47\textwidth]{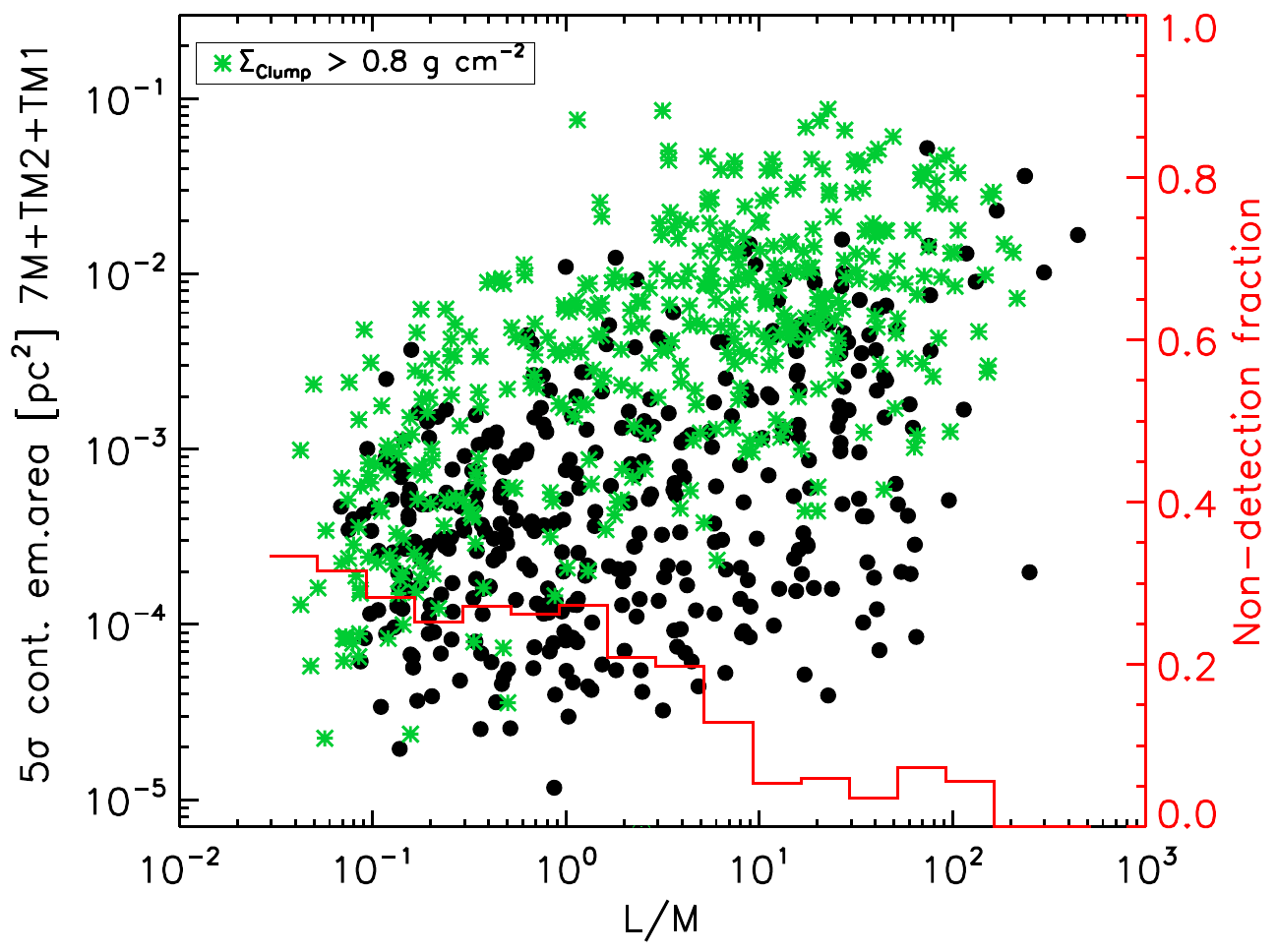} 
\caption{ALMA 1.38~mm total emission area above 5$\sigma$ in \fres\ images as a function of the clump $L/M$ from Table 1. The green symbols are for target clumps with $\Sigma_{\mathrm{Clump}}\geq0.8$ \gcmtwo. The red histogram reports the fractional distribution of the surface density  for ALMAGAL sources with no 5$\sigma$ detection (to be read on the red right $y$-axis).}
\label{roi-area_lm}
\end{center}
\end{figure}

To illustrate more clearly the degree to which the extent of dense ISM in the clumps fields depends on their evolutionary stage, we report in Fig.~\ref{roi-area_lm} the 1.38~mm ALMA 5$\sigma$ emission area as a function of the clumps, $L/M$. The two parameters appear indeed correlated, and the large scatter in the Y axis is this time due to the spread in surface density. Indeed in the figure we outline with green symbols the clumps with higher surface density, finding that the larger 1.38 mm emission areas are indeed found on average toward clumps with higher surface density. The correlation coefficient for the entire dataset in Fig. \ref{roi-area_lm} is \cpear=0.53, confirming a relatively good correlation. However, this is mostly driven by the higher $\Sigma_{\mathrm{Clump}}$ points for which we have \cpear=0.66 and for which the correlation is therefore stronger. The correlation for the subset with lower $\Sigma_{\mathrm{Clump}}$ is only moderate, with correlation coefficient dropping to \cpear=0.42.

Since the clump surface density does not appear to be an evolutionary indicator based on the $L/M$ \citep{Elia+2017}, we conclude from this preliminary global analysis of the continuum emission in ALMAGAL fields that the presence of dense ISM substructures in the target clumps depends both on the initial conditions for clumps fragmentation (as traced by the clump-averaged surface density) and the global evolutionary stage of the clumps as traced by the $L/M$ parameter. 

\subsection{The morphology of 1.38~mm emission in ALMAGAL}
\label{cont_morph}

Given this large variance in the morphological appearance of the 1.38~mm  continuum emission (Fig. \ref{continuum_maps}), we would like to verify whether there is a metric that is able to clearly categorize the morphology into different classes that could be then related to physical parameters. In the following, we  explore the relationship between the area of the emission with its perimeter and with the area of its convex hull.

\subsubsection{Area vs perimeter}
\label{area-per}

A first simple approach in the characterization of the morphology of the 1.38 mm continuum's spatial distribution, is to analyse the relationship between the area A and the perimeter P of the RoIs identified by thresholding images above the 5$\sigma$ level. %after distance normalization (see above in this section). 
In generalized terms, the two quantities are related as: 
\begin{equation}
    P=k A^{d/2}
\label{eq_perimarea} 
,\end{equation}
with $k$ depending on the specific geometric figure describing the RoI delimiting shape; for instance, in 2D we have  $k$=4 for a square or $k=2\sqrt{\pi}$ for a circle. 
%For an ellipse\juan{,} the k constant can be approximated by $k=\sqrt{(\pi/2)*(2r^2+1)/2}$, where $r$ is the aspect ration of the ellipse's axis.
% che poi sto k mi sa che l'ho derivato male....tanto non lo uso direttamente

The exponent $d$ is the dimension of the shape bounding the RoI; it is $d=1$ for linear contours in 2D space and $d=2$ for surfaces enclosing volumes in 3D. As we depart from regular Euclidean planar figures, replicating so-called generator patterns in a self-similar way at smaller and smaller scales to describe increasingly irregular contours, $d$ can assume non-integer values between 1 and 2 and in this we recognize the fractal dimension.

%Figure \ref{area_perim_roi} illustrates the perimeter-area plot for the largest 5$\sigma$ RoI identified in each ALMAGAL field. The distribution of points shows a continuum of situations, from morphologies that are in agreement with circular \juan{or} elongated shapes to more irregular shapes where the departure from circularity cannot be explained with elongated structures\juan{,} even with relatively large aspect ratios. In particular, we note that above a RoI area of $\sim5$ arcsec$^2$, the shape of the continuum \juan{gets} more and more irregular, clearly indicating the emergence of more patches of extended emission with increasingly irregular shapes. 

\begin{figure}[h]
\begin{center}
\includegraphics[width=0.48\textwidth]{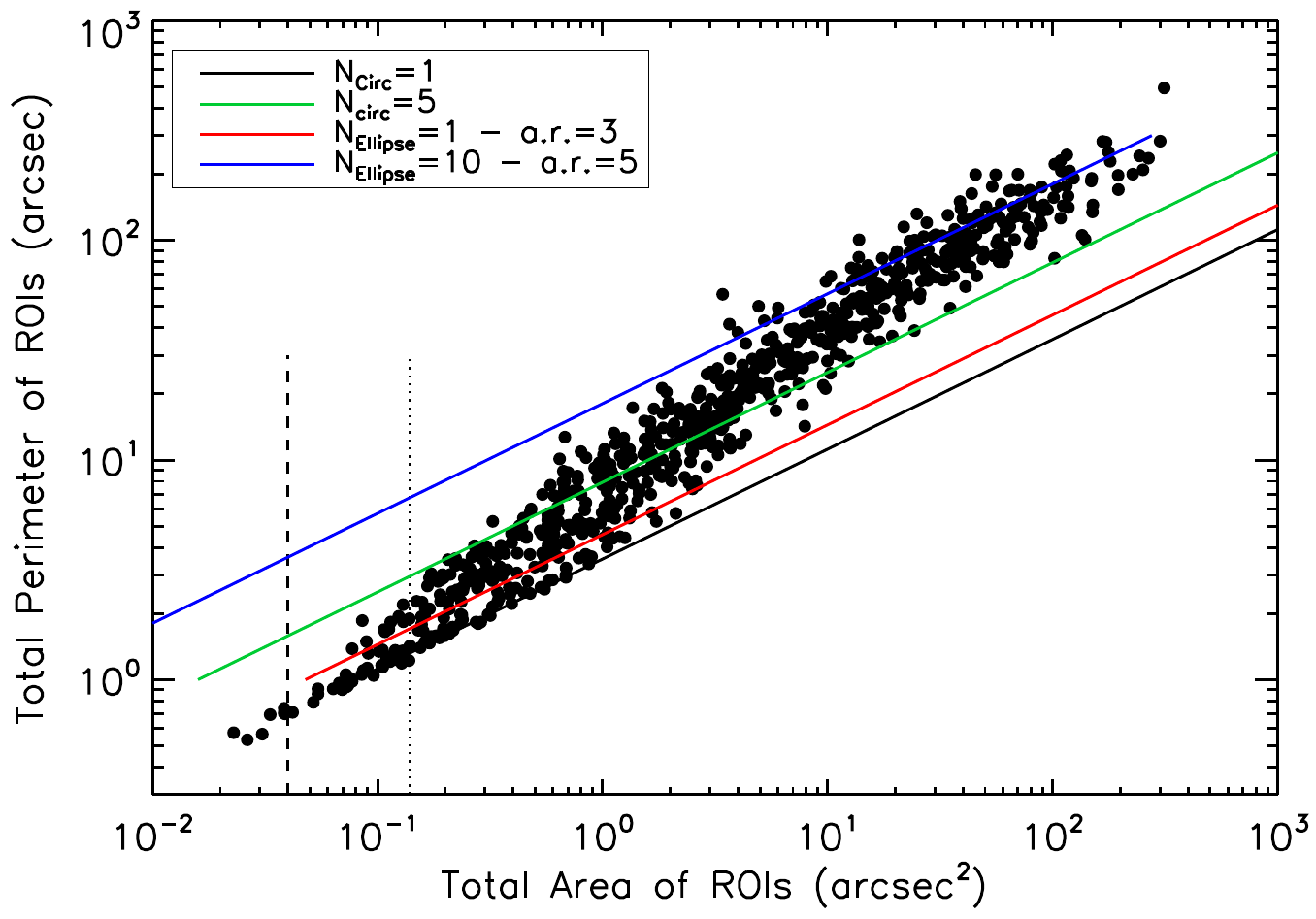} 
%\caption{Perimeter-Area relationship for the largest 5$\sigma$-RoI identified in each ALMAGAL field. The black and green lines represent Eq.\ref{eq_perimarea} in the case of a single circular shape or where the assumed perimeter is distributed for 5 circular shapes. The red and blue lines indicate the cases for a single ellipse with aspect ratio $r=3$, and 10 elliptical shapes of aspect ratio $r=5$.}
\caption{Perimeter-area relationship where the two parameters have been summed over all the 5$\sigma$-RoIs identified in each ALMAGAL field. The black and green lines represent Eq.\ref{eq_perimarea} in the case of a single circular shape or where the assumed perimeter is distributed for five circular shapes. The red and blue lines indicate the cases for a single ellipse with aspect ratio $\mathrm{a.r.}=3$ and ten elliptical shapes with $\mathrm{a.r.}=5$. The vertical dashed(dotted) lines show the median  area of the \fres\ beam for the $far/(near)$ sources (from \citealt{SanchezMonge+2025}). }
\label{area_perim_roi}
\end{center}
\end{figure}

Figure \ref{area_perim_roi} illustrates the perimeter-area plot for the 5$\sigma$ RoIs identified in each ALMAGAL field, where the two parameters have been summed over all RoI. To understand the meaning of the spread of the points in this plot, it is useful to mark locations that would be compatible with simple structures or their combination. For example, the black and red lines corresponds to Eq. \ref{eq_perimarea} in the case of one circle ($k=2\sqrt{\pi}$) or of one ellipse\footnote{For the ellipse we adopted an approximation for $k$ obtained from the approximation formula P$_{\rm ellipse}\sim \pi\sqrt{2(a^2+b^2)}$.} with aspect ratio $\mathrm{a.r.}=3$. We see that only in a small number of cases the emission area is compatible with a single circular or elliptical shape and only for small areas of a few arcsec$^2$, corresponding to about four to five times the maximum beam area represented by the vertical dashed line (from \citealt{SanchezMonge+2025}). As another example, if we assume that the total emission area is divided in, say, five equal circular RoIs then the correspondent locus in the plot is given by the green line; similarly, the blue line shows the locus in case the emission area was composed of ten equal elliptical RoIs with $\mathrm{a.r.}$=5.  

The distribution of points shows a continuum of situations. The limited vertical scatter for relatively small emission areas (left of the vertical dashed line) is a construction bias, as for these small areas, there is low maximum number of RoIs  it can be divided into and still be at least the size of the beam. The vertical scatter becomes much wider as we increase the total RoIs area. All clumps below and about the green line would be compatible with  small clusters of compact objects even for relatively large total emission areas (up to $\sim$100 sq. arcsec.). Clumps with emission showing a larger departure from the circularity could be compatible with even larger cluster of compact roughly circular objects, or a more limited number of filamentary patches idealized as the combination of ten high-a.r. ellipses and represented by the blue line. Of course, large deviations from a single circular RoI may also be compatible with a few patches of very irregular emission; we  present  an additional characterization in this respect below.

%, from morphologies that are in agreement with circular \juan{or} elongated shapes to more irregular shapes where the departure from circularity cannot be explained with elongated structures\juan{,} even with relatively large aspect ratios. In particular, we note that above a RoI area of $\sim5$ arcsec$^2$, the shape of the continuum \juan{gets} more and more irregular, clearly indicating the emergence of more patches of extended emission with increasingly irregular shapes. 

To quantify this behavior, we defined a parameter \deltacirc\ computed as the distance between each point and the black line (along the direction perpendicular to the line) in Fig. \ref{area_perim_roi}. 

\begin{figure}[h!]
\begin{center}
\includegraphics[width=0.47\textwidth]{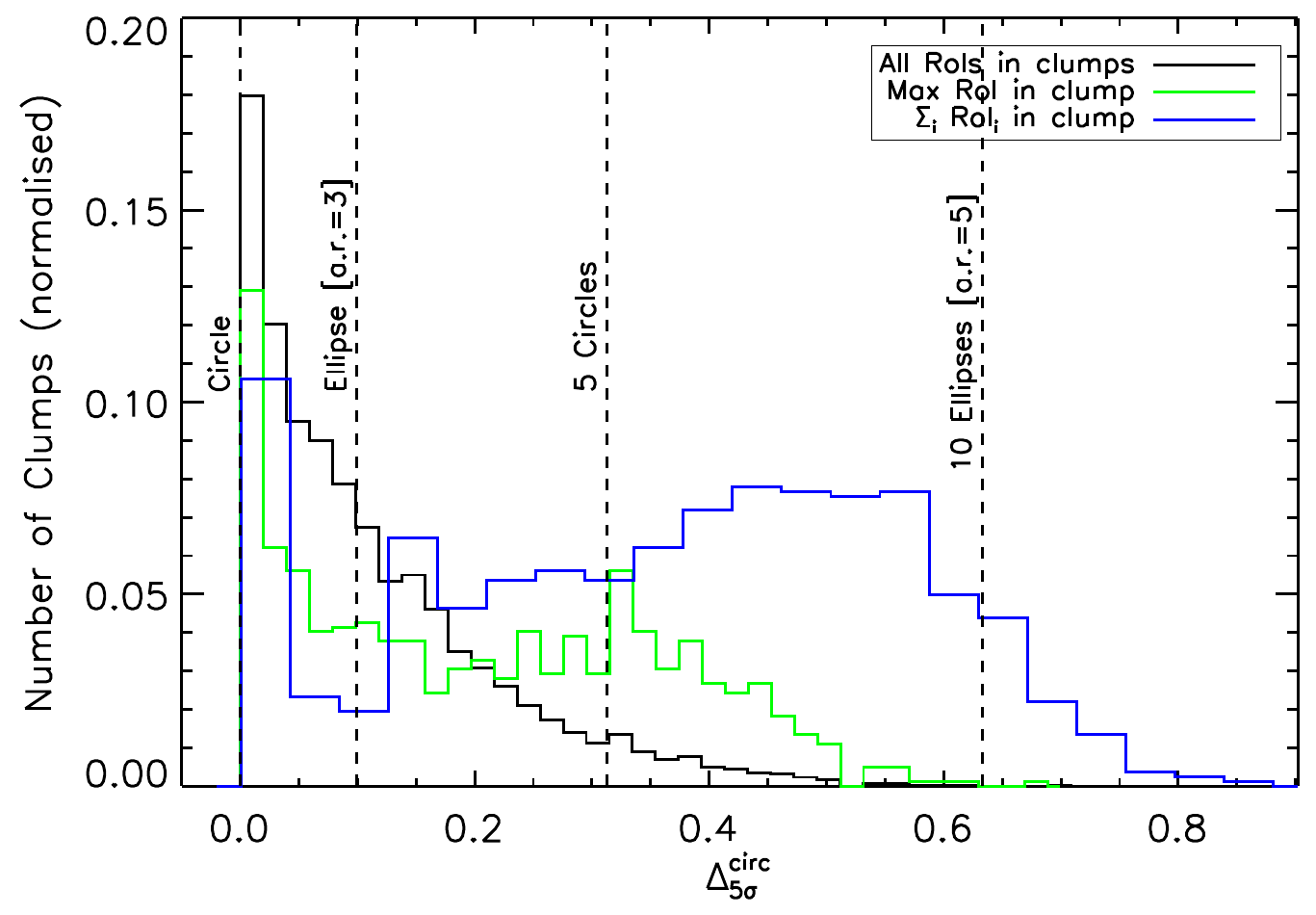} 
\caption{Normalized histograms of the \deltacirc\ parameter for the largest RoI in each field (black line) and for all 5$\sigma$ RoIs in the fields. The \deltacirc\ values corresponding to the loci indicated by the colored lines in Fig. \ref{area_perim_roi} are also reported here for reference.}
\label{histo_deltacirc}
\end{center}
\end{figure}

In Fig. \ref{histo_deltacirc}, we report the distribution of the \deltacirc\ parameters in our fields, computed individually for all the RoIs in the fields (black line), for the largest RoI in each field (green line), and for the sum of the RoIs in each field (blue line). Histograms are normalized to their integral to make their comparison meaningful. The distribution for all RoIs (black) shows a clear predominance of circular and slightly elongated shapes, with a monotonically decreasing trend toward larger values of \deltacirc. This might suggest that in many cases the secondary RoIs at the 5$\sigma$ level are dominated by small circular areas that are either fainter patches of extended emission or compact objects isolated with respect to the main emission area in the field.

When considering only the largest RoI in each field (green), the distribution is clearly different and shows a first component peaked around 0 (circular shape) decreasing down to about \deltacirc$\sim0.2$, followed by a second smaller and broader peak down to \deltacirc$\sim 0.5$. If, instead, we consider all the RoIs in each field and we compute the total perimeter and total area, the corresponding \deltacirc\ distribution is the blue line; in addition to this blue line, a narrow component skewed toward 0 (i.e., circular structure) shows an even broader (compared to the green distribution) component extending to $\sim0.8$. This metric seems therefore able to select fields that are showing either complex patterns of extended emission or collections of roughly circular  or elongated structures (or, mostly likely, both). Values of \deltacirc\ for all clumps are reported in Table 2.

\begin{figure}[h!]
\begin{center}
\includegraphics[width=0.47\textwidth]{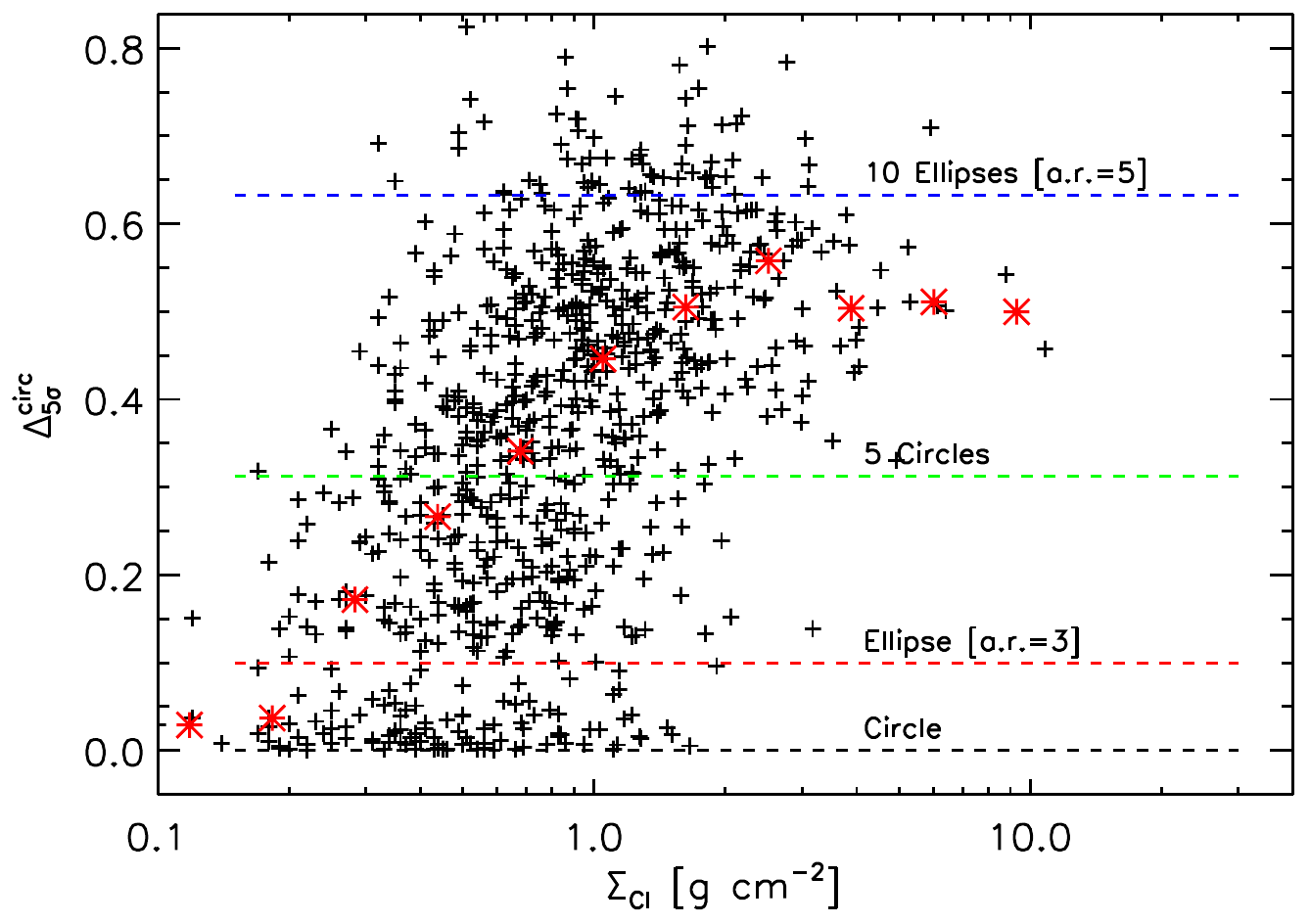} 
\caption{Departure from circularity, \deltacirc, of the total of the 5$\sigma$ ALMA emission RoIs in each field as a function of the clump surface density. The red asterisks are the medians of \deltacirc\ in logarithmic bins of surface density.}
\label{deltacirc_surfd_fig}
\end{center}
\end{figure}

It is interesting to verify whether these metrics are related to global clump properties. The relationship of the \deltacirc\ parameter with the clump-averaged surface density in Fig. \ref{deltacirc_surfd_fig} shows a positive correlation, although as usual with a large scatter around the median values computed in bins of $\Sigma_{\mathrm{Clump}}$ (the red asterisks). While clumps with $\Sigma_{\mathrm{Clump}}\leq$1\,\gcmtwo\ span the entire range of \deltacirc, almost the totality of denser clumps have \deltacirc\ higher than values compatible with circular or slightly elliptical shapes. Similarly to Fig. \ref{roi-area_surf_d}, we exclude the presence of remarkable distance biases in Fig. \ref{deltacirc_surfd_fig} as similar trends are found if the two parameters are plotted for sources in 1 kpc-bin distances (Fig. \ref{deltacirc_surfd_fig_d_bins}). 

\begin{figure}[h!]
\begin{center}
\includegraphics[width=0.47\textwidth]{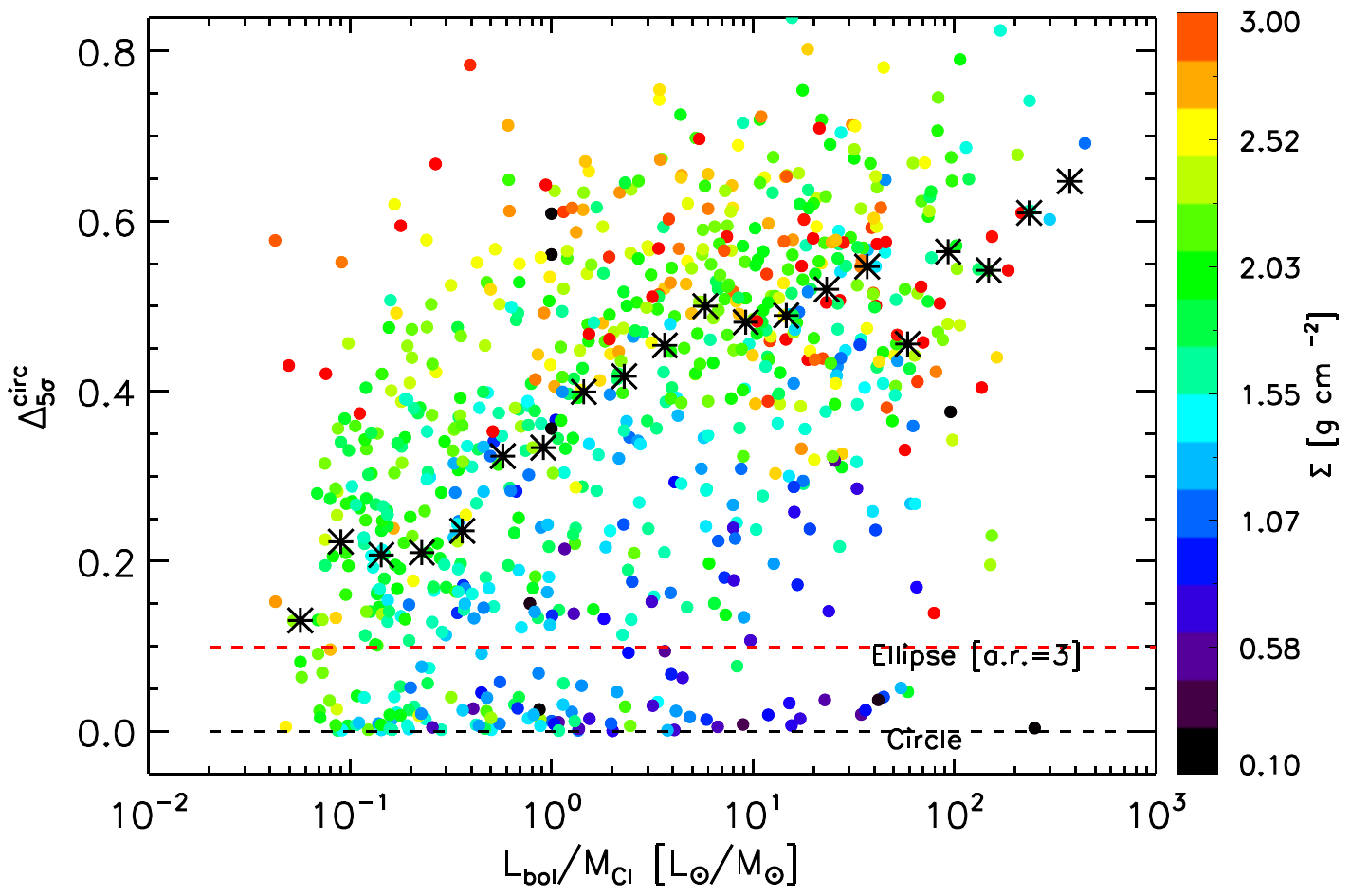} 
\caption{Departure from circularity, \deltacirc, of the total of the 5$\sigma$ ALMA emission RoIs in each field as a function of the clump, $L/M$. The color scale represents the clump surface density as reported in the lateral bar, while the black asterisks are the medians of \deltacirc\ in logarithmic bins of $L/M$ for the clumps with $\Sigma_{\mathrm{Clump}}\geq 1 $\gcmtwo.}
\label{deltacirc_lm_fig}
\end{center}
\end{figure}

A similar situation is found when relating \deltacirc\ with $L/M$. Figure \ref{deltacirc_lm_fig} shows that the scatter in the Y axis grows with evolution, with median-averaged values clearly rising with $L/M$, depicting the more and more frequent presence of complex morphologies with evolution. The points in the figure are color-coded with the clump surface density, and confirm (see also Fig. \ref{deltacirc_surfd_fig}) that this parameter has a strong role in modulating the scatter in the Y axis of Fig. \ref{deltacirc_lm_fig}. For example, if we restrict to the clumps with $\Sigma_{\mathrm{Clump}}\geq 1 $\gcmtwo\ (the cyan to red points), we are left with a moderate but significant  \deltacirc-$L/M$ trend with \cpear$\sim0.46$.

\subsubsection{Area versus convex hull}
\label{area-chull}

Another interesting metric to characterize the complexity of the emission morphology is offered by the comparison of the areas of the RoIs and the respective convex hull areas. For a bounded subset of the plane, the convex hull may be visualized as the shape enclosed by a rubber band stretched around the subset \citep{Berg+2008}. As it essentially represents the smallest convex geometric figure containing the emission RoIs, we expect the two areas to be very similar in case of very regular (roughly circular or elliptical) shapes. On the contrary,. the two areas will be different in case where the RoI is the result of the merging of two or more distinct regions, for instance, or in the case of hub-like filamentary structure. In the following, we define the parameter \qhull\ as the ratio $A_{RoI}/A_{C-Hull}$ between the area of the largest RoI and its convex hull area computed over the areas with signal above the 5$\sigma$ level.
%{\sf\textcolor{red}{I find that confusing. Usually you compare the area under convex hull with the enclocing CIRCLE. If the object is very elongated then this ratio is very large (or small depending on how you define the ratio). The RoI's of the objects are elliptical, as indicated in the discussion further above. If you compare with the ellipse, you can always adjust the aspect ratio a.r. such that the difference becomes minimal. The convex hull -- per definition -- is close to the minimum ellipse that covers the structure. In that sense, you may need to redo this part. --- Specifically, since you DO compare with the enclosing minimum circle in the analysis further down. --- So, I think this needs to be corrected here.}}

\begin{figure}[h!]
\begin{center}
\includegraphics[width=0.47\textwidth]{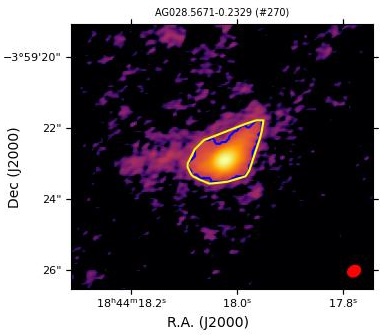} 
\includegraphics[width=0.47\textwidth]{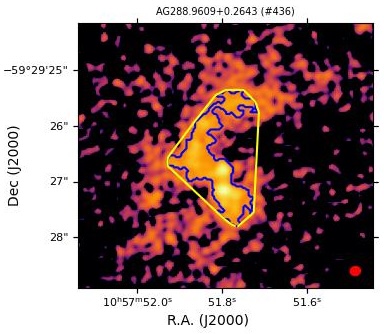} 
\caption{Two ALMAGAL fields showing different morphologies for similar 5$\sigma$ emission areas. The blue contours in each field correspond to the 5$\sigma$ noise level, while the yellow polygons represent the computed convex hull for the largest %of the 
RoI in the fields. We have \qhull\ is $\sim$0.85 for the field in the top map, and $\sim$0.5 for the field in the bottom map. In both plots, the red ellipse in the bottom-right corner is the beam. Source AG name and running number as in Table 1.}
\label{area-roi-chull}
\end{center}
\end{figure}

\begin{figure}[h!]
\begin{center}
\includegraphics[width=0.47\textwidth]{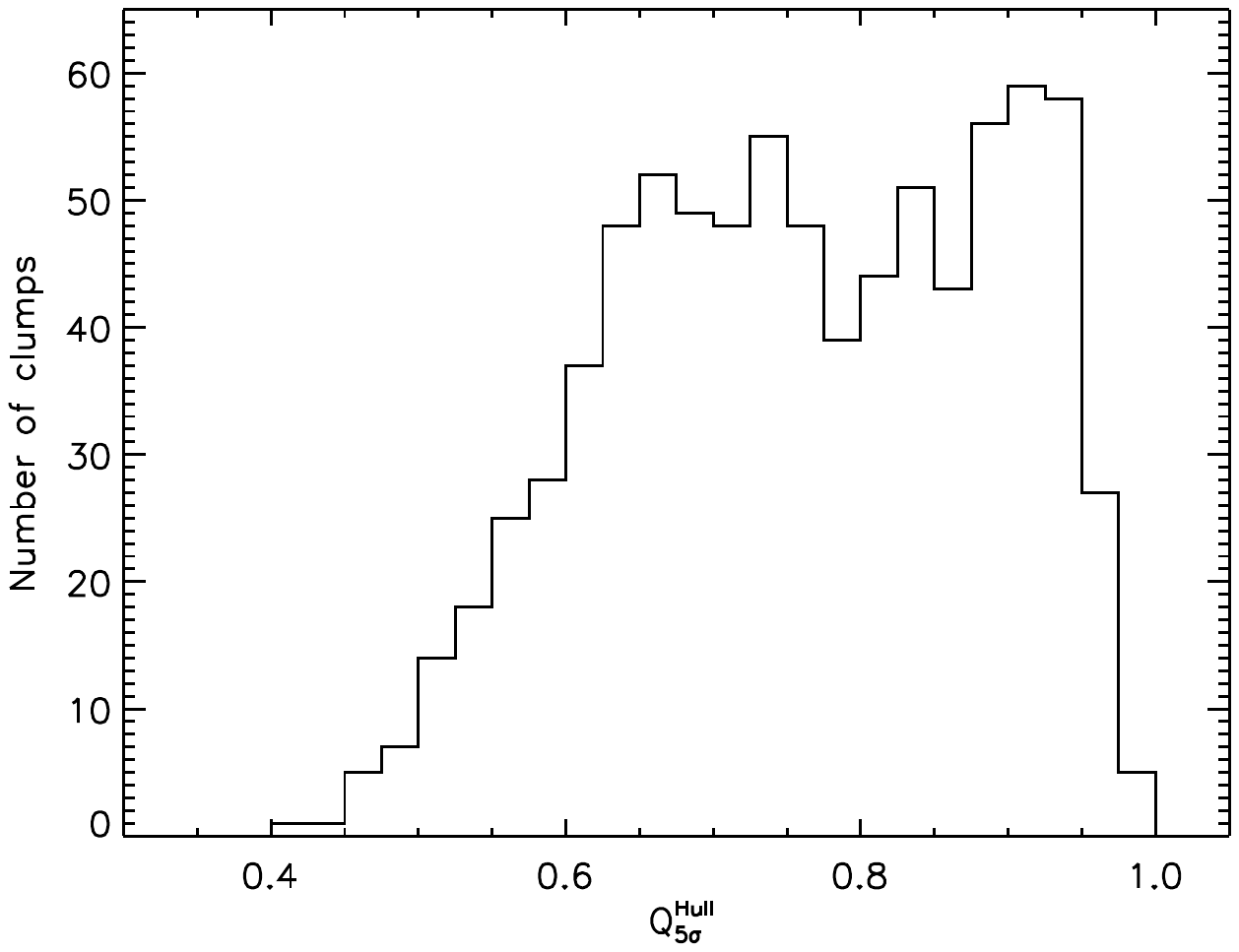} 
\caption{Distribution of the \qhull\ parameter for all ALMAGAL fields where 5$\sigma$ continuum emission is detected.}
\label{histo-roi-chull}
\end{center}
\end{figure}

Figure~\ref{area-roi-chull} shows an example of two extreme cases in which the emission area above 5$\sigma$ has a similar extent (inside the blue contour), but \qhull\ is relatively high ($\sim0.85$) for the relatively regular source (top), and drops to 0.5 for the more complex structure field (bottom). Figure~\ref{histo-roi-chull} reports the histogram of this parameter over the entire sample where 5$\sigma$ continuum emission is detected, and suggests that there may possibly be two peaks in the distribution, separated at a value of $\sim$0.8. We ran the Silverman test \citep{Silverman1981} to characterize multi-modality using kernel density estimates over the distribution of the original data, and hence not on binned histograms. The test confirms that the distribution reported in Fig. \ref{histo-roi-chull} has a 100\%\ probability of being better described by two modes instead of one. \qhull\ then offers a simple metric to classify fields based on the convex (or not) nature of their continuum emission. 

\begin{figure}[h!]
\begin{center}
\includegraphics[width=0.47\textwidth]{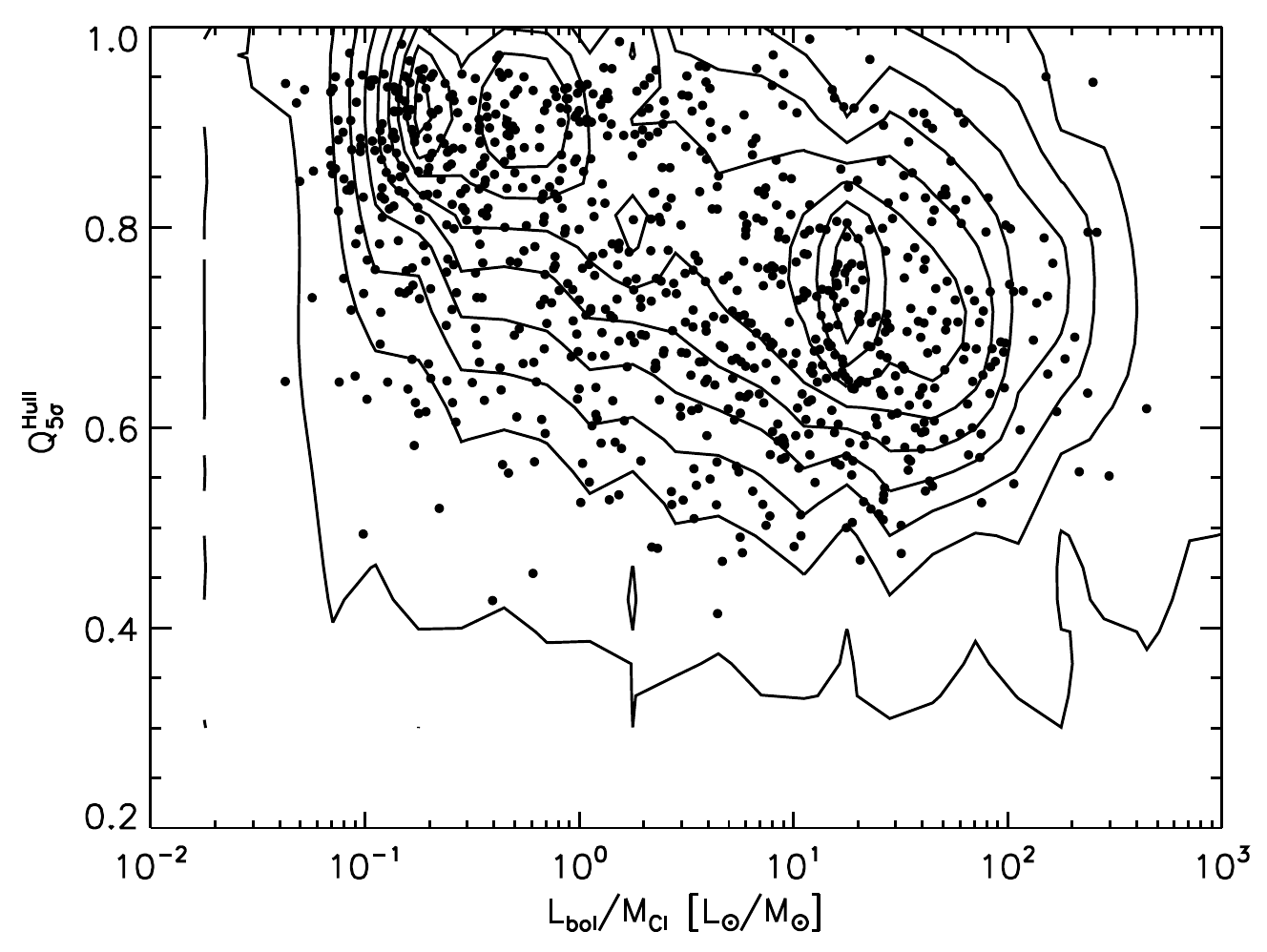} 
\caption{Distribution of \qhull\ for all ALMAGAL fields with respect to the evolutionary stage of the clump as traced by the $L/M$. Contours represent the normalized source density in steps of 10\%.}
\label{area-roi-chull-lm}
\end{center}
\end{figure}

\begin{figure}[h!]
\begin{center}
\includegraphics[width=0.47\textwidth]{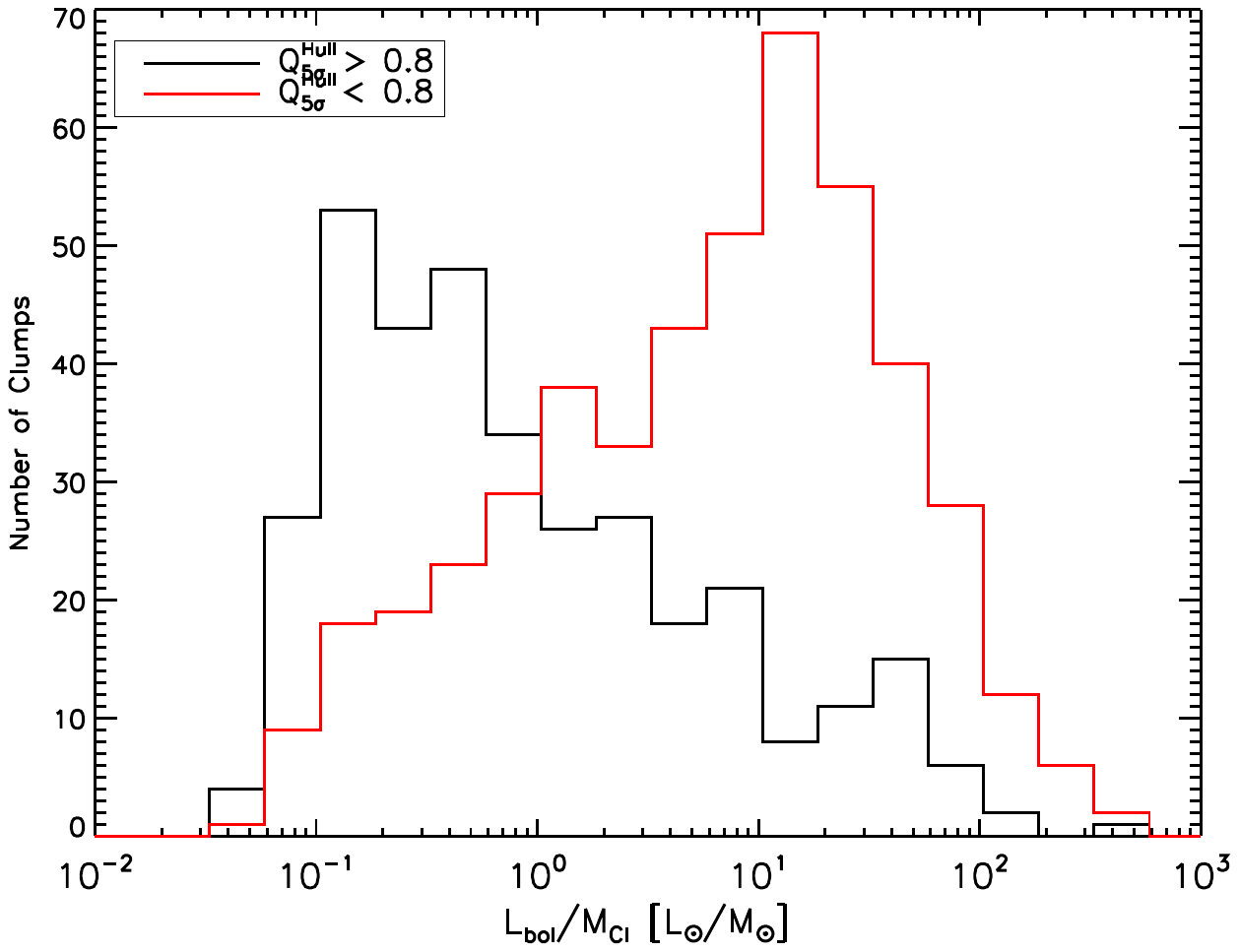} 
\caption{Distribution of the $L/M$ parameters for fields with \qhull\ parameter above (black line) or below (red line) the value of 0.8.}
\label{histo-area-roi-chull-lm}
\end{center}
\end{figure}

The relationship of this parameter with the clump's evolutionary stage in Fig. \ref{area-roi-chull-lm} shows a broadly decreasing trend with $L/M$, indicating that the emission morphology of the largest emission area in each field gets more and more complex as the field is more evolved. This is also clearly shown in Fig.~\ref{histo-area-roi-chull-lm}, where the distribution of $L/M$ is reported separately for the fields with ``regular'' (\qhull$\geq$0.8, black  line) and ``irregular'' (\qhull$\leq$0.8, red line) continuum emission morphologies. The $L/M$ for the two fields in Fig. \ref{area-roi-chull} is $\sim$0.05 for AG028.5671-0.2329 and $\sim26$ for AG288.9609+0.2643, respectively.

\begin{figure}[h!]
\begin{center}
\includegraphics[width=0.47\textwidth]{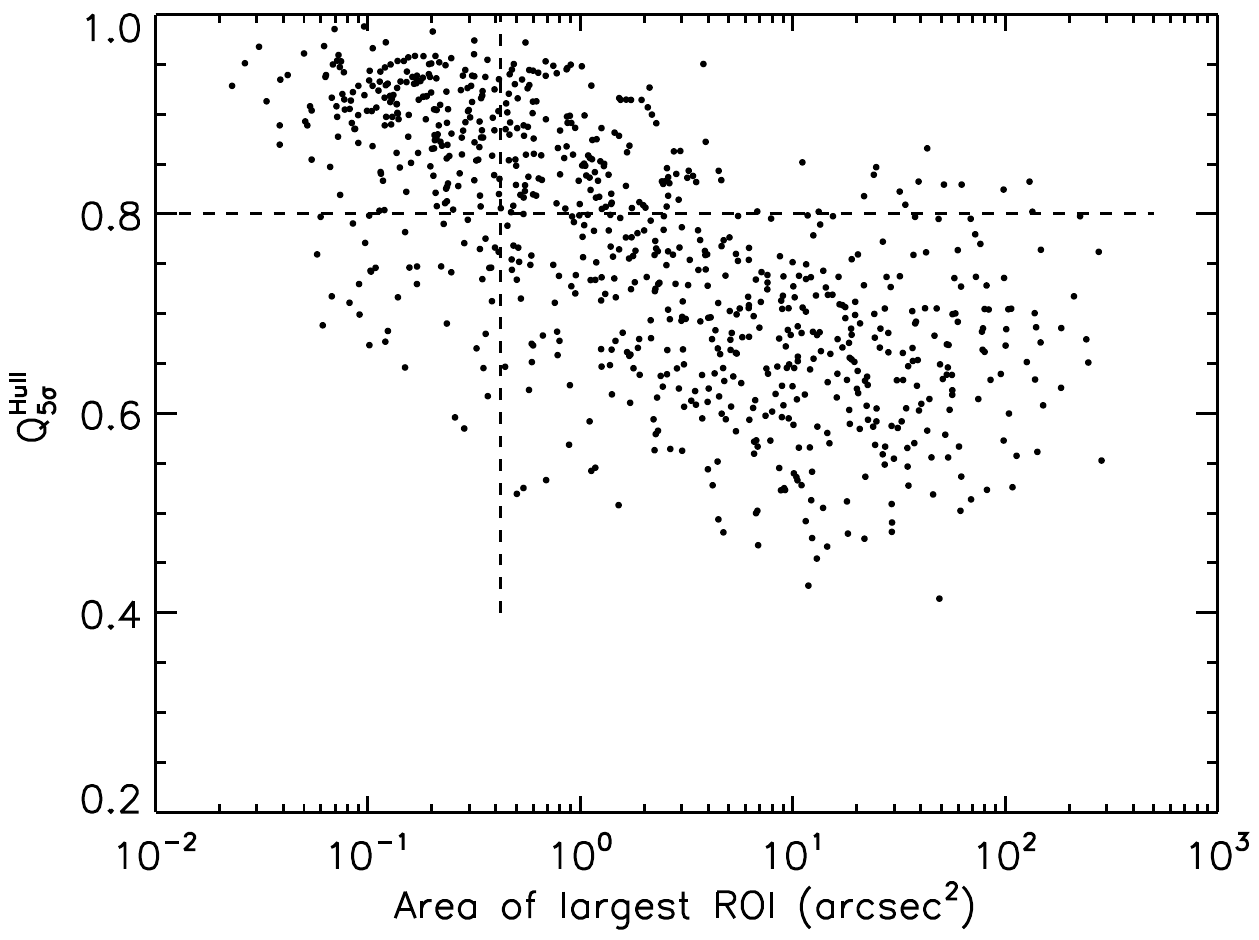} 
\caption{Distribution of the \qhull\ parameter as a function of the area of emission above 5$\sigma$. The horizontal dashed line marks the \qhull$\sim$0.8 threshold identified in Fig. \ref{histo-roi-chull} to distinguish roughly circular areas of emission from more complex morphologies. The vertical dashed line indicates the maximum area of the beam for all fields, showing that a significant fraction of approximately circular emission areas (above the horizontal line) have areas much larger than the beam.}
\label{area-vs-chull}
\end{center}
\end{figure}

There is, of course, a bias introduced by the finite resolution of the images. Fields with only compact beam-like emission patches will naturally be roughly circular in morphology, making this metric not informative. These are, however, only a portion of the fields with high \qhull. Figure \ref{area-vs-chull} illustrates the relationship between \qhull\ and the 5$\sigma$ emission area. The horizontal dashed line marks the \qhull$\sim$0.8 threshold identified in Fig. \ref{histo-roi-chull} to distinguish approximately circular areas of emission from more complex morphologies, while the vertical dashed line indicates the maximum area of the beam for all fields. If roughly circular areas were only due to beam-like or modestly resolved compact emission, we would not expect to see points in the top-right quadrant (defined by the two dashed lines) that instead contains $\sim44$\%\ of the fields with \qhull$\geq$0.8. We then conclude that the \qhull\ metric contains valuable information to classify the continuum emission morphology.

From a qualitative viewpoint, both the \deltacirc\ and the \qhull\ metrics suggest that a more complex dust spatial structure is observed as the clumps evolve. This is fully compatible with a concurrent contribution from increasing column density and/or increasing temperature of the dust with $L/M$ due to radiative (and possibly dynamical) feedback from the forming protostars. In both cases, the millimeter flux would increase, causing more and more material to become detectable above the noise. Figure \ref{roi-area_lm} shows that indeed the area of detectable emission increases with $L/M$ (or evolution), and Figs. \ref{deltacirc_lm_fig}, \ref{area-roi-chull-lm}, and \ref{histo-area-roi-chull-lm} show that this larger area deviates more and more from a roughly circular shape and gets more and more irregular and complex with $L/M$. In other words as the clumps evolve we do not find relevant occurrences of large and circular areas of emission. This certainly agrees with a dynamically evolving clump fragmentation, also influenced by increasing levels of feedback. Conversely, these findings do not support the notion of relatively massive seeds of fragmentation that regularly grow in extension with increasing column density or temperatures. In Table 2, available at the CDS, contains some of the parameters we have presented in this section; namely, for each field, the emission area above 5$\sigma$ level both in physical units and as fraction of the field of view, plus the \deltacirc\ and the \qhull\ parameters determined for the largest detected structure in each field.

\begin{table}
\label{morph_par_table}
\end{table}

\section{Forward outlook}
\label{outlook}

ALMAGAL is a game changer in areas like fragmentation statistics and morphologies, gas dynamics in infall and outflow motions in the clumps, chemistry, and systematic disk detections, among others. Here, we present a brief summary of the more immediate works that are already published or will be submitted and published  in the next six months. 
\cite{Wells+2024} published a first study of flow dynamics in a subsample of 100 ALMAGAL fields based on \hiico\ lines. \cite{Coletta+2025} presents the first release of a continuum compact source catalog, which includes the core physical properties and a first discussion of the core mass function in evolutionary terms. Elia et al. (in prep.) will discuss core properties in the context of the hosting clumps, while Schisano et al. (in prep.) will present an in-depth analysis of the morphological distribution of cores and their mass segregation. Wallace et al. (in prep.) will investigate the multiscale morphological properties of continuum emission, greatly extending the preliminary analysis presented in this paper. \cite{Mininni+2025} discusses the multiscale morphological relationship between the dust continuum and molecular line emission of CH$_3$CN, CH$_3$OH, H$_2$CO, HC$_3$N, and other tracers. Benedettini et al. (in prep.) analyses the relationship between the distribution of \vlsr\ of molecular lines and the dynamical properties of the hosting clumps gas. An investigation of the $^{13}$C/$^{12}$C Galactic gradient will be presented by Law et al. (in prep.), while Stroud et al. (in prep.) will discuss the evolutionary properties of continuum cores in the context of the association with methanol masers. Finally, Jones et al. (in prep.) and Allande et al. (in prep.) will present a first catalog of molecular lines, complete with the derived gas physical properties, toward the 1.38 mm cores of \citet{Coletta+2025}.

Extending the forward look toward 2026, we anticipate studies on complex organic molecules (Allande et al., in prep.), line-based physical and evolutionary studies based on machine-learning methodologies (Sanchez-Monge et al., in prep.) and cores virial analysis (Mininni et al., in prep.). Finally, the ALMAGAL data have also proved effective for preliminary studies of disks towards the brightest cores that led to follow-up studies that will be presented by Ahmadi et al. (in prep.).

A first complete public release of the \fres\ and \ires\ continuum images will be done after the acceptance of the first round of continuum-based papers introduced above, and we anticipate this to take place no later than December 2025. A first round of papers with a preliminary exploitation of the \fres\ and \ires\ spectroscopic datacubes is in relatively less advanced stage, so that their public release  takes place no later than December 2026.

\section{Conclusions}
\label{concl}
The ALMAGAL Large Program has mapped 1013 dense star-forming clumps in the 1.38 mm continuum and lines to obtain the first statistically significant demographic description of the fragmentation process that leads to the formation of dense cores. In this first paper, we present a complete review of the target clumps physical properties, with a first analysis of the morphological properties of the dust continuum emission. We find that metrics such as the perimeter and area ratio, along with the convex hull and area ratio, are effective in mapping the emergence of complex irregular morphologies proportionally to the evolutionary stage of the clumps and their surface density. 

\begin{acknowledgements}
S.M and the Team at INAF-IAPS gratefully acknowledge financial support from the European Research Council via the ERC Synergy Grant ``ECOGAL'' (project ID 855130). R.S.K.\ acknowledges financial support from the European Research Council via the ERC Synergy Grant ``ECOGAL'' (project ID 855130),  from the Heidelberg Cluster of Excellence (EXC 2181 - 390900948) ``STRUCTURES'', funded by the German Excellence Strategy, and from the German Ministry for Economic Affairs and Climate Action in project ``MAINN'' (funding ID 50OO2206). R.S.K. is grateful for support from the Harvard Radcliffe Institute for Advanced Studies and Harvard-Smithsonian Center for Astrophysics for their hospitality and support during his sabbatical. The team in Heidelberg also thanks for computing resources provided by the Ministry of Science, Research and the Arts (MWK) of {\em The L\"{a}nd} through bwHPC and the German Science Foundation (DFG) through grant INST 35/1134-1 FUGG and 35/1597-1 FUGG, and also for data storage at SDS@hd funded through grants INST 35/1314-1 FUGG and INST 35/1503-1 FUGG. S.W. gratefully acknowledges the Deutsche Forschungsgemeinschaft (DFG) for funding through SFB~1601 ``Habitats of massive stars across cosmic time’' (sub-project A5) and the ''NRW-Cluster for data-intensive radio astronomy: Big Bang to Big Data (B3D)'' funded through the ''Profilbildung 2020'' program of the Ministry of Culture and Science of the State of North Rhine-Westphalia, Germany.
A.S.-M.\ acknowledges support from the RyC2021-032892-I grant funded by MCIN/AEI/10.13039/501100011033 and by the European Union `Next GenerationEU’/PRTR, as well as the program Unidad de Excelencia María de Maeztu CEX2020-001058-M, and support from the PID2020-117710GB-I00 (MCI-AEI-FEDER, UE). L.B. gratefully acknowledges support by the ANID BASAL project FB210003. 
G.A.F gratefully acknowledges the Deutsche Forschungsgemeinschaft (DFG) for funding through SFB~1601 ``Habitats of massive stars across cosmic time’' (sub-project B2), the University of Cologne and its Global Faculty Programme. A.A. acknowledges support from the UK ALMA Regional Centre (ARC) Node which is supported by the Science and Technology Facilities Council [grant number ST/T001488/1]. C.B. and J.W. gratefully acknowledge funding from the National Science Foundation under Award Nos. 2108938 and 2206510, as well as CAREER 2145689 for C.B. R.K.  acknowledges financial support via the Heisenberg Research Grant funded by the Deutsche Forschungsgemeinschaft (DFG, German Research Foundation) under grant no. KU 2849/9, project no. 445783058. Part of this research was carried out at the Jet Propulsion Laboratory, California Institute of Technology, under a contract with the National Aeronautics and Space Administration (80NM0018D0004).
\\
This paper makes use of the following ALMA data: ADS/JAO.ALMA\#2019.1.00195.L. ALMA is a partnership of ESO (representing its member states), NSF (USA) and NINS (Japan), together with NRC (Canada), MOST and ASIAA (Taiwan), and KASI (Republic of Korea), in cooperation with the Republic of Chile. The Joint ALMA Observatory is operated by ESO, AUI/NRAO and NAOJ.
\end{acknowledgements}

\bibliographystyle{aa}
\bibliography{sergio_bib_2}
%\bibliography{sergio_bib_paper0a}
%\bibliography{sergio_bib.bib}

% switch to one column as per recommendations form A&A doc
\onecolumn

\begin{appendix}

\section{ALMAGAL Clumps Properties}
\label{appendix_clump_prop}
%here put the Tabel with ALMAGAL clump properties

Here, we  give a brief description the Table 1, available at the CDS, containing the consolidated properties of the ALMAGAL targets revised as explained in the main text:

\begin{itemize}
        \item Col. 1: an ordinal number of the field
        \item Cols. 2-3: unique ALMAGAL source name file ID together with the original designations either in the Hi-GAL band-merged catalog ("HBM" prefix) ot the RMS survey ("G" prefix) 
        \item Cols. 4-5: Galactic coordinates of the fields center 
        \item Col 6: the target membership to either "near" or "far" ALMAGAL subsamples (col. 6) defining the ALMA antenna configurations used and hence the minimum achieved angular resolution (0\asec,15 and 0\asec.3 respectively)
        \item Col. 7: revised \vlsr
        \item Col. 8: revised heliocentric distance
        \item Cols. 9-10: Clump mass and bolometric luminosity from \cite{Elia+2021}, or derived with the same methodology (see Sect. \ref{sample_homo} above) rescaled at the revised distance of Col. 8
        \item Cols. 11-12: cold dust temperature and bolometric temperature as in \cite{Elia+2021}, or derived with the same methodology. 
        \item Col. 13: clump surface density reported from \cite{Elia+2021}, with the difference that the area used for the computation is based on the measured clump size without deconvolving the \textit{Herschel} beam.
        \item Col. 14: Evolutionary class after \cite{Urquhart+2022}: Q (quiescent), P (protostellar), Y (young stellar object), and H (HII region). Cases where the methodology failed to identify a specific class are labeled "n/a".
\end{itemize} 

The last column may contain a series of strings describing some specificities about the revision process of the target properties. The strings have the following meanings:

\begin{itemize}
\item\textbf{-kda\_defnear}: source originally set at FAR kinematic distance solution by Mege et al. 2021 in absence of any KDA discriminator. In this work the NEAR distance decision is instead adopted for these sources, for consistency with the approach followed for the sources from the RMS sample.
\item\textbf{-V7}: the new ALMAGAL-based \vlsr\ differs from the original by more than 7\,\kms
\item\textbf{-D500}: the new ALMAGAL-based distance differs from the original by more than 500pc.
\item\textbf{-Dn2f}: the new distance determined from the \vlsr\ of ALMAGAL dense tracer lines, switches this source from the NEAR to the FAR sample. As originally belonging to the NEAR sample it was observed with a maximum 0.3 arcsec resolution, therefore now resulting in a sub-optimal linear resolution.
\item\textbf{-Df2n}: the new distance determined from the \vlsr\ of ALMAGAL dense tracer lines, switches this source from the FAR to the NEAR sample. As originally belonging to the FAR sample it was observed with a maximum 0.15 arcsec resolution, therefore now resulting in a super-optimal linear resolution.
\item\textbf{-D>8}: the new distance determined from the \vlsr\ of ALMAGAL dense tracer lines, puts the source beyond the distance limit used to select the original sample.
\item\textbf{-refit40k}: the dust temperature estimated from \textit{Herschel} fluxes has been updated here with respect to Elia et al. (2017, 2021) due to a 40K hard limit that was reached in the grey-body fit.
\item\textbf{-sed\_irr}: RMS source with a Hi-GAL counterpart in the \textit{Herschel} 5 bands, but the SED is irregular (not a single downward-facing concavity) and did not pass criteria to be included in Elia et al. (2017,2021). This criteria have been relaxed for the present work to estimate physical parameters.
\item\textbf{-sed\_rebuilt}: RMS source not in Elia et al. (2017, 2021), that needed a custom build of the \textit{Herschel} SED to estimate physical parameters in the present work. Possible reasons: source had multiple counterparts in one \textit{Herschel} band, or counterparts were not linked in a single SED.
\item\textbf{-sed\_sat}: RMS source not in Elia et al. (2017, 2021) due to saturation in some \textit{Herschel} bands. To estimate physical parameters for the present work the SED has been manually built allowing missing bands.
\item\textbf{-sed\_noband}: RMS source not in Elia et al. (2017, 2021) because of missing 350-500\um \textit{Herschel} bands due to source faintness. Physical parameters for the present work are estimated using three remaining bands
\item\textbf{-no\_props}: RMS source not in Elia et al. (2017, 2021) because has a \textit{Herschel} counterpart in 1 or 2 bands only. It was not possible to estimate physical parameters.
\end{itemize}

\eject

\section{Plots for distance bias characterization and other supporting material}
\label{dist_plots}

This section reports useful plots for the characterization of observational biases and to support the conclusions in the various relationships explored in this work. They are referenced where needed in the body of the paper.

%\begin{figure}[t]
%\begin{center}
%\includegraphics[width=0.47\textwidth]{emissionarea-vs-F350-d2_tm1_nsig5.pdf} 
%\caption{ALMA 1.38`mm total emission area above 5$\sigma$ in the \fres\ images as a function of the 350~\um\ monochromatic luminosity, obtained multiplying the Herschel 350\um\ flux by 4$\pi d^2$ (with distance in kpc). The relationships between the two parameters is therefore not biased by distance.}
%\label{roi-area_lum350}
%\end{center}
%\end{figure}

\begin{figure}[h]
\begin{center}
\includegraphics[width=0.47\textwidth]{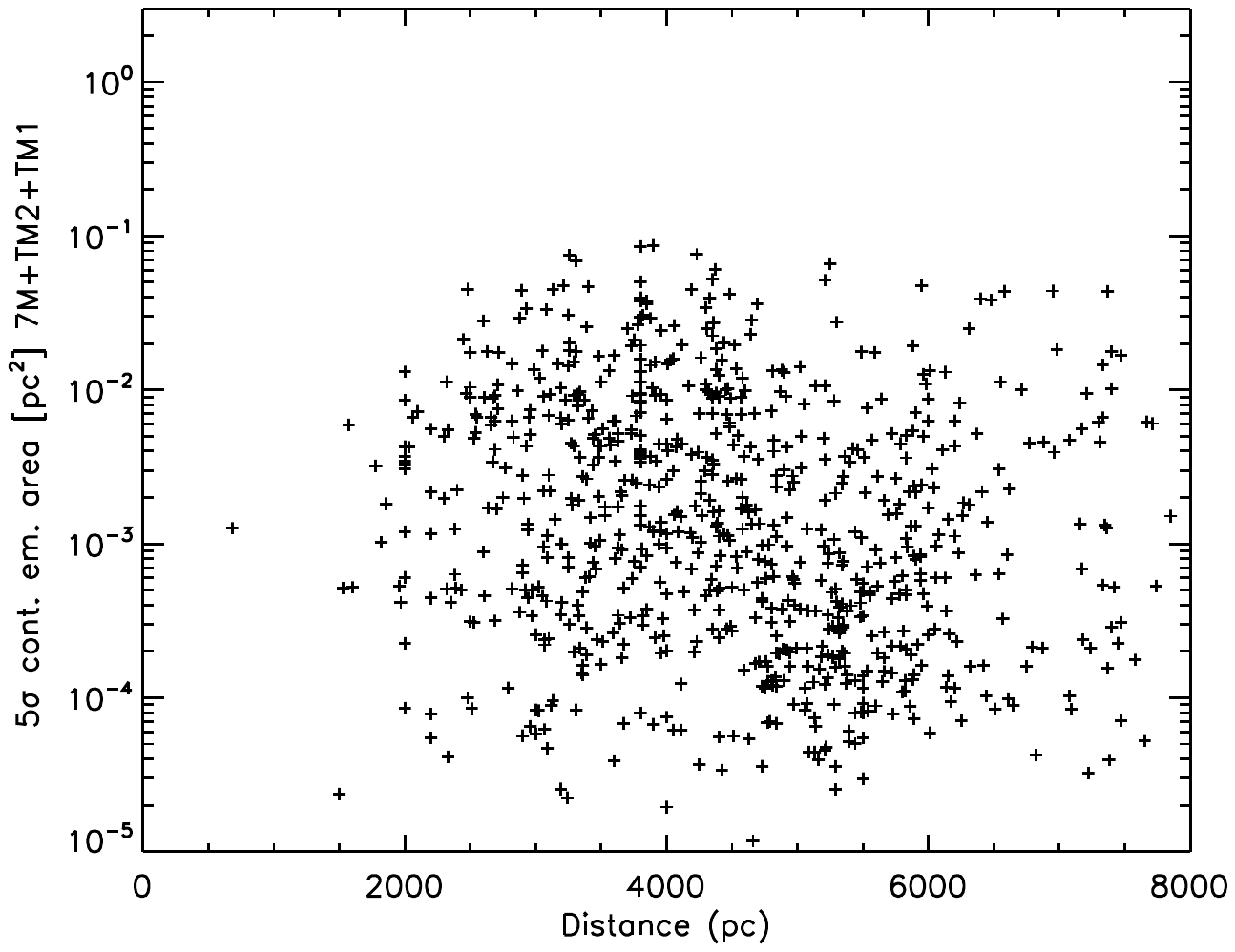} 
\caption{ALMA 1.38`mm total emission area above 5$\sigma$ in the \fres\ images as a function of source distance. No relationship can be seen.}
\label{roi-area_dist}
\end{center}
\end{figure}

\begin{figure}[h]
\begin{center}
\includegraphics[width=0.47\textwidth]{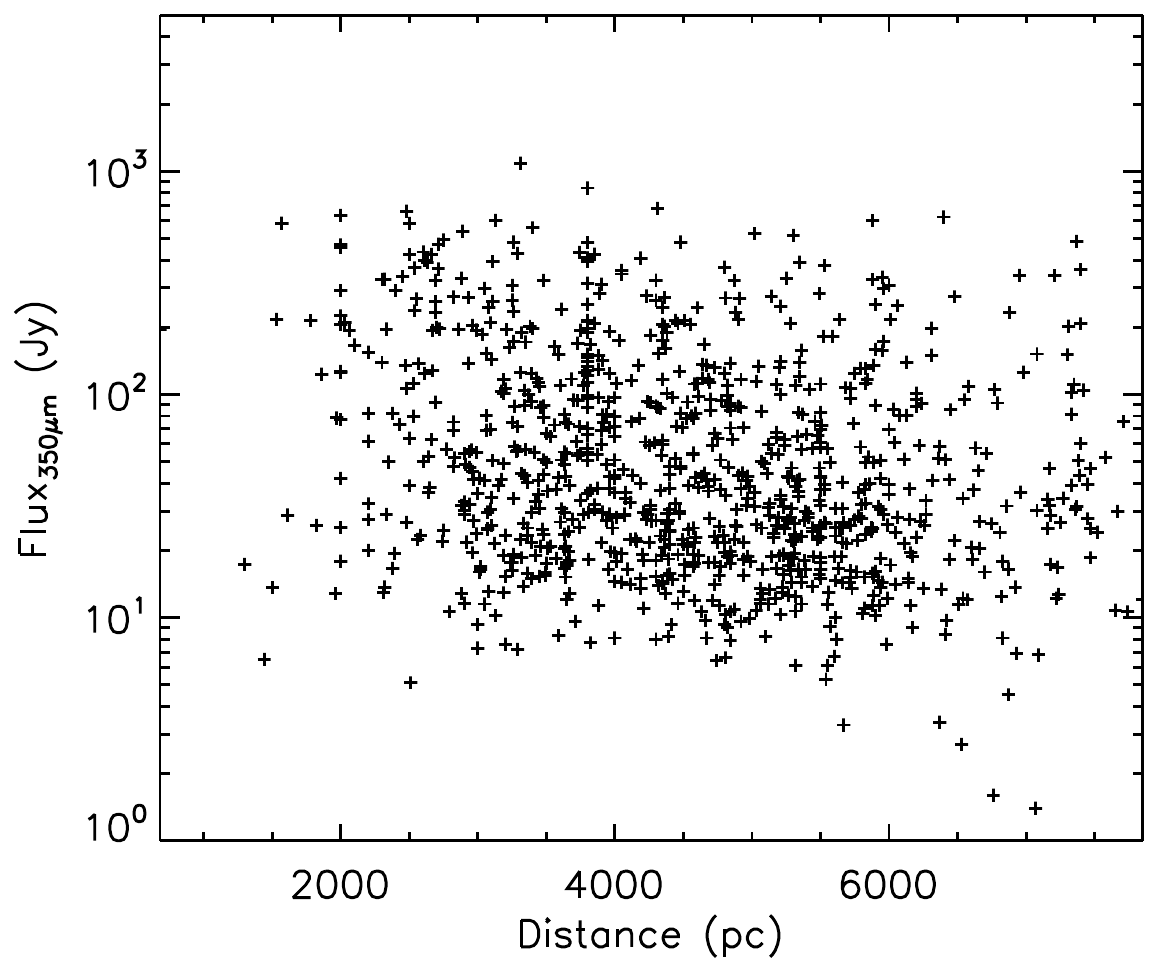} 
\caption{\textit{Herschel} 350\um\ flux for each clumps as a function of clump  distance. No relationship can be seen.}
\label{f350_dist}
\end{center}
\end{figure}

\begin{figure*}[h]
\begin{center}
\includegraphics[width=0.32\textwidth]{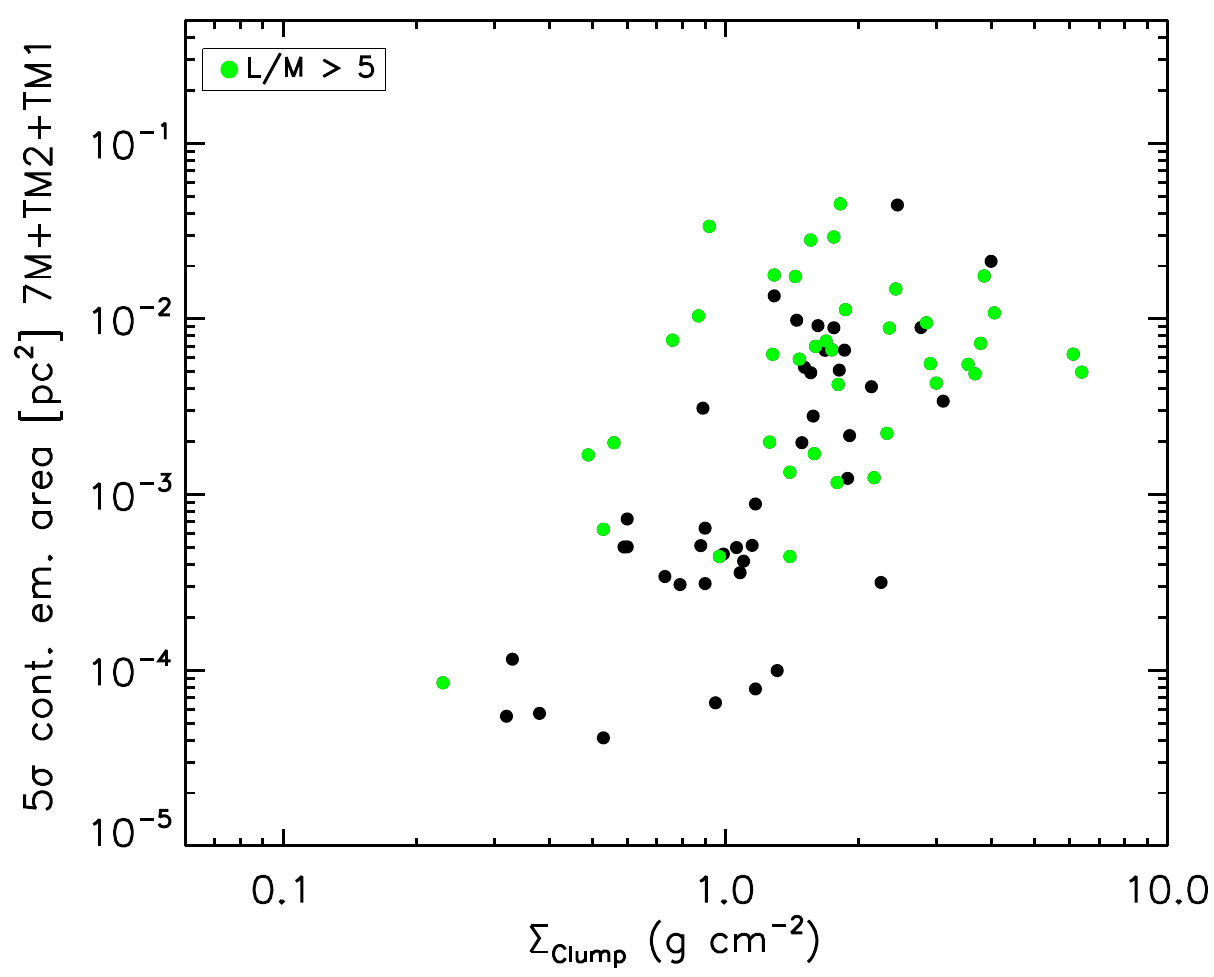} 
\includegraphics[width=0.32\textwidth]{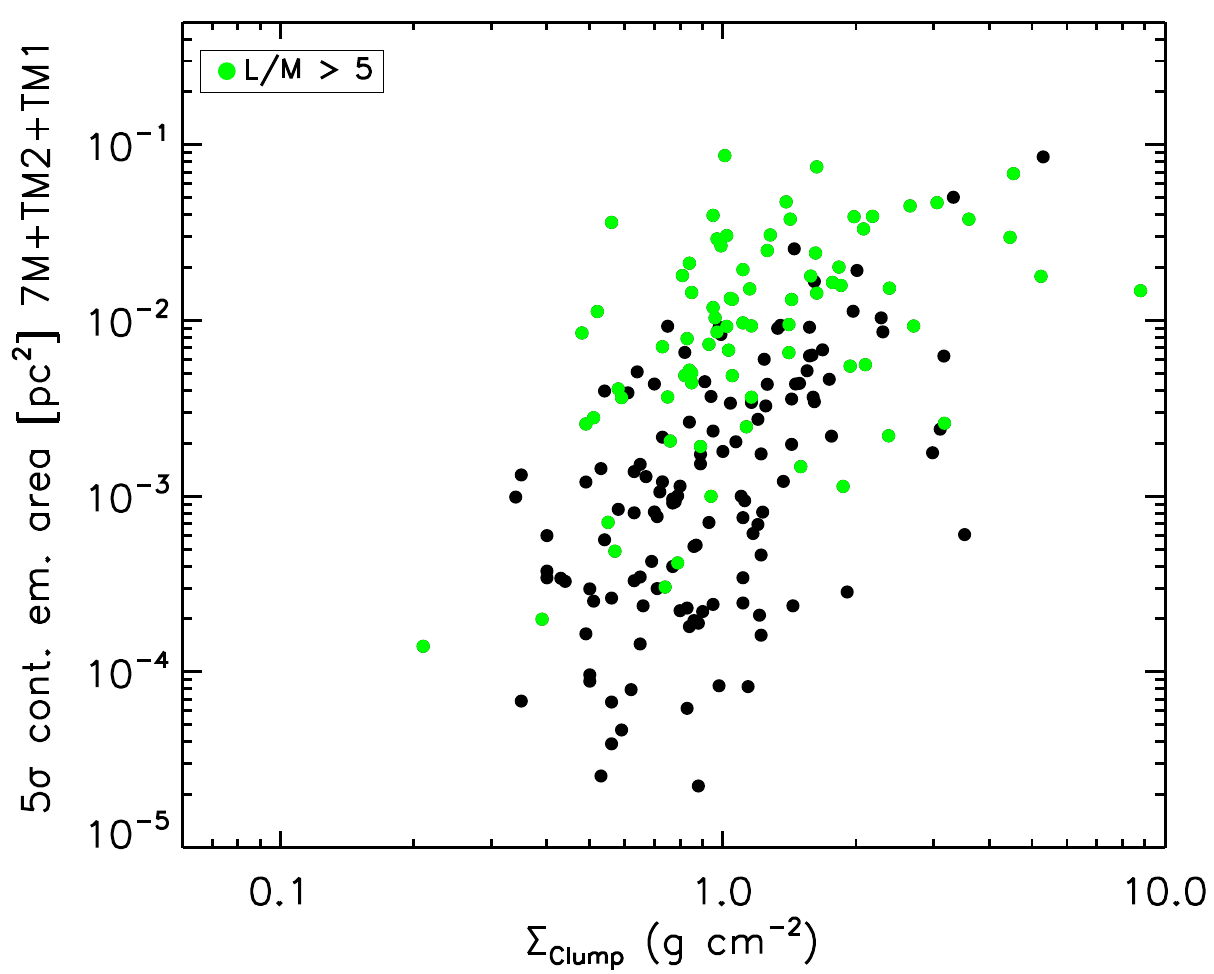} 
\includegraphics[width=0.32\textwidth]{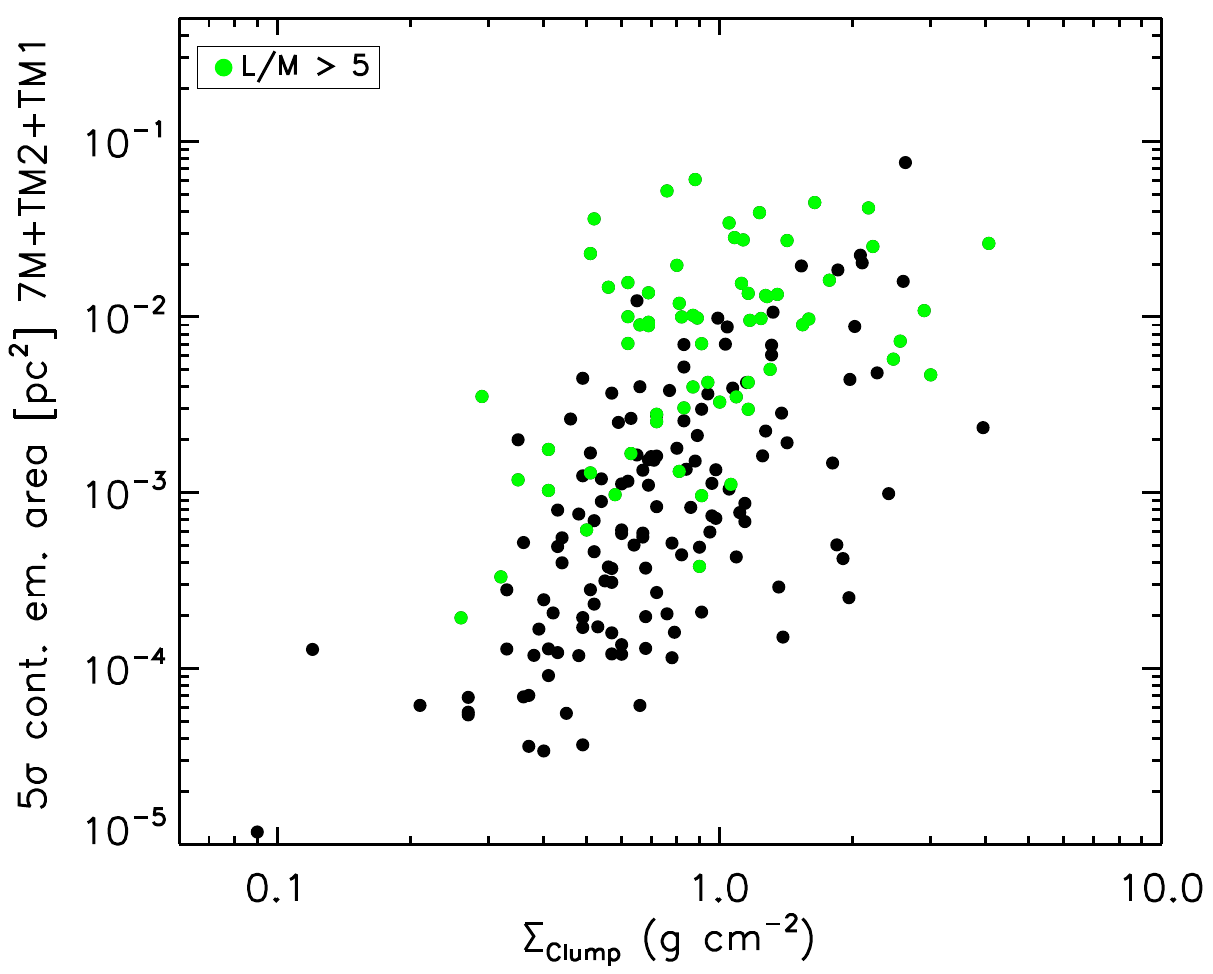}

\includegraphics[width=0.32\textwidth]{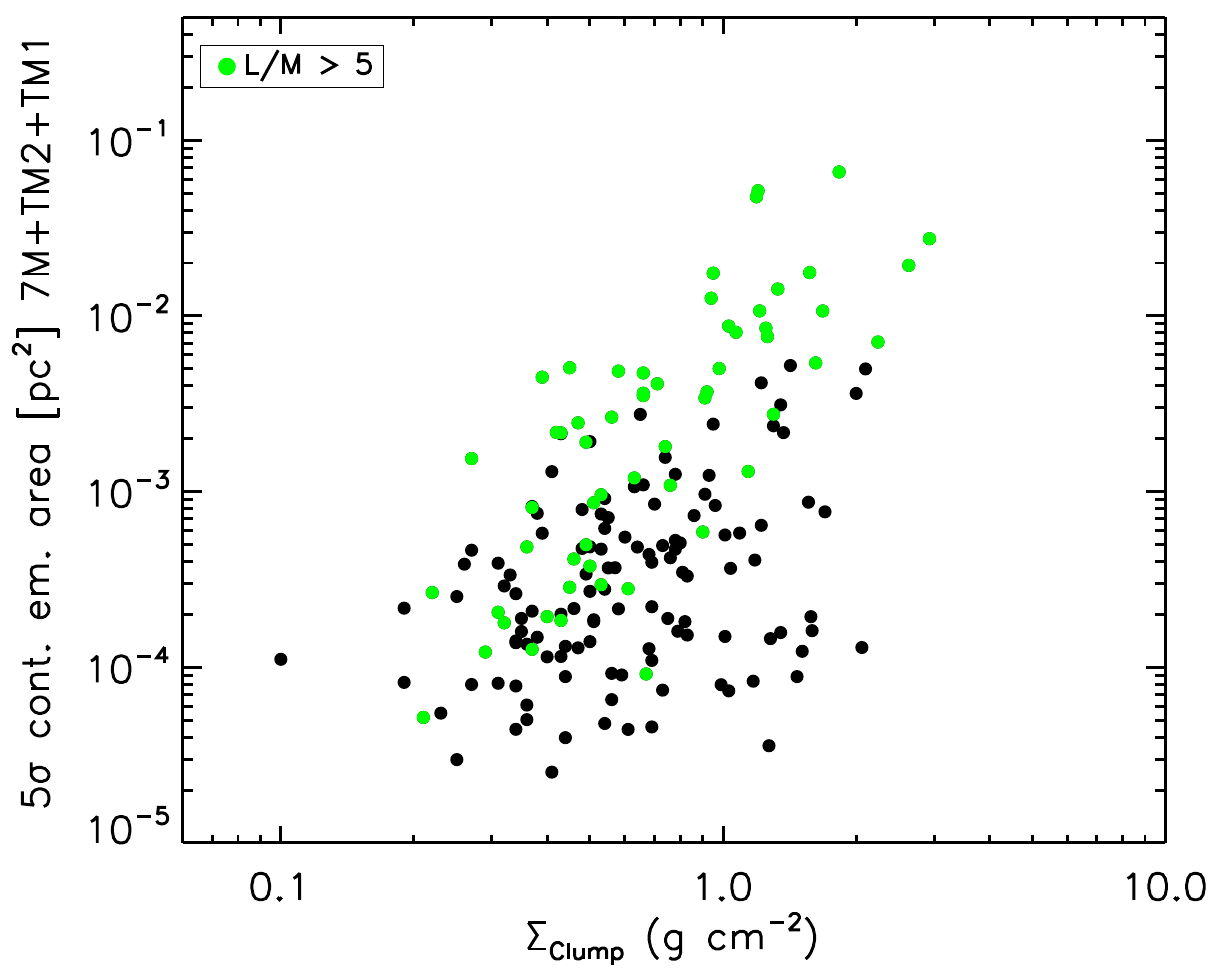} 
\includegraphics[width=0.32\textwidth]{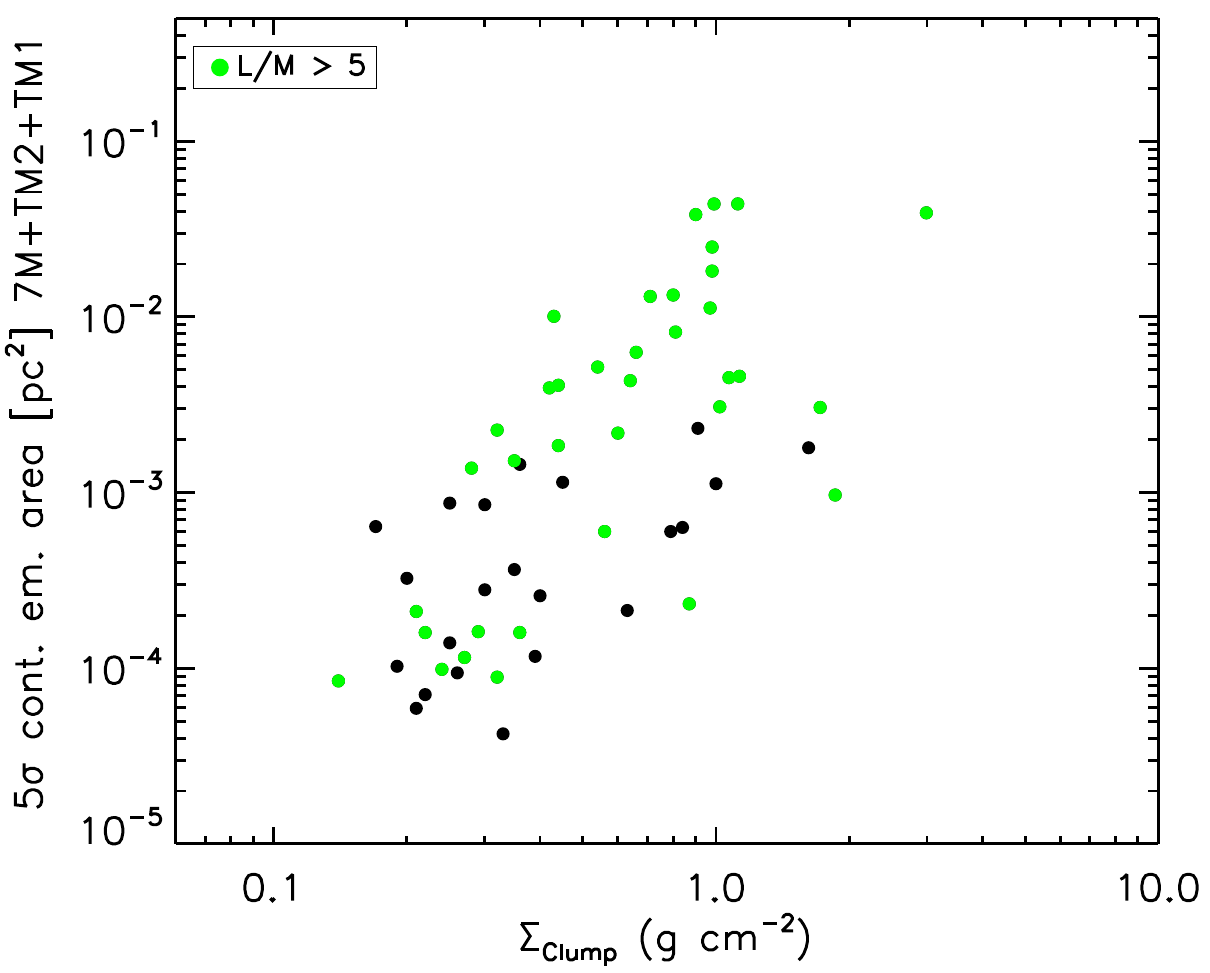} 
\includegraphics[width=0.32\textwidth]{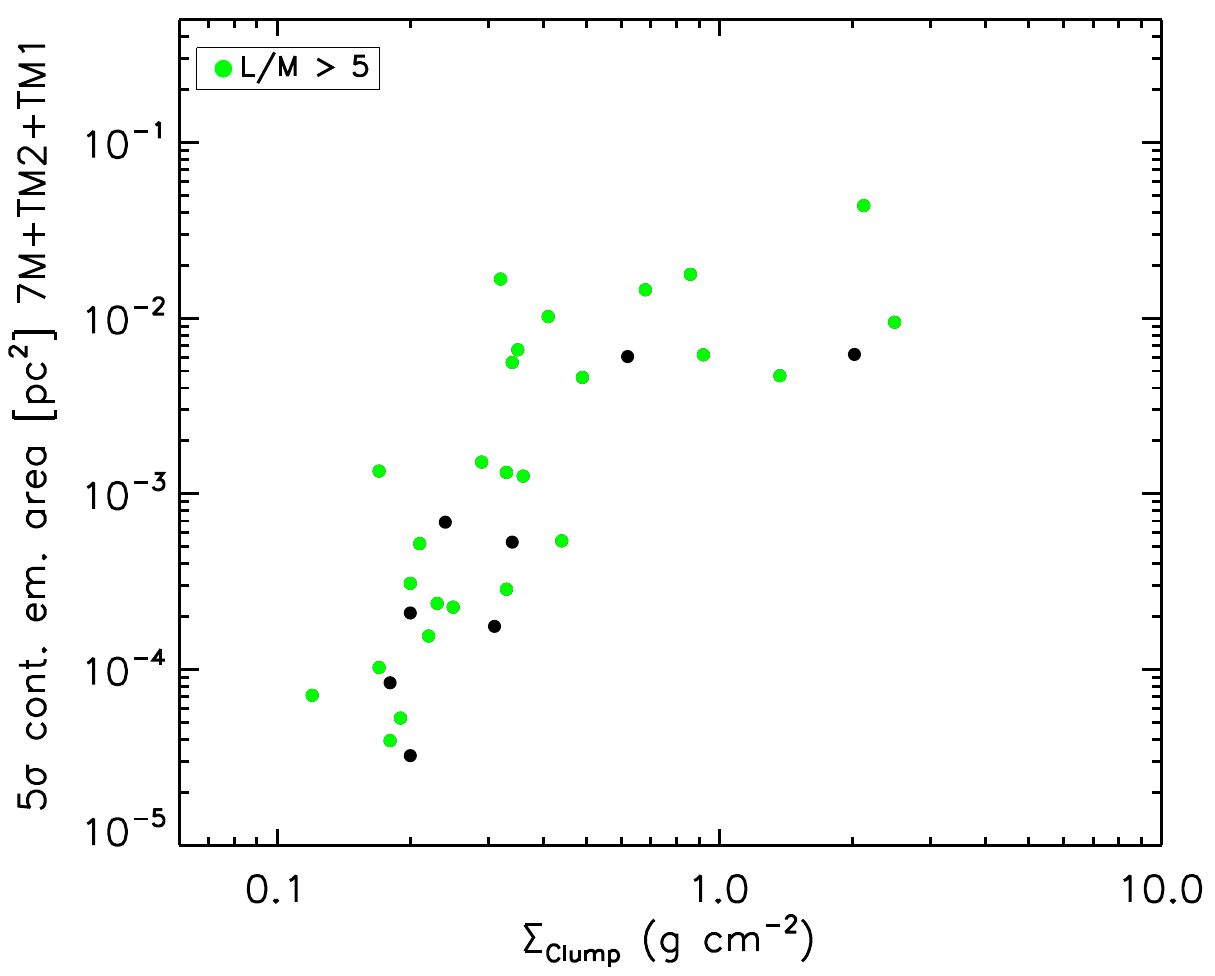}
\caption{1.38mm continuum emission area in 5$\sigma$ contour vs clump surface density, as in Fig.\ref{roi-area_surf_d}, but separately in different distance 1kpc-bins. Bins center distance is from 2.5 to 4.5kpc (top row, left to right), and from 5.5 to 7.5kpc (bottom row, left to right).}
\label{roi-area-surf_d_bins}
\end{center}
\end{figure*}

\begin{figure}[h]
\begin{center}
\includegraphics[width=0.47\textwidth]{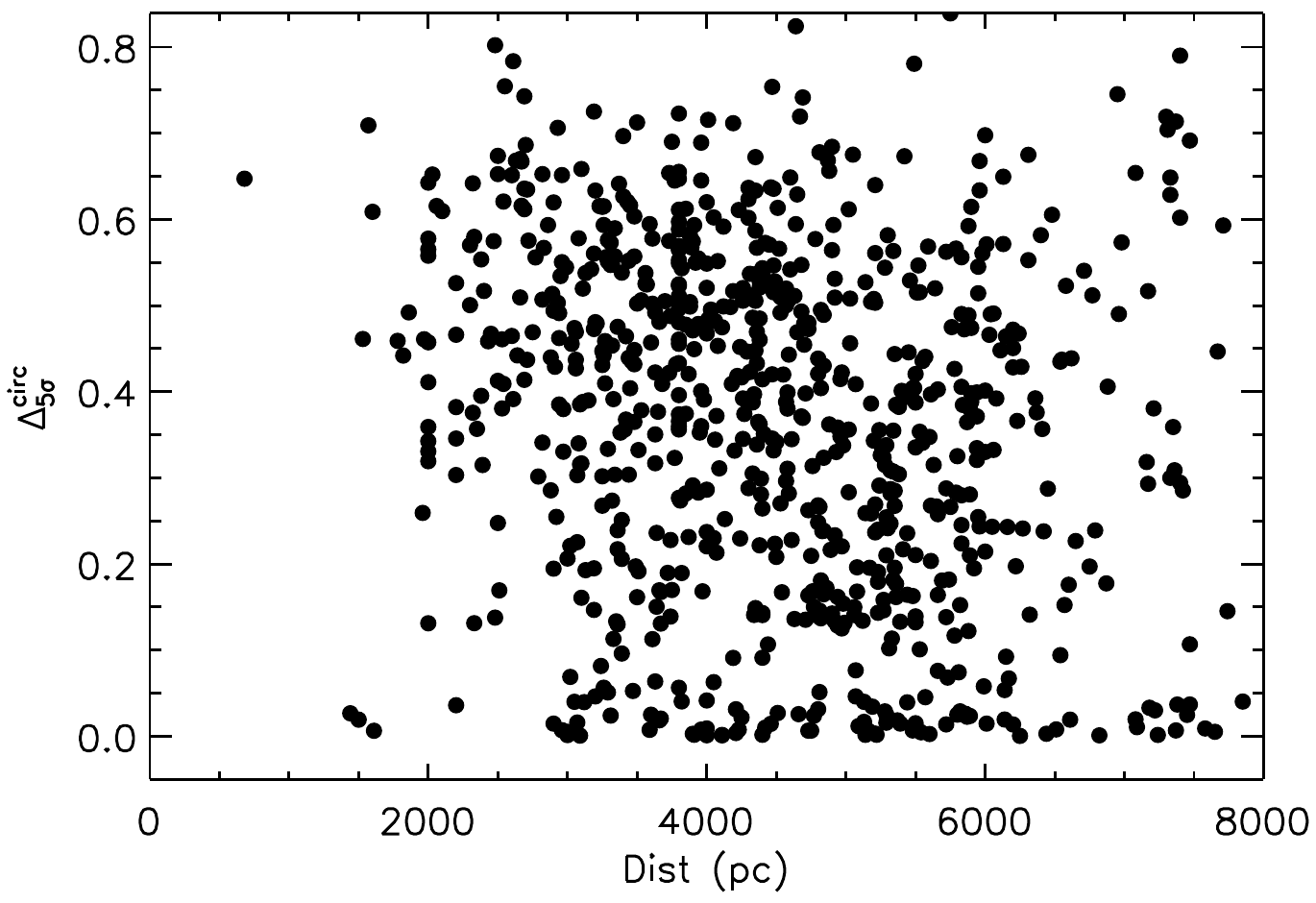} 
\caption{\deltacirc\ parameter for the total RoIs 5$\sigma$ area as a function of clump distance. No relationship can be seen.}
\label{deltacirc_dist}
\end{center}
\end{figure}

\begin{figure*}[h]
\begin{center}
\includegraphics[width=0.32\textwidth]{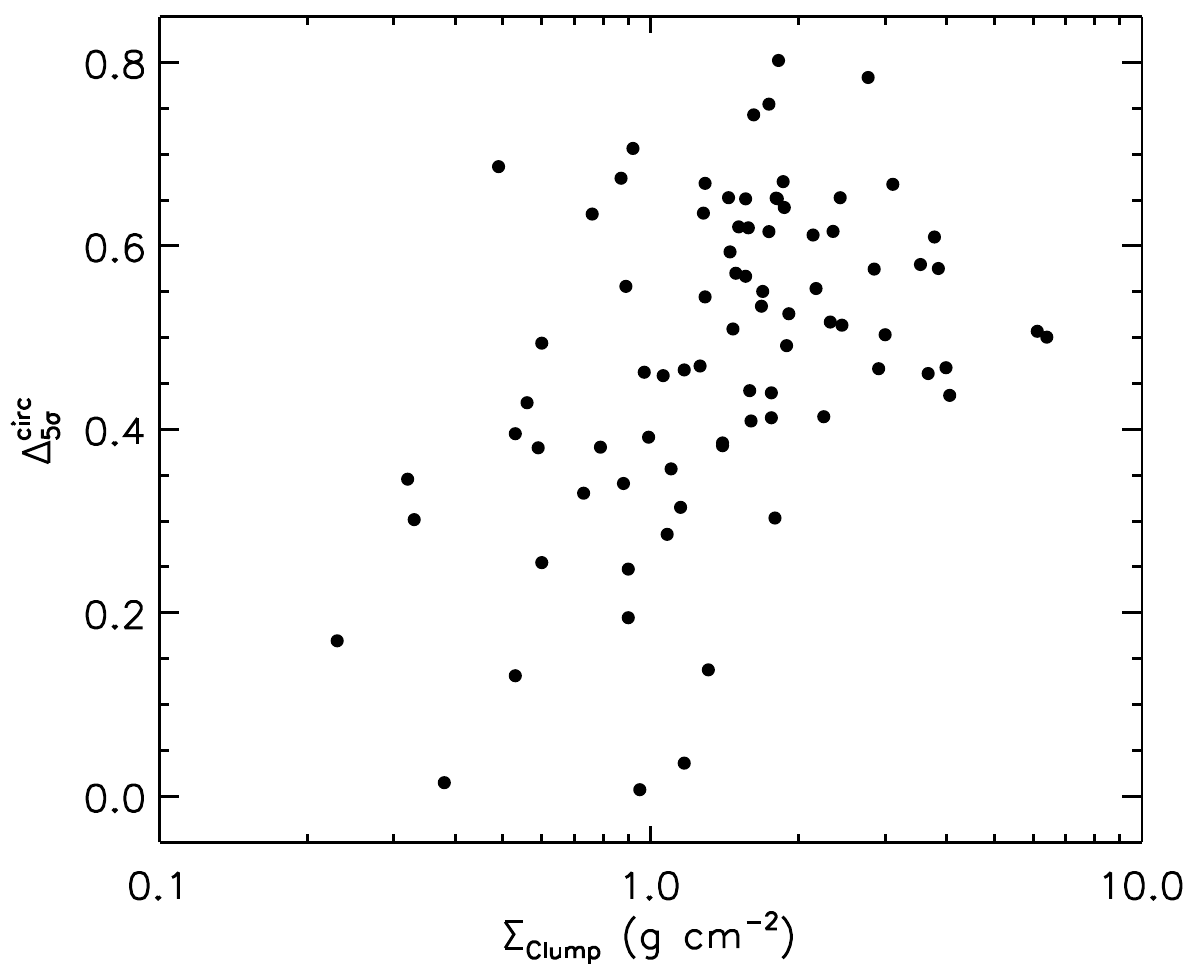} 
\includegraphics[width=0.32\textwidth]{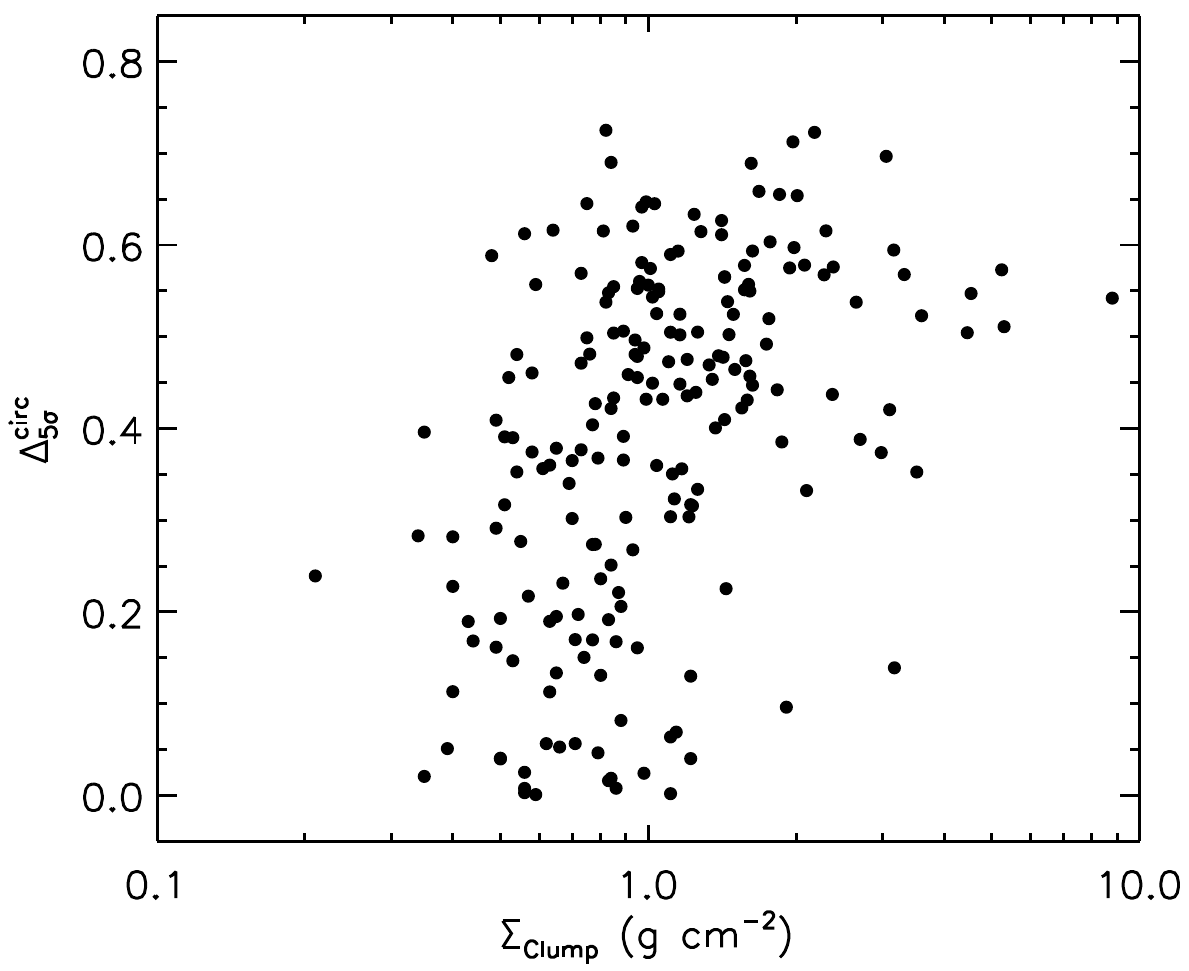} 
\includegraphics[width=0.32\textwidth]{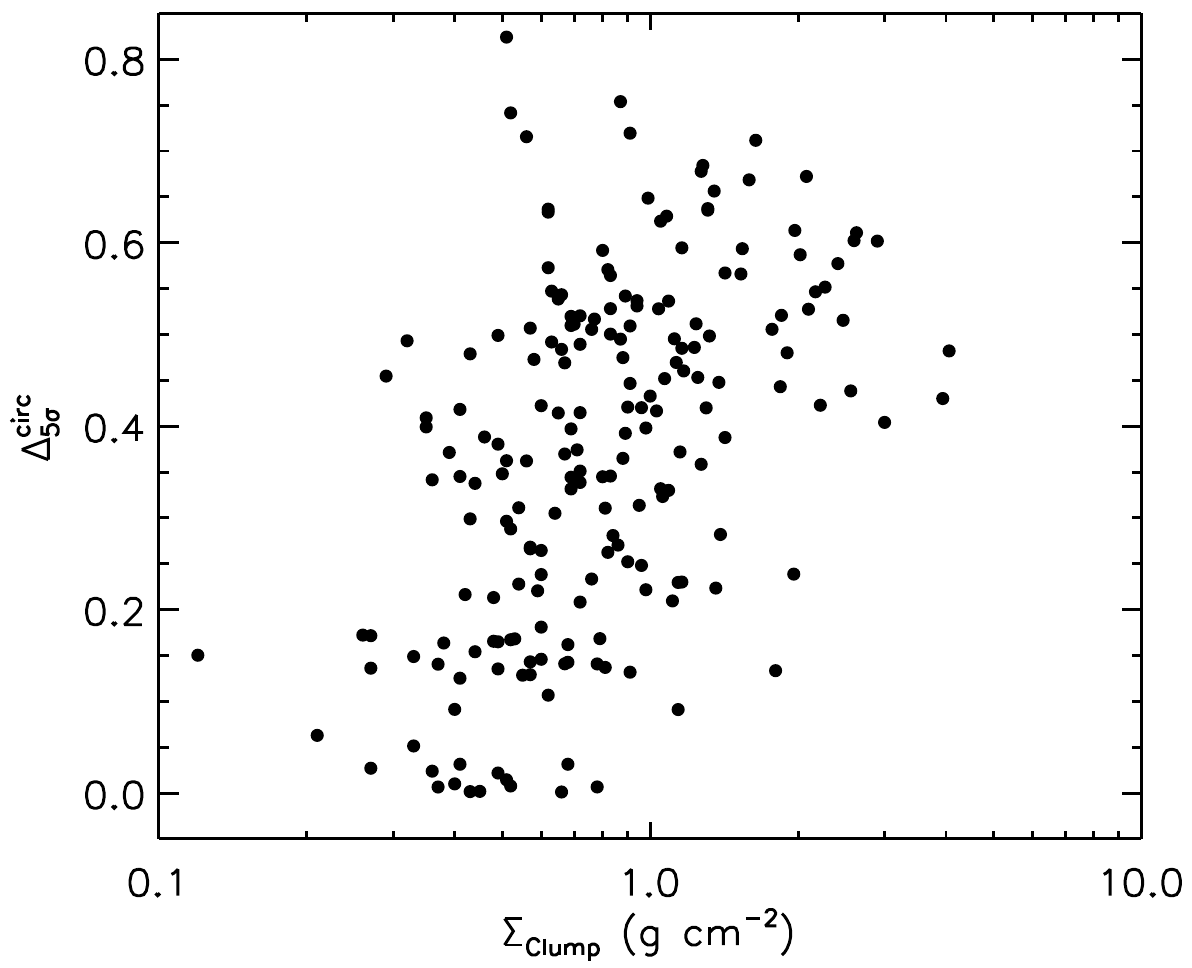}

\includegraphics[width=0.32\textwidth]{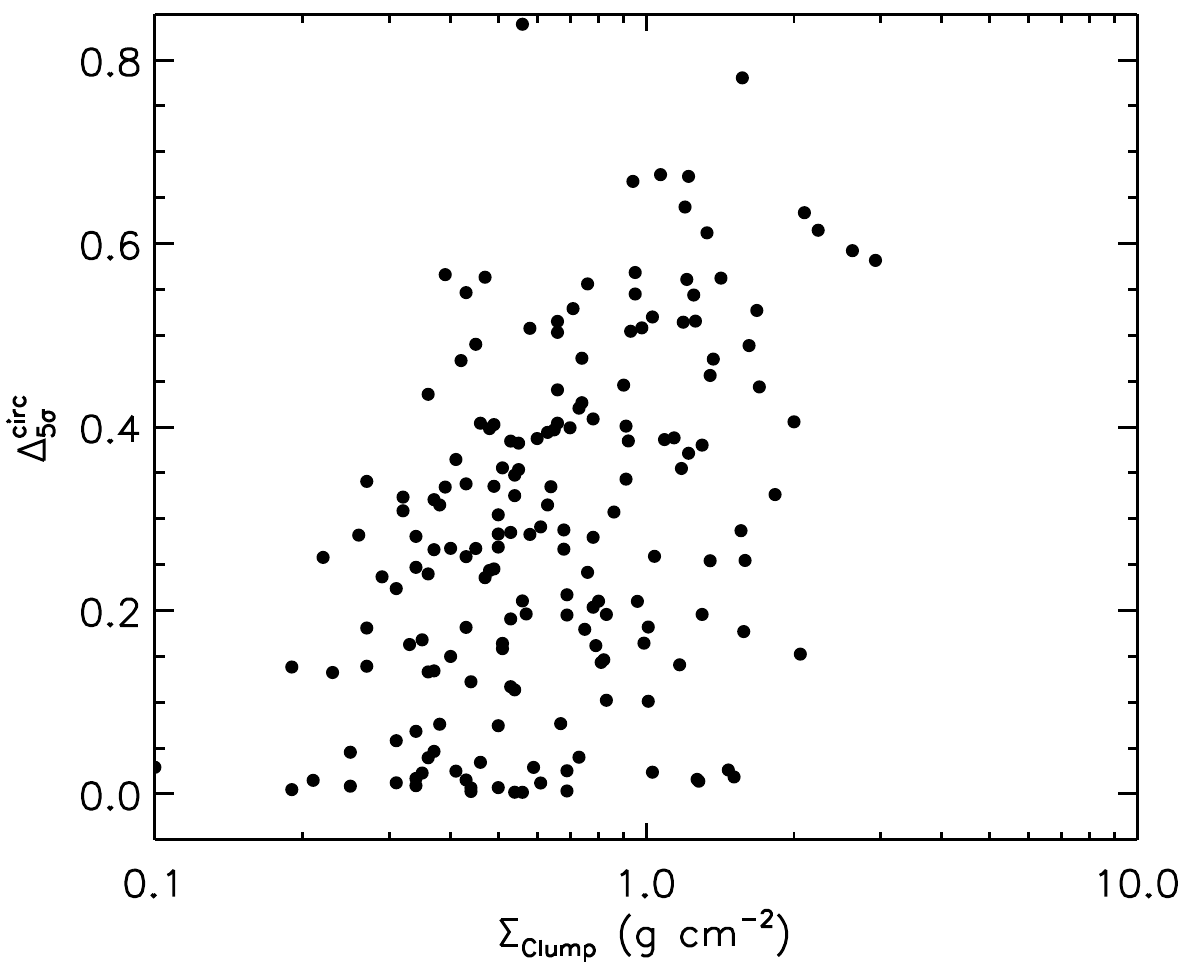} 
\includegraphics[width=0.32\textwidth]{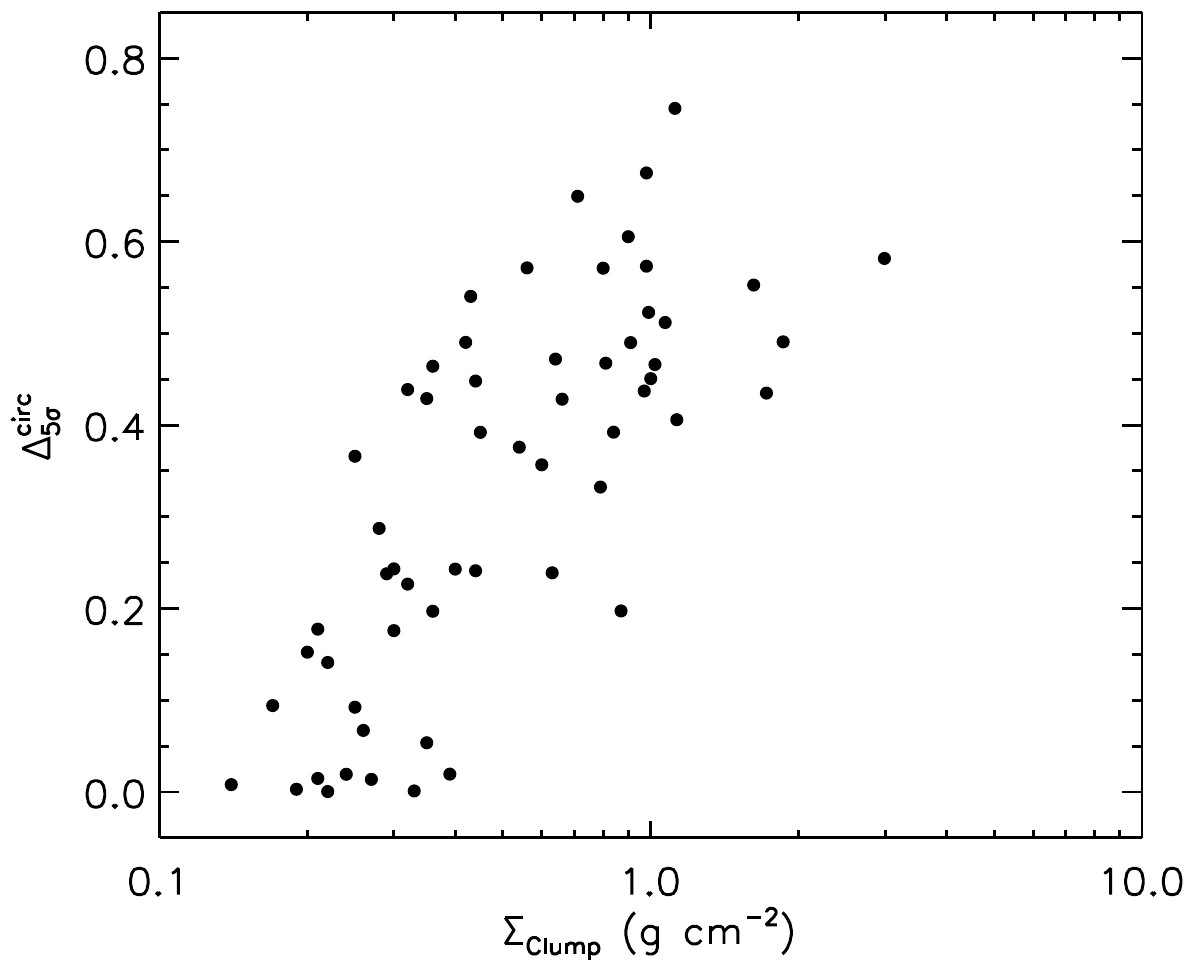} 
\includegraphics[width=0.32\textwidth]{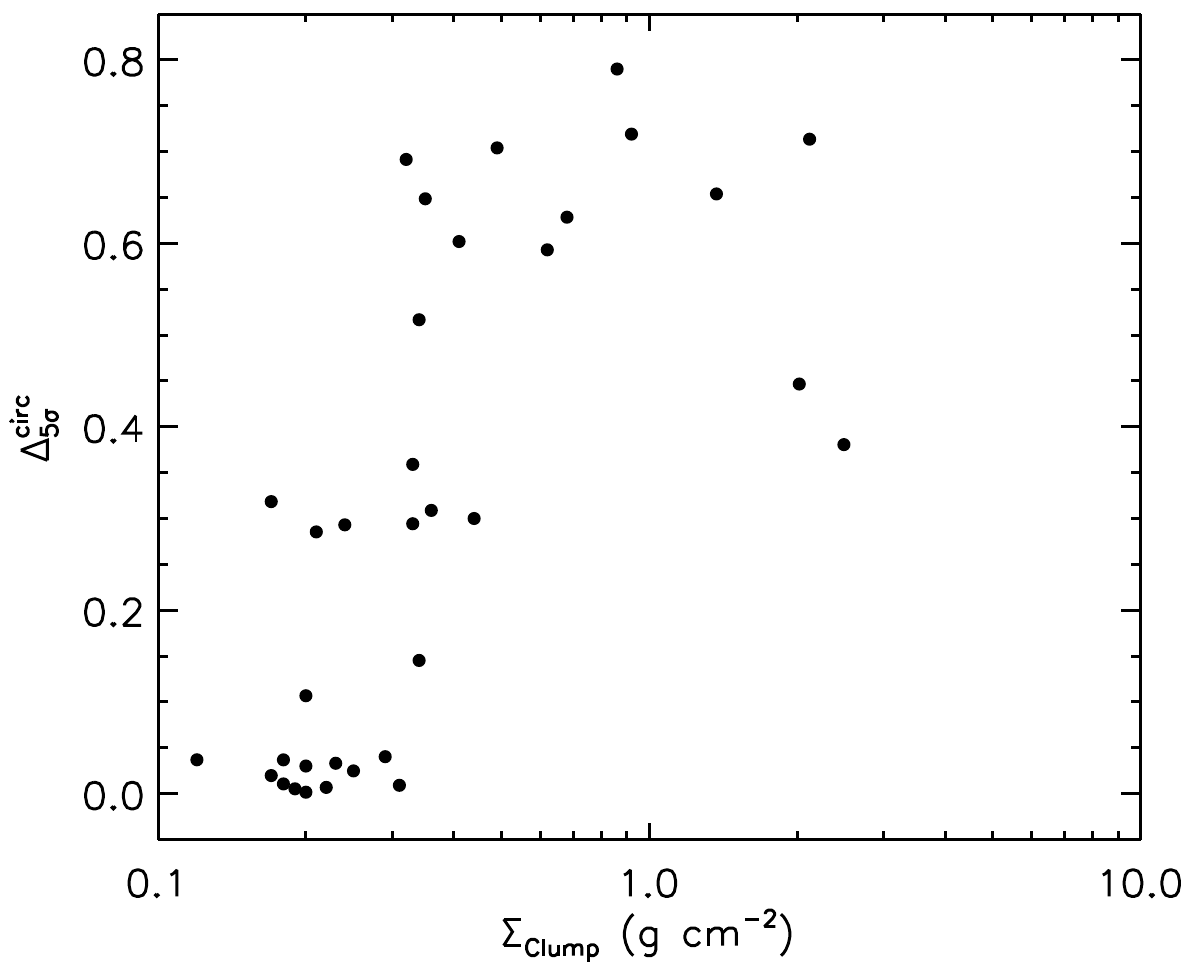}
\caption{Departure from circularity, \deltacirc, of the total of the 5$\sigma$ ALMA emission RoIs in each field as a function of the clump surface density, as in Fig.\ref{deltacirc_surfd_fig}, but separately in different distance 1kpc-bins. Bins center distance is from 2.5 to 4.5kpc (top row, leftto right), and from 5.5 to 7.5kpc (bottom row, left to right).}
\label{deltacirc_surfd_fig_d_bins}
\end{center}
\end{figure*}

\begin{figure}[h]
\begin{center}
\includegraphics[width=0.47\textwidth]{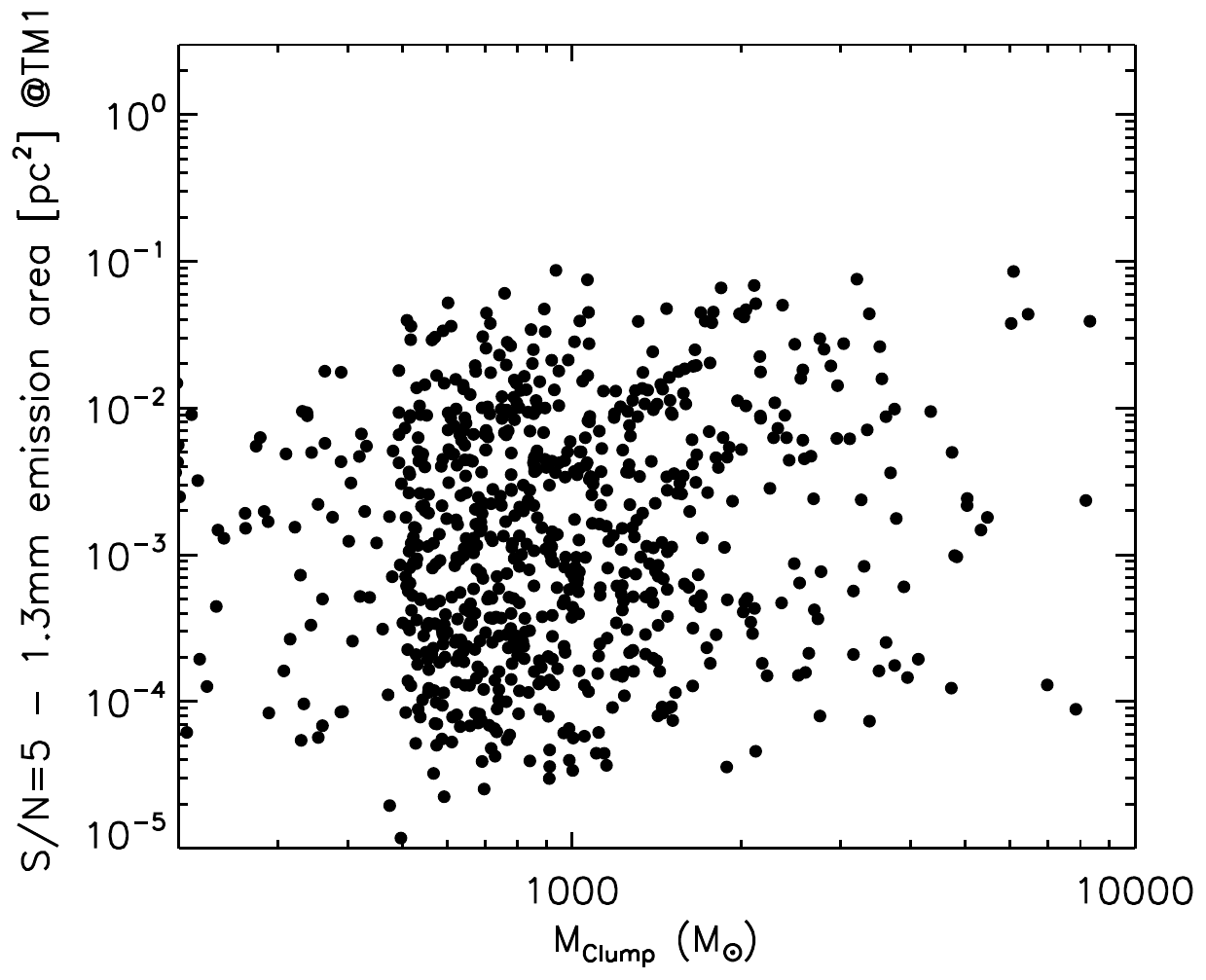} 
\caption{1.38mm continuum emission area in 5$\sigma$ contour of \fres\ images versus clump Mass. No correlation is present.}
\label{roi_area_mass}
\end{center}
\end{figure}

%\section{Morphological parameters of continuum emission}
%\label{appendix_morph_prop}
%
%In the following table we report for each field some of the morphological parameters we have derived in Sect. \ref{cont_morph}. Col. 1 reports the clump running number as in Table \ref{source_table}. Cols. 2-3 report the extent of the emitting area in the \fres\ continuum images in sq. parsecs and in percentage of the field of view. Cols. 4-5 report the \deltacirc\ and \qhull\ parameters for the largest patch of emission area detected above 5$\sigma$.

\end{appendix}

\end{document}